\documentclass{ws-ijmpe}
\usepackage[super,compress]{cite}
\usepackage[dvips,breaklinks]{hyperref}
\hypersetup{colorlinks,urlcolor=black,citecolor=black,linkcolor=black,filecolor=black}
\usepackage{breakurl}
\usepackage[super,compress]{cite}
\usepackage{enumerate}
\usepackage{subfig}
\usepackage{feynmp}
\usepackage{slashed}
\usepackage{rotating}
\usepackage{multirow}
\usepackage{psfrag}
\usepackage{tabularx}
\makeindex
\newcommand{\newc}{\newcommand}
\newc{\beq}{\begin{equation}}
\newc{\eeq}{\end{equation}}
\newc{\barr}{\begin{eqnarray}}
\newc{\earr}{\end{eqnarray}}
\newc{\bea}{\begin{eqnarray}}
\newc{\eea}{\end{eqnarray}}

\def\bfs{\mbox{\boldmath $\sigma$}}
\def\lambf{\mbox{\boldmath $\lambda$}}

	\def\sbf{\mbox{\boldmath $\sigma$}}

\begin{document}

\title{Symmetries in subatomic multi-quark systems}

\author{J.D. Vergados$^1$ and D. Strottman$^2$}

\address{$^1$Physics Department,  University of Ioannina, Ioannina, Gr 451 10, Greece
\footnote{e- mail: vergados@uoi.gr},$^2$LANL, Los Alamos, NM 87544, USA}
\maketitle

\begin{history}
\end{history}

\begin{abstract}
We discus the role of QCD (Quantum Chromodynamics) to low energy phenomena involving  the color-spin symmetry of the quark model. We then combine it with  orbital and  isospin symmetry to obtain wave functions with  the proper permutation symmetry, focusing on  multi quark systems.
\end{abstract}


\ccode{PACS numbers:}

\section{Motivation for contributing  to this volume}
It is with humble feelings that JDV is contributing to this review article in the memory of E. M. Henley. \\His first countenance with Ernie was through his book with H. Frauenfelder \cite{Henley74}. This book had a profound influence on JDV. First, as its title indicates, because of its unifying view of particle and nuclear physics, and second due to its fine interplay of theoretical and experimental physics and, above all, its excellent  exposition of the  symmetries present in subatomic physics, especially the discreet symmetries. No wonder the 3nd edition of this book is still one of the most popular books in the field of particle and nuclear physics. \\ JDV met him for the first time in the office of H. Primakoff, when he was an Assistant Professor of Physics at the University of Pennsylvania. He also met him again when he visited  the University of Washington for a year in 1983, on subbatical from the University of Ioannina in Greece at the invitation of Gerry Miller. During that time Ernie, Takamitsu Oka and JDV collaborated on a number of physics projects, which will be discussed later. This has perhaps been the most enjoyable collaboration in his life, combining Ernie's insight on symmetries, Takamitsu's  talent in arithmetic manipulation , the result of the well known Japanese tradition, and JDV's background on Group Theory. Beyond the specific research done together,  JDV was impressed by Ernie's deep breadth of physics, his kind personality and his wide view of science and education, combined with excellent managerial skills, which he obtained as he was moving up the academic ladder. The latter proved  extremely useful, when JDV became a Rector of the  Universityof Ioannina 8 years later.\\
Motivated by the above,   some aspects  of the subject he enjoyed the most, namely symmetries,  will be explored in this article, focusing on the role of Quantum Chromodynamics (QCD) at low energies and, in particular, on understanding the behavior of multiquark systems .
\\
\section{Introduction}

Multiquark systems, like the pentaquark and tetraquark, have already been found to exist. Why not multi quark systems in he nucleus? So we will examine the possibility of the presence of six quark clusters in the nucleus, a much more complex problem. Such clusters, if present with a reasonable probability in the nucleus, may contribute  to various proceses, like neutrinoless double beta decay mediated by heavy neutrinos or other exotic particles. In conventional nuclear physics the relevant nuclear matrix elements are suppressed due to the presence of the nuclear hard core.
In this presence of such clusters, however,   the interacting quarks are in the same hadron. So one can have a contribution even in the
case of of a $\delta$-function interaction \cite{JDV85}. Symmetries, of course, play a crucial role in reliably  estimating the probability
of finding such six quark clusters in the nucleus .

\section{Quantum chromodynamics $(QCD)$}
\begin{figure}[h!]
\begin{center}
\includegraphics[width=10cm]{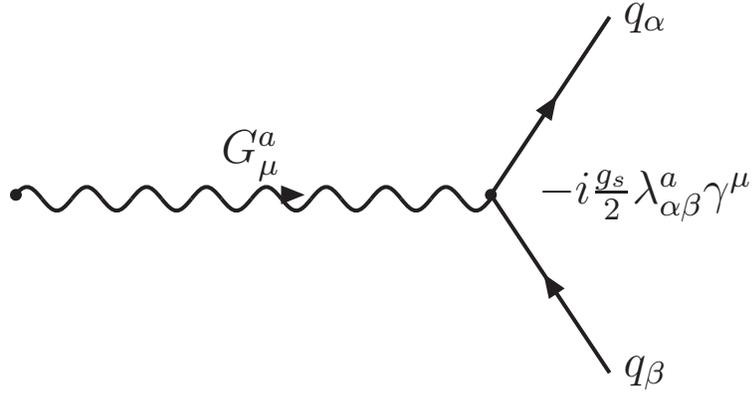}
\caption{\footnotesize \it
A Feynman diagram indicating the gluon mediated interaction} \index{gluon exchange}
\label{Fig:GluonInt}
\end{center}
\end{figure}

\index{quantum chromodynamics}
\index{QCD}
It is well known (see, e.g. \cite{Langacker16}, \cite{Vergdos17}, \cite{Vergdos16}) that the strong interaction is governed by the group $SU_c(3)$ of the standard model with a set of traceless generator generators, \index{Gell-Mann matrices} the Gell-Mann matrices $\lambda^a$, such that $tr(\lambda^a)^2=2$, $a=1\cdots 8$. It is also known  that associated with this symmetry we have a set of 8 gauge bosons\footnote{Sometimes the Lorentz index is understood and omitted and the gluons are written in the form of a $3\times3$ matrix $G^{\mu}_{\lambda}$ with $\mu$ and $\lambda$ color degrees of freedom taking values, e.g. $r,\,g,\,b$. The two diagonal matrices have the form diag(1,-1,0) and $(1/\sqrt{3})$diag(1,1,-2).} $G^a_{\lambda}$, with   $\lambda$ a Lorentz index and $a=1\cdots 8$, the {\bf gluons}. The interactions mediated by the gluons is given in Fig. \ref{Fig:GluonInt}.

\begin{figure}[h!]
\begin{center}
\subfloat[]
{
\includegraphics[width=0.4\textwidth]{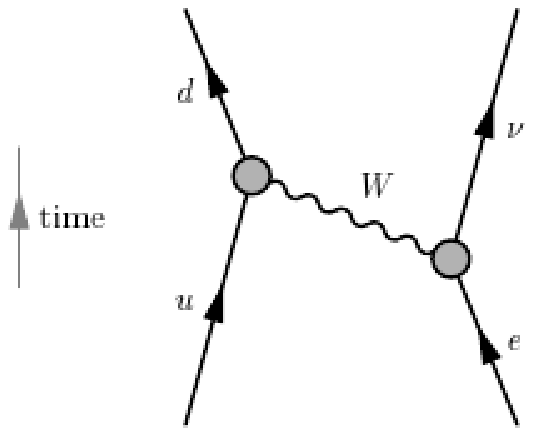}
}
\subfloat[]
{
\includegraphics[width=0.5\textwidth]{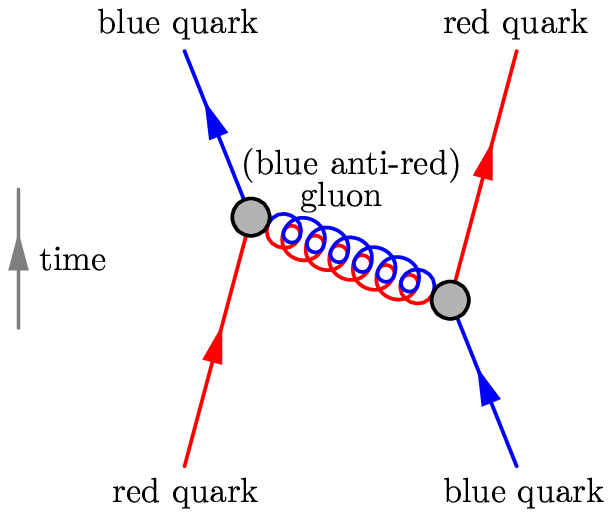}
}
\caption{\footnotesize \it
A comparison of a $W$ exchange (weak interaction) (a) and a gluon exchange (strong interaction) (b). Note the propagation of color via the gluons.}
\label{Fig:WQCD}
\end{center}
\end{figure}

 A comparison of the gluon exchange and the $W$ exchange in weak interactions is given in Fig. \ref{Fig:WQCD}

\begin{figure}[h!]
\begin{center}
\subfloat[]
{
\includegraphics[width=0.4\textwidth,height=0.2\textwidth]{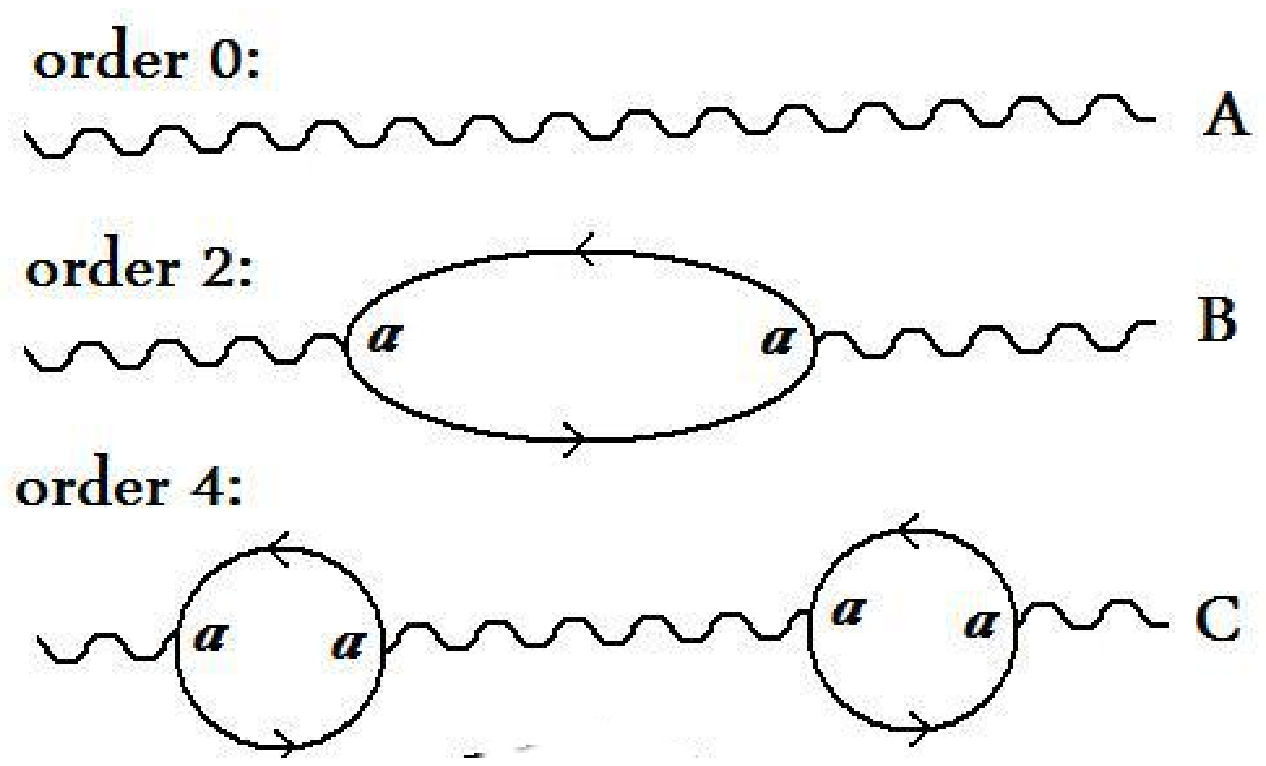}
}
\subfloat[]
{
\includegraphics[width=0.5\textwidth,height=0.15\textwidth]{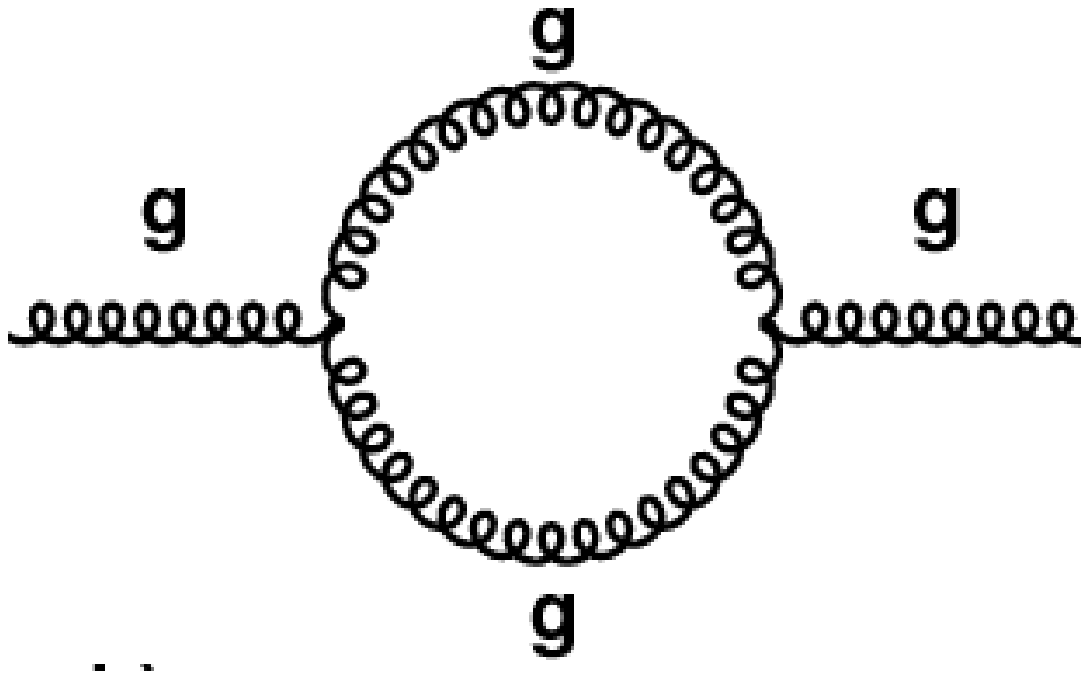}
}\\
\caption{\footnotesize {\it
In the one gluon exchange of Fig. \ref{Fig:WQCD}(b) at some point  becomes cheaper to create quark-antiquark pairs out of the vacuum (a) forming loops. As a result the interaction between  the quarks is screened and decreased. This is analogous to QED and leads to a decrease in the interaction between to charged particles in dielectrics. In QCD, however, we also have virtual gluon creation out of the vacuum. For some cubic and quartic couplings, which are present in a non Abelian theory, this is exhibited in (b). Such virtual gluons  lead to anti-screening, \index{screening} \index{anti-screening} i.e. the interaction increases, "color dielectric". This has dramatic effects  at low energies (large distances). As a result we have confinement of colored particles.}}
\label{Fig:VacuumBreak}
\end{center}
\end{figure}

\begin{figure}[h!]
\begin{center}
\subfloat[]
{
\rotatebox{90}{\hspace{2cm} $\alpha_s \rightarrow$}
\includegraphics[width=0.5\textwidth]{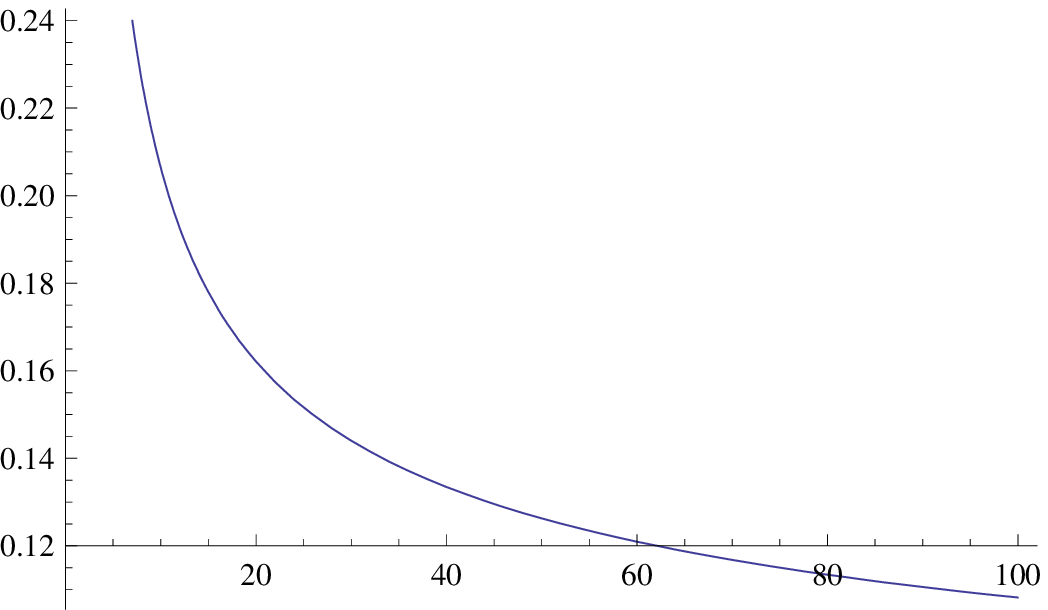}
}
\subfloat[]
{
\rotatebox{90}{\hspace{3 cm} $\alpha_s \rightarrow$}
\includegraphics[width=0.4\textwidth]{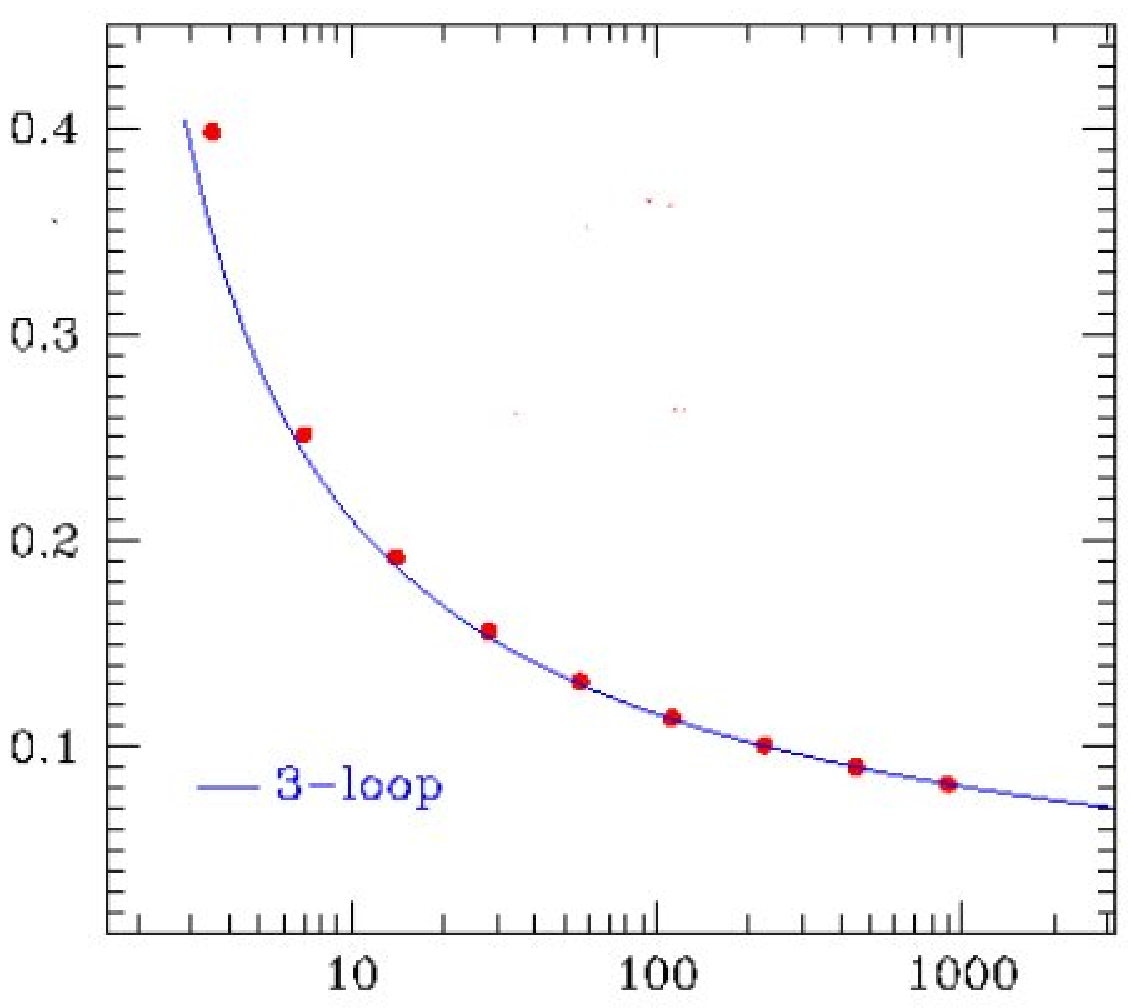}
}\\
{\hspace{2cm} $E\rightarrow$ GeV}

\caption{\footnotesize \it
\\The strong coupling strength deceases as a function of the energy scale. At large scale it becomes very weak (asymptotic freedom), while at small scales it becomes very strong (imposed slavery). Shown is the prediction of the simple formula (Eq. \ref{Eq:alphath})   (a) and the experimental data of the ALPHA collaboration (b).}
\label{Fig:gStrong}
\end{center}
\end{figure}

 The gluons remain massless even after  the spontaneous symmetry breaking \index{spontaneous symmetry breaking}. Thus this symmetry remains unbroken. Since, however, it is not Abelian, it does not lead to any long range force. The emerging theory is called {\bf Quantum Chromodynamics}. This theory is described by the Lagrangian:
\beq
{\cal L}_c=-\frac{1}{4}F^{j}_{\mu \nu}F^{j\mu \nu}+\sum_r {\bar q}_{r\alpha}\gamma^{\mu} \left (D_{\mu}\right )^{\alpha}_{\beta}q_{r}^{\beta},
\eeq
where $r$ is a  quark flavor index, $\alpha$ and $\beta$ are color indices and $D^{\mu}$ is the covariant derivative
\beq
\left (D_{\mu}\right )^{\alpha}_{\beta}=\partial_{\mu}\delta^{\alpha}_{\beta}-i g_s G^{j}_{\mu}\frac{1}{2}\left (\lambda^{j}\right )^{\alpha}_{\beta}
\eeq
 and
\beq
F^{j}_{\mu \nu}=\partial_ {\mu}G^j_{\nu}-\partial_ {\nu}G^j_{\mu}-g_s  f_{jk\ell}G^{k}_{\mu}G^{\ell}_{\nu},
\eeq
where $g_s$ is the croup gauge coupling constant\footnote{ We use here $g_s$ instead of $g_3$ commonly used in the standard model.}  and  $f_{ijk}$ are the structure constants of $SU(3)$ given by:
\beq
\left [\lambda^j, \lambda^k \right]= 2 i f_{j k \ell}\lambda ^{\ell}.
\eeq
Note that, in contrast to the other unbroken Abelian $U_{EM}(1)$ symmetry, the presence of the SU(3) $F^2$ term leads to three and four point gluon self couplings \index{gluon self-couplings}, which is due to the couplings $f_{jk\ell}$. This effect results in some technical difficulties in $QCD$ as compared to the electrodynamics.
\begin{itemize}
\item The gluons tend to polarize the medium. In other words it becomes energetically cheaper to create quark anti-quark pairs out of the vacuum (see Fig. \ref{Fig:VacuumBreak}).
\item The polarization of the vacuum \index{vacuum polarization} is not like the dipole creation familiar from dielectrics, which tends to decrease the interaction between two charges. Instead it behaves more like quadrupoles in dielectrics, hard to make in macroscopic scale, which tend to increase the interaction between the colored quarks (see Fig. \ref{Fig:gStrong}). In other words at high energies, or small distances, the quarks behave almost like being free, i.e. we have {\bf asymptotic freedom}. On the other hand at low energies (large distances of the order of fm) the interaction becomes very strong. Thus we encounter {\bf confinement}, perpetual quark slavery. \index{confinement} \index{quark confinement} \index{quark slavery}
\item Since the interaction becomes strong at low energies, one cannot invoke perturbation theory. So multi-gluon exchange or pair creation out of the vacuum diagrams  become important (see Fig. \ref{Fig:VacuumBreak}).
\end{itemize}
The strength of the interaction depends on the energy and the number $n_f$ of active quark pairs:
\beq
 \alpha_s(E)=\frac{4 \pi}{\left (-33+2 n_f\ln{\frac{E^2}{\mu^2}}\right)},
 \label{Eq:alphath}
 \eeq
 $$ \mbox{ where }\{\begin{array}{l}n_f=\mbox{the number of quarks active  pair production (up to 6)}\\\mu=\mbox{experimentally determined scale}\approx0.2\mbox{GeV}\end{array}.$$\\
Another common parametrization is given by
\beq
 \alpha_s(q^2)=\frac{ \alpha_s(\mu^2)}{1+\beta \alpha_s(\mu^2) \left (\ln{\frac{q^2}{\mu^2}}\right)},\, \beta=\frac{33-2 n_f}{12\pi}.
 \label{Eq:alphath1}
 \eeq
These formulas yield $ \alpha_s=0.12$ at $q^2=(100\mbox{MeV})^2$.
\index{asymptotic freedom}
\section{The color structure of the one gluon exchange potential involving quarks}
\label{sec:ColorStructure}
\begin{figure}[h!]
\begin{center}
\subfloat[]
{
\includegraphics[width=0.7\textwidth]{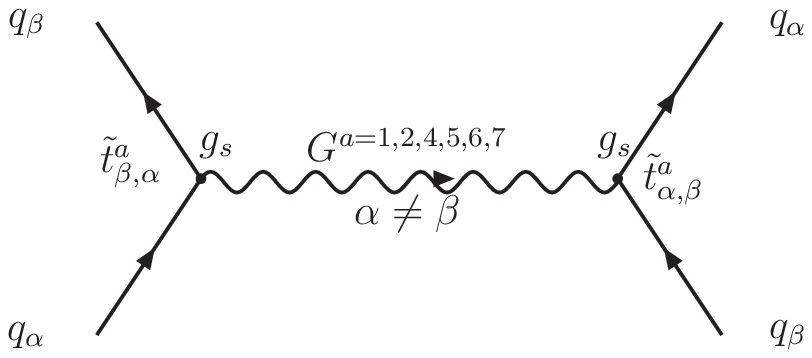}
}\\
\subfloat[]
{
\includegraphics[width=0.5\textwidth]{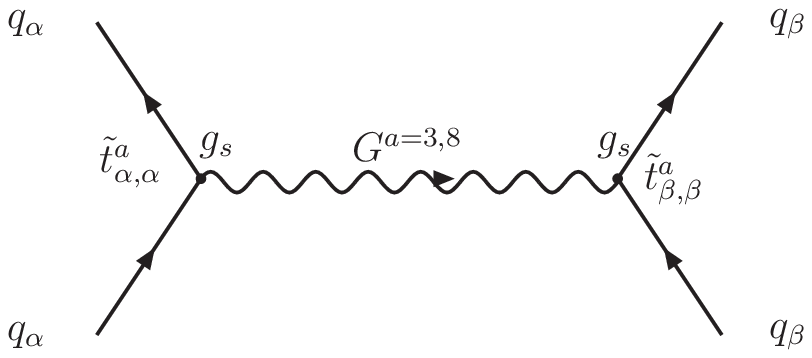}
}
\subfloat[]
{
\includegraphics[width=0.5\textwidth]{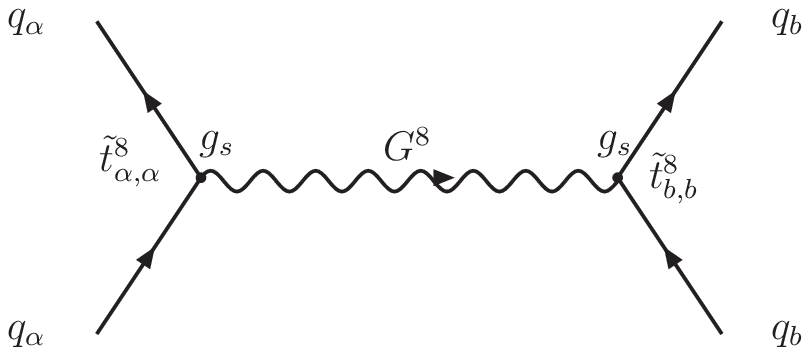}
}
\\
\caption{\footnotesize \it
\\The strong interaction mediated by gluons, with the Lorentz index suppressed. Shown are the color changing gluon interaction ($\tilde{t}^a_{\alpha \beta},\,\alpha\ne \beta$, which destroys a quark $\beta$ and converts to a quark $\alpha$) (a), the color conserving interaction involving quarks other than $b$ ($\tilde{t}^{3,8}_{\alpha \beta},\alpha,\beta=r,g$) (b) and the color conserving interaction in which at least one of the quarks is blue ($\tilde{t}_{\alpha \alpha}^{3,8},\,\alpha=r,g,b$) (c). Note that all operators are normalized to unity $tr(\tilde{t}^a \tilde{t}^b)=\delta_{a,b}$}
\label{Fig:2qint}
\end{center}
\end{figure}

Let us suppose\footnote{For more details and explanations on this and the next two sections see a recent textbook \cite{Vergdos17}.} that the  two quarks are $q_{\alpha}(1)q_{\beta}(2)$, where $\alpha$ and $\beta$ are color indices, taking values $r,g,b$, and (1) (2) label the particles. For simplicity of notation we will drop the particle index, with the understanding that particle one will be first and particle second, by writing $q_{\alpha}(1)q_{\beta}(2)\Leftrightarrow \alpha\beta=rr,rg,\mbox{etc}$.
The Feynman diagrams leading to the interaction between two quarks are exhibited in Fig. \ref{Fig:2qint}.

Ignoring the overall strength (color charge, $g_r=g_s=4 \pi \alpha_s$ ) and omitting the gluon propagator,  the interaction between similar colors is $$ \langle rr|V|rr\rangle=\langle gg|V|gg\rangle=\left (\frac{1}{6}+\frac{1}{2}\right )=\frac{2}{3}$$
 The first contribution comes from the exchange of the  gluon $G^8$ and the second one comes from $G^3$ (see Fig. \ref{Fig:2qint}b). Similarly from the exchange of the  gluon $G^{8}$ (see Fig. \ref{Fig:2qint}c) we get
$$ \langle bb|V|bb\rangle=\frac{1}{6} 2^2 =\frac{2}{3}$$
In a similar fashion we get
$$ \langle gr|V|gr\rangle=\langle rg|V|rg\rangle=\left (\frac{1}{6}-\frac{1}{2}\right )=-\frac{1}{3}\mbox{(see Fig. \ref{Fig:2qint}b) }$$
$$ \langle rb|V|rb\rangle=\langle br|V|br\rangle=\frac{1}{6} (-2) =-\frac{1}{3}\mbox{(see Fig. \ref{Fig:2qint}c) }$$
Furthermore
$$\langle \alpha \beta|V|\beta \alpha\rangle=1,\, \alpha \ne \beta \mbox{(see Fig. \ref{Fig:2qint}a) }$$
In this last case the off diagonal gluon is exchanged, $G^{1,2,4,5,6,7}$. All other matrix elements are zero.\\ The same results hold in the case of two antiquarks.

 The above product of two quarks constitutes a basis in color space, $\underline{3}\otimes \underline{3}$, which yields a reducible nine dimensional representation of $SU(3)$. Diagonalizing this interaction in the nine dimensional space one finds thee states with eigenvalue $-\frac{4}{3} $ and six states with eigenvalue $\frac{2}{3} $  as follows:
$$ -\frac{4}{3}  \Leftrightarrow \tilde{b}=\frac{1}{2}\left(rg-gr \right ),\tilde{g}=\frac{1}{2}\left(rb-br \right ),\tilde{r}=\frac{1}{2}\left(gb-bg \right )$$
$$ \frac{2}{3} \Leftrightarrow \frac{1}{2}\left(rg+gr \right ),\,\frac{1}{2}\left(rb+br \right ),\,\frac{1}{2}\left(gb+bg \right ),\,rr,\,gg,\,bb .$$
The first is an irreducible three-dimensional \index{irreducible SU(3) representations} \index{color anti-triplet} \index{color sextet} representation (anti-triplet) and the second is a six dimensional representation (sextet), which is symmetric in the color indices.

One uses the anti-triplet and the sextet as a basis and thus the color interaction takes the form
$$\lambf_1.\lambf_2=\frac{1}{2}\left( C(\lambda,\mu)-2C(1,0)\right)$$
where $\lambda$ and $\mu$ specify the irreducible representation of $SU(3)$, which are non negative integers.
$ C(\lambda,\mu)$ is the value of the Casimir operator of $SU(3)$, given by:
$$ C(\lambda,\mu)=\frac{2}{3}\left(\lambda^2+\mu^2+\lambda \mu +3(\lambda+\mu)\right)$$
Thus:
$$\mbox{triplet:}(\lambda,\mu)=(1,0)\rightarrow C(\lambda,\mu)=\frac{8}{3}$$
$$\mbox{anti-triplet:}(\lambda,\mu)=(0,1)\rightarrow C(\lambda,\mu)=\frac{8}{3}\rightarrow\lambf_1.\lambf_2=-\frac{4}{3}$$
$$\mbox{sextet:}(\lambda,\mu)=(2,0)\rightarrow C(\lambda,\mu)=\frac{20}{3}\rightarrow\lambf_1.\lambf_2=\frac{2}{3}$$

Let us now examine the interaction between quarks and antiquarks. The antiquark viewed as the antisymmetric combination of two quarks given above. Than we can evaluate the matrix element of the gluon between the relevant three quark states, noting that one of the quarks of the antisymmetric combination is a spectator (not interacting). Thus we find the non vanishing  independent ME involving different colors are:
$$ \langle r \bar{b}|V|  r \bar{b}\rangle=\frac{1}{3},\,\langle r \bar{g}|V| r \bar{g}\rangle=\frac{1}{3},\,\langle g \bar{b}|V|  g \bar{b}\rangle=\frac{1}{3}$$
For the similar color combination we get:
$$ \langle r \bar{r}|V|  r \bar{r}\rangle=\langle g \bar{g}|V|  g \bar{g}\rangle=\langle b \bar{b}|V|  b \bar{b}\rangle=-\frac{2}{3},$$ $$\langle r \bar{r}|V|  g \bar{g}\rangle=\langle r \bar{r}|V|  b \bar{b}\rangle=\langle g \bar{g}|V|  b \bar{b}\rangle-1$$
Note the change in sign compared to the $q-q$ interaction. This usually is attributed to the "opposite color charge of anti-quarks", but note the notational difference in the last equation. We have seen, however, that it comes as a transformation from the picture of anti-quarks as an antisymmetric combination of quarks.\\
The product of a quark and anti-quark transforms under $SU(3)$ like $\underline{3}\otimes \underline{3}^{*}$, which is reducible. One can easily see that the above above six combinations $q_{\alpha} \bar{q}_{\beta},\, \alpha\ne\beta$ transform as the six members of the octet. Diagonalizing the above matrix in the case of similar color indices we find the eigensolutions:
$$\epsilon_0=-\frac{8}{3}\leftrightarrow r_0=\frac{1}{\sqrt{3}}( |r \bar{r}\rangle+ |g \bar{g}\rangle+b \bar{b}\rangle),$$
$$\epsilon_1=\frac{1}{3}\leftrightarrow r^1_8=\frac{1}{\sqrt{3}}( |r \bar{r}\rangle- |g \bar{g}\rangle),$$
$$\epsilon_2=\frac{1}{3}\leftrightarrow r^2_8=\frac{1}{\sqrt{6}}( |r \bar{r}\rangle+ |g \bar{g}\rangle-2b \bar{b}\rangle)$$
The first transforms like the singlet, while the other two are the additional members of the octet. The similarity with the expressions of the $SU(3)$  should not come as a surprise.

In summary we have seen that: \index{quark interactions}
  $$\underline{3}\otimes \underline{3}^*=\underline{1}+\underline{8}$$
	As a result we can write the quark antiquark interaction as :
$$\lambf_1.\lambf_2=\frac{1}{2}\left( C(\lambda,\mu)-C(1,0)-C(0,1)\right)$$
$$\mbox{singlet:}(\lambda,\mu)=(0,0)\rightarrow C(\lambda,\mu)=1\rightarrow\lambf_1.\lambf_2=-\frac{8}{3}$$
$$\mbox{octet:}(\lambda,\mu)=(1,1)\rightarrow C(\lambda,\mu)=\frac{18}{3}\rightarrow\lambf_1.\lambf_2=\frac{1}{3}$$

Before concluding this section we should mention that in the context of the  one gluon exchange
 potential between the above allowed  color combinations is proportional to $ \lambf_1.\lambf_2 {\alpha_s}$

\section{Approximations at low energies-Interaction potentials between quarks}
\label{sec.potentials}
 We know that the quarks do not appear free and all the observed hadrons are colorless. So all experimental information regarding quarks is necessarily indirect and complicated manifestations of chromodynamics. It looks as though our only access to electrodynamics came from Van der Waals forces between molecules. \\	
  We have already seen that due to anti-screening the interaction between quarks become very strong at low energies so perturbative techniques are not going to be effective. So some approximations have to be made. It common to assume that the quarks have a mass, which is cannot be directly determined since they are never free. They are obtained from fits of the appropriate spectra of hadrons viewed as bound states of quarks. This is achieved by assuming a confining potential, which is attractive, with a strength that increases with the distance between the interacting quarks. The most popular are the linear, $V(r)\propto r$, logarithmic $V(r)\propto \ln(r/a)$ and  quadratic  $V(r)\propto r^2$ The latter is going to be discussed below (see next section). On top of this one supposes an interaction between the quarks
	
In this section
we are going to derive the effective potential between quarks in the one gluon approximation in the non relativistic limit up to including terms of second order in the quark momenta, .i.e of order of $p^2/m_q^2$. To this end we express the 4-spinor forms into matrix elements involving two component wave functions. Clearly  this approximation will be applicable to heavy quarks or assuming a constituent  masses, about a third of the nucleon mass, for light quarks.
\subsection{One gluon exchange potential in a process involving only baryons}
\index{gluon exchange}
The non relativistic reduction of the one gluon exchange amplitude \cite{HOV86} (see Fig. \ref{Fig:WQCD}) leads to the effective
2-body operator:
\barr
\tilde{V}&=&-\frac{1}{(2 \pi ) ^3} \lambf_1.\lambf_2\delta ( {\bf { \tilde q}}_1+{\bf { \tilde q}}_2 ) \frac{4 \pi \alpha_s}{{\bf { \tilde q}}_1^2}\nonumber\\
&&\left [ 1- \frac{1}{(2 m_q^2}\left ( {\bf {\tilde Q}}_1.{\bf {\tilde Q}}_1 +i \bfs_1. ({\bf {\tilde q}}_1\times {\bf {\tilde Q}}_2)
+i \bfs_2. ({\bf {\tilde q}}_2\times {\bf {\tilde Q}}_1 ) \right . \right .\nonumber\\
 &-& \left . \left . (\bfs_1 \times {\bf {\tilde q}}_1). (\bfs_2 \times {\bf {\tilde q}}_2) \right ) \right ],
\label{Eq:1p1p1p1p}
\earr
where $  \lambf_1.\lambf_2$ the SU(3) invariant, ${\bf {\tilde q}}_i={\bf p}^{'}_i-{\bf p}_i$ and ${\bf {\tilde Q}}_i={\bf p}^{'}_i+{\bf p}_i$, $i=1,2$.
It is traditional to perform a Fourier transform and go to the coordinate space. Thus we get
\beq
V=-\alpha_s \lambf_1.\lambf_2 \left[\frac{1}{r}+\frac{1}{(2 m_q)^2} \left(\frac{2}{3}V_S+\frac{1}{3}V_T + \frac{1}{r}V_{QQ}+ \frac{1}{r^3}V_{qQ} \right )\right],
\eeq
with $ \alpha_s$ treated as a  parameter to be fitted to the spectra. It is assumed to be of order 1, i.e. about five times larger than a typical value used in high  energy physics (see Fig. \ref{Fig:gStrong}). Furthermore
$$V_S=4 \pi \delta ({\bf r}) \bfs_1 . \bfs_2,$$
$$V_T=3 \bfs_1.\hat{{\bf r}}. \bfs_2.\hat{{\bf r}}-\bfs_1.\bfs_2,$$
$$V_{QQ}=\nabla _{r_1}(\leftarrow ).\nabla _{r_2}(\leftarrow ) +\nabla _{r_1}(\rightarrow ).\nabla _{r_2}(\rightarrow)
-\nabla _{r_1}(\leftarrow ).\nabla _{r_2}(\rightarrow  )$$ $$-\nabla _{r_1}(\rightarrow ).\nabla _{r_2}(\leftarrow   ).$$
$$V_{qQ}=(\sigma_1+\sigma_2).\left (\ell (\rightarrow )-\ell (\leftarrow )\right )$$ $$-\frac{1}{2} \left((\bfs_2-\bfs_1).(i~r\times \nabla _{R}(\leftarrow )-(\bfs_2-\bfs_1).(i~r\times \nabla _{R}(\rightarrow ) \right ).$$
In the above expressions the arrows in parenthesis indicate the direction (bra or ket) on which the non local operator
acts. As usual $\ell=\ell_2-\ell_1 $ is the relative orbital angular momentum, ${\bf r}={\bf r_2}-{\bf r_1}$ and
${\bf R}=\frac{1}{2}({\bf r_2}+{\bf r_1})$. The relative orbital angular momentum gives no contribution
since the relevant matrix element is diagonal (this part of the operator does not depend on the CM coordinates).
\subsection{One gluon exchange potential in processes involving the  creation of a $q{\bar q}$ pair.}
\begin{figure}[h!]
\begin{center}
\includegraphics[width=0.7\textwidth]{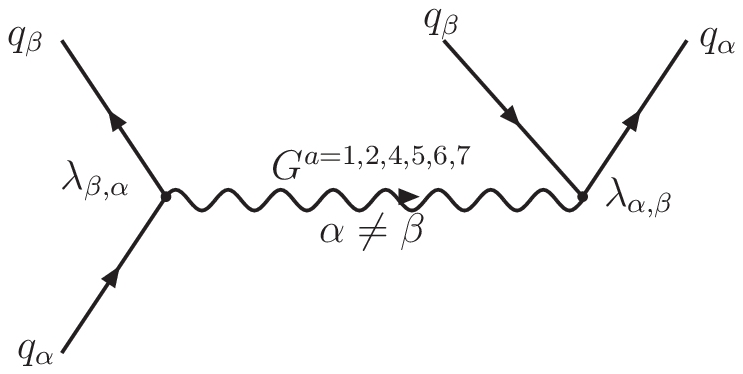}
\caption{\footnotesize \it
\\ A typical diagram involving the creation of a $q-\bar{q}$ pair by a gluon (for notation see Fig. \ref{Fig:2qint})}
\label{Fig:qbarqint}
\end{center}
\end{figure}

In this case in the place of the diagrams of Fig. \ref{Fig:2qint} one obtains a new diagram by replacing a a $q-q$ line by a $q-\bar{q}$ pair as in  Fig. \ref{Fig:qbarqint}.
The non relativistic reduction of the one gluon exchange amplitude \cite{HOV86}, resulting from this diagram, leads to the effective
2-body operator:
\beq
\tilde{V}=-\frac{1}{(2 \pi ) ^3} \lambf_1.\lambf_2\delta ( {\bf {\bf q}}_1-{\bf {\bf Q}}_2 ) \frac{4 \pi \alpha_s}{{\bf { \bf q}}_1^2} \left[ \frac{\bfs_2.{\bf Q}_1}{2 m_q}+i\frac{1}{2 m_q}{\bf q}_1. (\bfs_2\times \bfs_1)\right]
\label{Eq:1p2p1h}.
\eeq
Where the subscripts 1 and 2 refer to the left and right legs of the diagram \ref{Fig:qbarqint}. Here ${\bf q}_1=  {\bf p}_1-{\bf p}'_1$, ${\bf Q}_1=  {\bf p}_1+{\bf p}'_1$, ${\bf Q}_2=  {\bf p}_2+{\bf p}_2$. Momentum conservation reads $ {\bf p}_1= {\bf p}'_1+ {\bf p}_2+ {\bf p}'_2$ or ${\bf q}_1={\bf Q}_2$

\begin{figure}[h!]

\begin{center}
\subfloat[]
{
\includegraphics[width=0.45\textwidth]{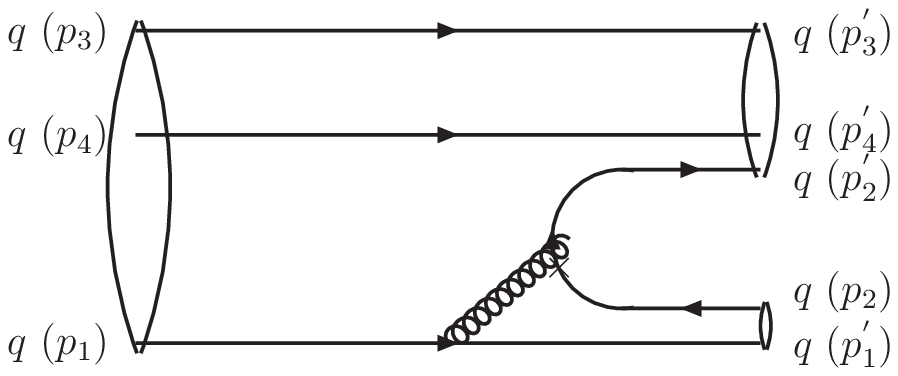}
}
\subfloat[]
{
\includegraphics[width=0.45\textwidth]{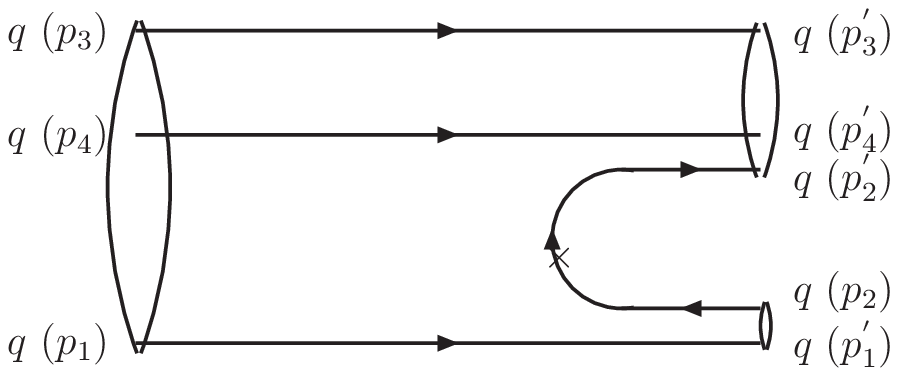}
}
\caption{\footnotesize \it
The 1 gluon exchange potential acting between the quarks labeled $q(p_1)$, $q(p^{'}_1)$, $q(p_2)$ and $q(p^{'}_2)$, while  the other quarks do not participate in the interaction (they are spectators)   (a). The same effect can be accomplished by creating  a $q {\bar q}$ pair  out of the vacuum,  marked by $\times$, (b), in which case  the pair is normally  in an $S=1,\ell=1,J=0$ state ($^3P_0$ model). The process exhibited describes  the baryon meson coupling. If the initial baryon is heavier than the final, it provides  the decay amplitude $B_1\rightarrow B_2 M$  (the lens like curves indicate a baryon or meson bound state). The same diagram is applicable in any process in which the interaction causes an $1q\rightarrow 2q-1{\bar q}$ transition. For example, if the middle horizontal line is missing and the top arrow is reversed, it describes the decay of one meson into two mesons.}
\label{Fig:1G3P0}
\end{center}
\end{figure}

 Furthermore this diagram, combined with a number of "spectator" (non interacting) quarks, can lead to processes involving only color singlet states (see Fig. \ref{Fig:1G3P0}), as, e.g., meson decay into two mesons \cite{HOV88}. Another approach is to consider the $^3P_0$, which is derived from the quark string model \cite{Perkins71}, yielding a $q-\bar{ q}$ pair, "created out of the vacuum".
\subsection{One gluon exchange potential in processes involving  meson spectra}
\begin{figure}[h!]
\begin{center}
\includegraphics[width=0.8\textwidth]{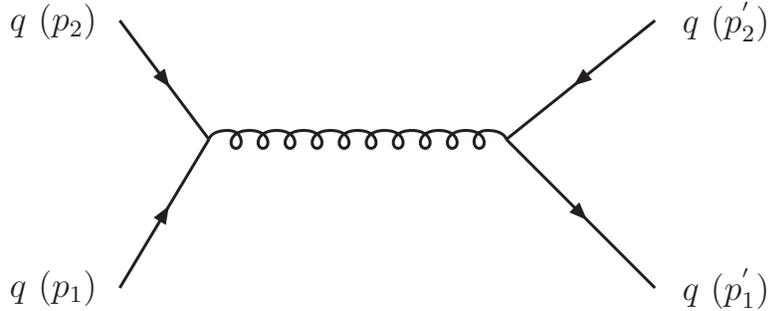}
\caption{\footnotesize \it
The one gluon effective potential relevant for meson spectra.}
\label{Fig:gantiq}
\end{center}
\end{figure}

The non relativistic reduction of the one gluon exchange amplitude (see Fig. \ref{Fig:gantiq}) in this case leads to the effective potential:
\barr
\tilde{V}&=&-\frac{1}{(2 \pi ) ^3} \lambf_1.\lambf_2\delta ( {\bf { \tilde Q}}_1-{\bf { \tilde Q}}_2 ) \frac{4 \pi \alpha_s}{{\bf { \tilde Q}}_1^2} \nonumber\\&&\left\{ \bfs_1.\bfs_2+\frac{1}{2} \frac{1}{(2 m_q)^2}\left [ \bfs_2.({\bf{ \tilde q}}_1\times {\bf { \tilde Q}}_1)+\bfs_1.({\bf { \tilde q}}_2\times {\bf{ \tilde Q}}_1)\right ]\right . \nonumber\\
 &+& \frac{1}{4}\frac{1}{(2 m_q)^2}\left [2 { \tilde Q}_1^2-{ \tilde q}_1^2-{ \tilde q}_2^2 +({\bf { \tilde q}_1}-{\bf{ \tilde q}_2})\times {\bf { \tilde Q}}_1). (\bfs_1\times \bfs_2)
\right\}.
\label{Eq:1p1h1p1h}
\earr
\section{Low energy formalism}
\label{sec:lowenergy}
\index{low energy approximation}
One would like to make suitable approximations to microscopically study the usual baryons, e.g., the proton, neutron, $\Delta$ resonances etc , and mesons, pions, $\rho$-mesons,$J/\Psi$,  $Y$ and their excitations  ($\Psi'$,  $Y'$ etc), etc called quarkonia.\\
Many phenomenologically realistic models exist, see e.g. the pioneer early works \cite{CJJKW74}-\cite{Eichten80}. For purposes of illustration we will consider the much simpler approach using a harmonic oscillator basis.\\
The kinetic energy part is easy to evaluate in momentum space, especially, if a convenient harmonic oscillator basis is adopted, since in this case one can separate the relative from the center of mass motion. The spin isospin quantum numbers can be treated in the usual way for the three body (baryon) and the two body system (meson). So we will concentrate on the orbital parts.
\subsection{The orbital part at the quark level}
\index{orbital part}
Orbital wave functions in momentum space are expressed in terms of Jacobi coordinates:
\beq
\psi_{{\bf P}_{M}}= \sqrt{2 E_{\pi}}\left (2 \sqrt{2} \right )^{1/2} \left ( 2 \pi \right ) ^{3/2} \delta \left ( \sqrt{2}{ \bf Q}_{M}-{\bf P}_{M} \right ) \phi_{M}( \mbox{ \boldmath $\rho$} ),
\eeq
\beq
\psi_{{\bf P}_B}= \left (3 \sqrt{3} \right )^{1/2} \left ( 2 \pi \right ) ^{3/2} \delta \left ( \sqrt{3}{ \bf Q-P}_{B} \right ) \phi_B(\mbox{ \boldmath $\xi$},\mbox{ \boldmath $\eta$}),
\eeq
where ${\bf P}_{M}$, and ${\bf P}_B$  are the momenta of the meson and the  baryon respectively and
\beq
\mbox{ \boldmath $\xi$}=\frac{1}{\sqrt{2}}({\bf p}_1-{\bf p}_2)~,~
\mbox{ \boldmath $\eta$}=\frac{1}{\sqrt{6}}({\bf p}_1+{\bf p}_2-2{\bf p}_3)~,~
Q=\frac{1}{\sqrt{3}}({\bf p}_1+{\bf p}_2+{\bf p}_3),
\eeq
\beq
\mbox{ \boldmath $\rho$}=\frac{1}{\sqrt{2}}({\bf p}_1-{\bf p}_2)~,~Q_{M}=\frac{1}{\sqrt{2}}({\bf p}_1+{\bf p}_2),
\eeq
where $\bf{p}_i,~i=1\cdots3$ are the momenta of the three quarks of the baryon, or $\bf{p}_i,~i=1,2$ are the momenta of the quark and anti-quark in the meson.

 The above wave functions were normalized in the usual way:
 \beq
 \langle \psi_{{\bf P}_B}|\psi_{{\bf P}'_B}\rangle =(2 \pi) ^3 \delta({\bf P}_B-{\bf P}'_B)
~,~\langle \psi_{{\bf P}_{M}}|\psi_{{\bf P}'_{M}}\rangle = 2E_{M} (2 \pi)^3 \delta({\bf P}_{M}-{\bf P}'_{M}).
\label{wf}
\eeq
The harmonic oscillator \index{harmonic oscillator} wave functions describing the relative coordinates are well known. Thus, e.g. for $0s$ astes take the form:
\beq
 \phi_B(\mbox{ \boldmath $\xi$},\mbox{ \boldmath $\eta$})=\phi(\xi)\phi(\eta),
\eeq
\barr
\phi(\xi)&=&\phi(0) e^{-(b^2_B\xi^2)/2},\,\phi(0)=\sqrt{\frac{b^3_B}{\pi \sqrt{\pi}}} \mbox { etc },\nonumber\\ \phi_M(\rho)&=&\phi_M(0)e^{-(b^2_M\rho^2)/2},\,\phi_M(0)=\sqrt{\frac{b^3_M}{\pi \sqrt{\pi}}}.
\earr
\subsection{The kinetic energy part}
\index{kinetic energy part}
The mass of the quarks is assumed to be the constituent quark mass, about a third of the mass of the nucleon for the light quarks $u,\,d$ and $s$. For the heavy quarks one extracts their masses from the mass of the corresponding mesons $J/\Psi$ and $Y$.\\
\subsection{The confining potential}
\index{confining potential}
We will assume a confining potential of the form:
\beq
V_c= -\lambf_1 .\lambf _2 k(\bf{r}_1-\bf{r}_2)^2,
\eeq
where $ \lambf_1 .\lambf _2 $ is the two-body part of the $SU(3)$ Casimir operator, quadratic $SU(3)$ invariant $g(\lambda,\mu)$, given by
$$
\lambf_1 .\lambf _2 =C_{(\lambda,\mu)},$$ $$ C_{(\lambda,\mu)}=\quad\frac{1}{2}\left (g(\lambda,\mu)-2 g(1,0) \right ), \,
g(\lambda,\mu)=\frac{2}{3}\left (\lambda^2+\mu^2 +\lambda \mu+3 (\lambda+ \mu)\right ).
$$
The integers $\lambda$ and $\mu$ characterize the $SU(3)$ representation, e.g.,
$$(\lambda, \mu)= (1,0),\,(0,1),\,(2,0),\,(1,1)$$ for the fundamental ${\underline3}$ (triplet), ${\underline3}^*$ (anti-triplet), ${\underline6}$ (sextet) and the ${\underline 8}$ (octet or adjoined) representations respectively.
 $k=\omega^2 m_q$ is a constant to be fitted by yielding the correct value of the mass of the nucleon  for a given size parameter $a$ (see below).
\subsection{Fitting the strength of the confining potential}
\index{confining potential}
The strength $k$ of the confining potential can be determined by considering the nucleon as a three quark system. One must first compute the mass of the baryons. To this end we consider:
\begin{itemize}
\item The kinetic energy :
\beq
\epsilon_q=\frac{1}{2 m_q}\left ({\bf p}_1^2+{\bf p}_2^2+{\bf p}_3^2 \right ), \quad m_q=\frac{m_p}{3}.
\eeq
For harmonic oscillator wave functions in the internal variables we find in the rest frame of the baryon:
\barr
\langle\psi_N|\epsilon_q|\psi_N\rangle&=&\langle\psi_N|\frac{1}{2 m_q}\left ( \frac{1}{2}\left ({\bf p}_1-{\bf p}_2 \right )^2+ \frac{1}{6}\left ({\bf p}_1+{\bf p}_2-2{\bf p}_3\right )^2\right )|\psi_N\rangle\nonumber\\
&=&\frac{1}{2 m_q}\left( \frac{3}{2}\frac{1}{b^2}+ \frac{3}{2}\frac{1}{b^2} \right)=\frac{9}{2 m_p}\frac{1}{b^2}.
\earr
\item The average  confining potential.\\
In this case we find $ME=4/3 \langle V_c\rangle$, with $\langle V_c\rangle$ the expectation of the radial part of the potential. The negative value is due to the fact that only the color isotriplet pairs contribute ($C_{(0,1)}=-4/3$). It is straightforward to show  that
\beq
\langle V_c\rangle
=3 \frac{1}{2} k b^2,\quad b=\mbox{the nucleon size }\Rightarrow ME=2 k b^2.
\eeq
\end{itemize}
We thus arrive at the equation:
\beq
m_p=\frac{9}{2 m_p b^2}+2 k b^2
\eeq
or
\beq
ka^2=\frac{m_p}{2}\left (-\frac{9}{2 m_p^2b^2}+1 \right ).
\eeq
For $a=1$fm we find:
$$ kb^2=0.05 m_p=50 \mbox{ MeV}\rightarrow k=50\frac{\mbox{MeV}}{\mbox{fm}^2}.$$
\section{Matrix elements involving two quarks}
We need not worry about the 1-body terms arising  from the kinetic terms, since it is trivial to compute them for any hadron.
\subsection{The confining potential}
\index{confining potential}
The confining potential is spin independent and does not change color. Furthermore the two quark matrix elements are independent of isospin. Thus:
$$
\langle (n_1l_1,n_2l_2)L;[f_{cs}](\lambda,\mu)S;JI|V_c|(n^{\prime}_1 l^{\prime}_1,n^{\prime}_2l^{\prime}_2)L^{\prime};[f^{\prime}_{cs}](\lambda^{\prime},\mu^{\prime})S^{\prime};JI\rangle=$$ $$\delta_{L,L^{\prime}}\delta_{S,S^{\prime}}\delta_{\lambda,\lambda^{\prime}}\delta_{\mu,\mu^{\prime}}\delta_{[f_{cs}],[f_{cs}^{\prime}]} (-)C_{\lambda,\mu}\langle(n_1l_1,n_2l_2)L|V_{c}|(n^{\prime}_1 l^{\prime}_1,n^{\prime}_2l^{\prime}_2)L\rangle,
$$
where $[f_{cs}](\lambda,\mu)S$ is the two particle  spin -color wave function with permutation  symmetry $[f_{cs}]$, color $(\lambda,\mu)$ and spin $S$. Similarly for $[f^{\prime}_{cs}](\lambda^{\prime},\mu^{\prime})S^{\prime}$. $J$ is the total angular momentum,   $\langle(n_1l_1,n_2l_2)L|V_{c}|(n^{\prime}_1 l^{\prime}_1,n^{\prime}_2l^{\prime}_2)L\rangle$ is the radial integral involving the potential and
$$C_{(\lambda,\mu)} =-\frac{4}{3} \mbox{ for }(\lambda,\mu)=(0,1)\mbox{ (anti-triplet)}$$ $$C_{(\lambda,\mu)} = \frac{2}{3}\mbox{ for }(\lambda,\mu)=(2,0) \mbox{ (sextet)}.$$
We will see below that we have the following possibilities:
$$ [f_{cs}]=[2]\mbox{ (symmetric)}\rightarrow (\lambda,\mu)S=(2,0)1,(0,1)0$$ $$[f_{cs}]=[1^2]\mbox{ (antisymmetric)}\rightarrow (\lambda,\mu)S=(2,0)0,(0,1)1$$
\subsection{The one gluon exchange potential}
\index{gluon exchange potential}
 Even though the above operator in the coordinate space is non local, one can still compute the needed matrix elements
 by taking gradients of the state vectors and employing the standard Racah algebra techniques. We find it more
 convenient, however, to work in momentum space. Furthermore most of the elementary interactions are derived in
 momentum space. So in many other applications one would like to have developed a formalism permitting
 exploitation of the advantages of momentum space.

 We first introduce the dimensionless variables ${\bf Q}_i=\frac{1}{\sqrt{2}}{\bf {\tilde Q}}_i b $, ${\bf q}_i=\frac{1}{\sqrt{2}}{\bf {\tilde q}}_i b $, $i=1,2$, with $b$ being the harmonic oscillator size
parameters . Then the above operator associated with baryons only (see Eq. (\ref{Eq:1p1p1p1p})) can be brought into a more convenient form:
 \barr
\tilde{V}&=&-\frac{1}{(2 \pi ) ^3}\lambf_1.\lambf_2 \frac{1}{2 \sqrt{2}} \delta ( {\bf q}_1+{\bf q}_2 )
 \frac{4 \pi \alpha_s}{2 b}\frac{1}{{\bf q}_1^2}\\
\nonumber
&&\left[ 1- 2 \kappa \left (  {\bf Q}_1.{\bf Q}_1 +i \sigma_1. ({\bf q}_1\times {\bf Q}_2)
+i \sigma_2. ({\bf q}_2\times {\bf Q}_1 )- (\sigma_1 \times {\bf q}_1). (\sigma_2 \times {\bf q}_2) \right ) \right],
\earr
with  $\alpha_s=\frac{g_r^2}{4 \pi}$ and $\kappa=\frac{1}{(2 m_q b)^2}$. In terms of tensor operators it is
conveniently rewritten as follows:
\barr
\tilde{V}&=&-\frac{1}{(2 \pi ) ^3} \lambf_1.\lambf_2\frac{1}{2 \sqrt{2}} \delta ( {\bf q}_1+{\bf q}_2 )
 \frac{4 \pi \alpha_s}{2 b}\frac{1}{{\bf q}_1^2}\nonumber\\
&&\left\{ 1- 2 \kappa \left (  {\bf Q}_1.{\bf Q}_1 -\sqrt{2} \sigma_1. [({\bf Q}_2\times {\bf q}_1)]^1
- \sqrt{2} \sigma_2. [({\bf Q}_1\times {\bf q}_2 )]^1-\right . \right .\nonumber\\
&& \left. \left. \left ( \frac{2}{3} {\bf \sigma}_1. {\bf \sigma}_2+
\frac{1}{3}T ( {\bf \sigma}_1,{\bf \sigma}_2,{\hat q}_1  )  \right ) q_1^2\right ) \right\},
\earr
 with $T$ the tensor operator defined by $$T ( {\bf \sigma}_1,{\bf \sigma}_2,{\hat q})=3 {\bf \sigma}_1.{\hat q}{\bf \sigma}_2.{\hat q}-{\bf \sigma}_1. {\bf \sigma}_2=\sqrt{\frac{6}{5}} \sqrt{4 \pi} Y^2({\hat q}).[{\bf \sigma}_1\times  {\bf \sigma}_2]^2.$$
In other words the operator in momentum space indeed takes a very simple form.\\
The same procedure can be applied in the case of the operators of Eqs (\ref{Eq:1p2p1h}) and (\ref{Eq:1p1h1p1h}).

At this point we should mention that we find it convenient to  evaluate the matrix elements of the two body strong interaction using the color-spin symmetry.
\subsection{A simple interaction} 
For illustration  purposes we will consider the simple spin color interaction
\beq
V_{12}=\sbf_1.\sbf_2 \lambf_1.\lambf_2.
\eeq
In the case of two quarks the color spin state is of the form
\beq
[2]=(2,0) s=1,(0,1) s=0,\,[2]=(2,0) s=0,(0,1) s=1,\,
\eeq
We know that
$$\langle S|\sbf_1.\sbf_2|S\rangle=2S(S+1)-3,\, \langle (\lambda,\mu)|\lambf_1.\lambf_2|(\lambda,\mu)\rangle=\left \{\begin{array}{rr}\frac{2}{3},&(\lambda,\mu)=(2,0)\\-\frac{4}{3},&(\lambda,\mu)=(0,1)\\ \end{array} \right .$$
(see section \ref{sec:ColorStructure}).
Thus
$$\langle[2] |V_{12}|[2]\rangle=\left \{ \begin{array}{rr}\frac{2}{3},&(\lambda,\mu)=(2,0),s=1\\4,&(\lambda,\mu)=(0,1),s=0\\ \end{array}\right .,\langle[1^2] |V_{12}|[1^2]\rangle=\left \{ \begin{array}{rr}-2,&(\lambda,\mu)=(2,0),s=1\\-4,&(\lambda,\mu)=(0,1),s=0\\ \end{array}\right .$$
In the case of a quark-antiquark  the color spin state is of the form
\beq
[2,1^4]=(1,1) s=1,(1,1) s=0,(0,0)s=1,\,[1^6]=(0,0) s=0
\eeq
$$\langle S|\sbf_1.\sbf_2|S\rangle=2S(S+1)-3,\, \langle (\lambda,\mu)|\lambf_1.\lambf_2|(\lambda,\mu)\rangle=\left \{\begin{array}{rr}\frac{1}{3},&(\lambda,\mu)=(1,1)\\-\frac{8}{3},&(\lambda,\mu)=(0,0)\\ \end{array} \right .$$
Thus
\beq
\langle[21^4] |V_{12}|[21^4]\rangle=\left \{ \begin{array}{rr}\frac{1}{3},&(\lambda,\mu)=(1,1),s=1\\-1,&(\lambda,\mu)=(1,1),s=0\\ -8/3,&(\lambda,\mu)=(0,0),s=1\\\end{array}\right .,\,\langle[1^6] |V_{12}|[1^6]\rangle=8
\eeq
In the the spin color symmetry is redundant. The quantum numbers $(\lambda,\mu),,\,S$ are adequate.

The above expressions must be multiplied with the matrix element of the radial part of the operator $V(|{\bf r_1-r_2}|)=\frac{ \alpha_s}{|{\bf r_1-r_2}|}$. The radial functions can be obtained by solving Scr$\ddot{o}$inger's equation with a confining potential as discussed in section \ref{sec:lowenergy}. In most applications only $s$-states are considered. In the present calculation  we will employ harmonic oscillator and the MIT bag model\cite{MITbag1}  \cite{MITbag2} wafe functions, in order to get analytic expressions for the Coulomb type interaction.\\
In the case of  $0s$ harmonic oscillator wave function  with size parameter $b$ one finds that 
\beq
I{\mbox{{\tiny HO}}}=\langle V(|{\bf r_1-r_2}|)\rangle=(\alpha_s\sqrt{\frac{2}{\pi}}\frac{1}{b}
\label{Eq:IHO}
\eeq
In the case of the MIT bag model\cite{MITbag1}  \cite{MITbag2} the nucleon s- wave function is of the form:
\beq
R(x)=\sqrt{\frac{2}{a-\sin{a}\cos{a}}}\frac{\sin{x}}{x},\, x=kr,\,0\leq x \leq a
\eeq
where $k$ is the nucleon momentum.\\
In this case the radial integral of the above potential is
$$
I=\frac{1}{(4 \pi)^2}\int \int R(x_1)R(x_2)\frac{\alpha_s k}{|{\bf x}_2-{\bf x}_1|}d^3{\bf x}_1 d^3{\bf x}_2$$ $$=\frac{2\alpha_s k}{a-\sin{a}\cos{a}}\int_0^a dx_1  \int_0^a d x_2 x_1\sin{x_1} x_2\sin{x_2}\left(\frac{1}{x_1}H(x_1-x_2)+\frac{1}{x_2}H(x_2-x_1)\right )
$$ 
where $H(x)$ is the Heavyside step function. Thus:
\beq
I_{\mbox{{\tiny bag}}}=k \alpha _s  \frac{(a (\cos (2
   a)+2)-3 \sin (a) \cos
   (a))}{a-\sin (a) \cos (a)}, a=k R_c
	\label{Eq:MITbag}
	\eeq
where $R_c$ is the bag radius.
\section{Symmetries in multi-quark systems}
Multi-quark systems are Fermions containing more than three quarks  and mesons containing more pairs than one quark one anti-quark .  Such systems have been examined from the point of view of symmetries a long time ago, see, e.g., \cite{Stro78,Stro79, multiq80a}. See also \cite{multiq80b, pentaq03, pentaq05}. The symmetries that were found useful were the combined spin color symmetry $SU_{cs}(6)$ the orbital symmetry and the flavor symmetry. Such multi-quarks  were later discovered experimentally, the penta-quark \cite{pentaq15,pentaq16a} and the tetra-quark \cite{fourq10,fourq14}. On the latter see  also the recent reviews \cite{EPP16},\cite{CCLZ16},\cite{ALS16},\cite{ARS17}, .

In the LHCb experiment \cite{pentaq15} what is observed is the decay of a particle with the quantum numbers of $\Lambda_b$ as follows:
\beq
\Lambda_b\rightarrow J/\psi+p +K^{-}
\eeq
The data seem to be consistent with a pentaquark of the type $(udb)c\bar{c}$, with the decay process at the quark level being $b\rightarrow du\bar{u}$ or $ub\rightarrow du$ weakly.

Motivated with this experimental evidence we will examine the general  case of a pentaquark system,  from the point of view of symmetries, in the expectation that structures may be found in the future.

In the case of multiquark systems one finds useful to employ the following symmetries:\\
i) The  orbital symmetry. This is described in terms of a Young tableaux [f]. \\
ii) the isospin symmetry. In this case the corresponding Young tableaux [f]$_I$ has at most two rows,[f]$_I=[f_1,f_2]$, and the total isospin is $I=\frac{1}{2}(f_1-f_2)$. This symmetry can be handled by the usual angular momentum techniques\\
iii) the spin color group $SU_{cs}(6) \supset SU_s(2)\otimes SU_c(3)$\\

\section{Symmetries involved in the case of pentaquarks}
In the case of pentaquarks the space part of the wave function is trivial, since it assumed to be made of $0s$  single particle states and, thus, it is completely symmetric.\\
For the group theoretical description of the pentaquarks we refer the reader to previous work \cite{SOSTRO},\cite{multiq80a}\\
Now since the orbital symmetry is trivial, the wave function becomes antisymmetrized if the Young tableaux [f]$_{cs}$ is the conjugate of that  of the isospin [f]$_I$, i.e. obtained from that by exchanging rows and columns.\\ Regarding isospin we use\\
 i) $SU_I(2)$  for $u$ and $d$ quarks, \\
ii) $U_1^n$ for $n=s,\, c, \,b$ quarks.\\
\subsection{Configurations  involved in the case of pentaquarks}
We will distinguish the following possibilities:\\
Configurations of the type:\\
i) $q^4\bar{q}$, $q=u,d$. Then we will consider wave functions of good isospin $[I_1\otimes I_2]^I$, where $I_2=1/2$ is the isospin of the antiquark and $I_1$ the isospin of the 4-quark state. We have:
$$I_1=0\Leftrightarrow [2,2]\mbox{symmetry},I_1=1\Leftrightarrow [3,1]\mbox{symmetry},I_1=2\Leftrightarrow [4] \mbox{symmetry}$$
For completely antisymmetric wave functions the color-spin symmetries are obtained by interchanging the rows and columns of the Young diagrams, i.e.
$$[2,2],\, [2,1^2],\, [1^4].$$
respectively. The associated color quantun number must be $(\lambda,\mu)=(1,0)$ or $\underline{3}$, so that combined with the color of the antiquark $\underline{\bar{3}}$ to yield a color singlet state $(0,0)$. Thus, since the color-spin symmetry of the antiquark is redundant, the wave functions can be cast in the form
\beq
\left[ [2,2](s_1=1 (10)), I_1=0 \times( s_2=1/2 ,(0,1)),I_2=1/2\right  ]^{I=1/2,s={1/2,3/2}}_{(0,0)}
\eeq
\beq
\left[ [2,1^2](s_1=0, (10)) I_1=1 \times (s_2=1/2, (0,1)),I_2=1/2\right  ]^{I=1/2,s={1/2}}_{(0,0)},
\eeq
\beq
\left[ [2,1^2](s_1=0, (10)) I_1=1 \times (s_2=1/2, (0,1)),I_2=1/2\right  ]^{I=3/2,s={1/2}}_{(0,0)},
\eeq
\beq
\left[ [2,1^2](s_1=1 (1,0)), I_1=1 \times (s_2=1/2 ,(0,1)),I_2=1/2\right  ]^{I=1/2,s={1/2,3/2}}_{(0,0)},
\eeq
\beq
\left[ [2,1^2](s_1=1, (1,0)), I_1=1 \times (s_2=1/2, (0,1)),I_2=1/2\right  ]^{I=3/2,s={1/2,3/2}}_{(0,0)}
\eeq
\beq
\left[ [2,1^2](s_1=2 (1,0)), I_1=1 \times (s_2=1/2 ,(0,1)),I_2=1/2\right  ]^{I=1/2,s={3/2,5/2}}_{(0,0)},
\eeq
\beq
\left[ [2,1^2](s_1=2, (1,0)), I_1=1 \times (s_2=1/2, (0,1)),I_2=1/2\right  ]^{I=3/2,s={3/2,5/2}}_{(0,0)}
\eeq
\beq
\left[ [1^4](s_1=1 (1,0)), I_1=2 \times (s_2=1/2 ,(0,1)),I_2=1/2\right  ]^{I=3/2,s={1/2,3/2}}_{(0,0)},
\eeq
\beq
\left[ [1^4](s_1=1, (1,0)), I_2=1 \times (s_2=1/2, (0,1)),I_2=1/2\right  ]^{I=5/2,s={1/2,3/2}}_{(0,0)}
\eeq
We have 7 states of isopin 1/2, 7 states of isospin 3/2 and two states of isospin 5/2. Since the strong interactions conserve isospin, states with different isospin do not get admixed.

ii) $q^4\bar{q}_a$, $q=u,d$,  $q_a=s,c,b,t$. Then we proceed as above noting that the antiquark is an iospin singlet, $I=I_1$ with  $I_1$ the isospin of the 4-quark state. We have:
\beq
\left[ [2,2](s_1=1 (10)), I_1=0 \times( s_2=1/2 ,(0,1)),I_2=0\right  ]^{I=0,s={1/2,3/2}}_{(0,0)}
\eeq
\beq
\left[ [2,1^2](s_1=0, (10)) I_1=1 \times (s_2=1/2, (0,1)),I_2=0\right  ]^{I=1,s={1/2}}_{(0,0)},
\eeq
\beq
\left[ [2,1^2](s_1=1 (1,0)), I_1=1 \times (s_2=1/2 ,(0,1)),I_2=0\right  ]^{I=1,s={1/2,3/2}}_{(0,0)},
\eeq
\beq
\left[ [2,1^2](s_1=2 (1,0)), I_1=1 \times (s_2=1/2 ,(0,1)),I_2=0\right  ]^{I=1,s={3/2,5/2}}_{(0,0)},
\eeq
\beq
\left[ [1^4](s_1=1 (1,0)), I_1=2 \times (s_2=1/2 ,(0,1)),I_2=0\right  ]^{I=2,s={1/2,3/2}}_{(0,0)},
\eeq
Since the strong interactions conserve isospin, these states do not admix with those of case i).
There exist $2\times n$ isospin 0 and 2 states and $5\times n$ isospin 1 states, where n is the number of the different  antiquarks considered. Again states of different isospin  do not get admixed.

iii) $q^3 q_b\bar{q}_a$, $q=u,d$,  $q_a,q_b=s,c,b,t$. Then we proceed as above noting that in the case of $a$ and $b$ flavors the isospin of each is zero and their combined isospin is zero.  Thus the possible isospins are those contained in the $q^3$ part, i.e.  $3/2\Leftrightarrow[3]$ and  $1/2\Leftrightarrow [21]$, i.e. $f_{cs}[1^3]$ and $f_{cs}[21]$ respectively. The spin-color possibilities of the $q_b\bar{q}_a$ combination are: $(\lambda,\mu)s=(1,1)1,(1,1)0,(0,0)1$ and (0,0)0. So among the components of $[2,1]$ relevant for yielding a color singlet state are  $(11)1/2,\,(11)3/2,\, (0,0)1/2$. The relevant components of the $[1^3]$ are $(0,0)1/2,\, (0,0)3/2$. The spin-color symmetries of the  $q_b\bar{q}_a$ are $[2,1^4]$ and $[1^6]$, but they are redundant. In any case
one can easily find that the $[2,1^4]$ contains
 $(\lambda,\mu)s=(1,1)1,(1,1)0,(0,0)1$. while the $[1^6]$ contains only the (0,0)0. Thus we can have:
\beq
\left[ [2,1](s_1=1/2, (1,1)), I_1=1/2 \times [2,1^4] (s_{a,b}=0, (1,1)),I_{ab}=0\right  ]^{I=1/2,s={1/2}}_{(0,0)}
\eeq
\beq
\left[ [2,1](s_1=1/2, (1,1)), I_1=1/2 \times [2,1^4] (s_{a,b}=1, (1,1)),I_{ab}=0\right  ]^{I=1/2,s={1/2,3/2}}_{(0,0)}
\eeq
\beq
\left[ [2,1](s_1=3/2, (1,1)), I_1=1/2 \times [2,1^4] (s_{a,b}=0, (1,1)),I_{ab}=0\right  ]^{I=1/2,s={3/2}}_{(0,0)}
\eeq
\beq
\left[ [2,1](s_1=3/2, (1,1)), I_1=1/2 \times [2,1^4] (s_{a,b}=1, (1,1)),I_{ab}=0\right  ]^{I=1/2,s={1/2,3/2,5/2}}_{(0,0)}
\eeq
\beq
\left[ [2,1](s_1=1/2, (0,0)), I_1=1/2 \times [2,1^4] (s_{a,b}=1, (0,0)),I_{ab}=0\right  ]^{I=1/2,s={1/2,3/2}}_{(0,0)}
\eeq
\beq
\left[ [2,1](s_1=1/2, (0,0)), I_1=1/2 \times [1^6] (s_{a,b}=0, (0,0)),I_{ab}=0\right  ]^{I=1/2,s={1/2}}_{(0,0)}
\eeq
\beq
\left[ [1^3](s_1=1/2, (1,1)), I_1=3/2 \times [2,1^4] (s_{a,b}=1, (1,1)),I_{ab}=0\right  ]^{I=3/2,s={1/2,3/2}}_{(0,0)}
\eeq
\beq
\left[ [1^3](s_1=3/2, (0,0)), I_1=3/2 \times [1^6] (s_{a,b}=0, (1,1)),I_{ab}=0\right  ]^{I=3/2,s={3/2}}_{(0,0)}
\eeq
These  states can, in principle, admix with the states of class i) with the same isospin.

iv) $q^2 q_{c};q_{a}q_{b},\, q_{c},q_{a},q_{b}=s,c,b,t$. The classification of the states $q_{a}q_{b}$ is exactly as in the previous case. The total isospin is 0 or 1.\\
The allowed $q^2 q_{c}$ combinations are 
$$ \left [q^2 q_{c}(\lambda_1,\mu_1)s'_1 I\otimes (1,0)\frac{1}{2}\right ]^{(\lambda,\mu),s1,I}$$
where $(\lambda,\mu)=(1,1),(0,0)$, $s'_1 \otimes\frac{1}{2}\rightarrow s_1$ and
$$I=0\Rightarrow (\lambda_1,\mu_1,s'_1)=(2,0)1,(0,1)0,\,I=1\Rightarrow (\lambda_1,\mu_1,s'_1)=(2,0)0,(0,1)1$$
The two sets of states are combined as in the previous case iii).

The remaining flavor combinations are not of interest from the point of symmetries and they are not going to be considered,

At this point we should mention that one can evaluate the matrix elements of the two body strong interaction knowing the one and two particle
coefficients of fractional parentage (CFP, see below). For the isospin part these are trivial, i.e. the usual Clebsch-Gordan coefficients $I_1\times I_2\rightarrow I$, indicated symbolically as $C^I_{I_1,I_2}$. For the color spin the 1-particle CFP's can by found in the work of So and Strottman \cite{SOSTRO}. Those of interest to the pentaquark systems are given in tables \ref{tab:spc6cfp10}-\ref{tab:spc6cfp11}. 
\begin{table}[h!]
\begin{center}
\caption{ The one particle CFP's involving the spin color symmetry, needed in the case of a 4-quark state,  with color symmetry (10) and spin 1.}
\label{tab:spc6cfp10}
$$
\begin{array}{|c|c|c|}
\hline
[f_1](\lambda_1,\mu_1)S_1&[f](\lambda,\mu)S&\mbox{CFP}\\
\hline
\mbox{$[2,1](1,1)\frac{1}{2}$}&[2,1^2](1,0)1&\sqrt{\frac{2}{3}}\\
\mbox{$[2,1](1,1)\frac{3}{2}$}&[2,1^2](1,0)1&\frac{1}{\sqrt{6}}\\
\mbox{$[2,1](0,0)\frac{1}{2}$}&[2,1^2](1,0)1&\frac{1}{\sqrt{6}}\\
\hline
\end{array}\quad
\begin{array}{|c|r|c|}
\hline
[f_1](\lambda_1,\mu_1)S_1&[f](\lambda,\mu)S&\mbox{CFP}\\
\hline
\mbox{$[2,1](1,1)\frac{1}{2}$}&[2^2](1,0)1&\frac{1}{\sqrt{3}}\\
\mbox{$[2,1](1,1)\frac{3}{2}$}&[2^2](1,0)1&-\frac{1}{\sqrt{3}}\\
\mbox{$[2,1](0,0)\frac{1}{2}$}&[2^2](1,0)1&\frac{1}{\sqrt{3}}\\
\hline
\end{array}
$$
$$
\begin{array}{|c|c|c|}
\hline
[f_1](\lambda_1,\mu_1)S_1&[f](\lambda,\mu)S&\mbox{CFP}\\
\hline
\mbox{$[1^3](1,1)\frac{1}{2}$}&[1^4](10)1&\sqrt{\frac{2}{3}}\\
\mbox{$[1^3](0,0)\frac{3}{2}$}&[1^4](10)1&\frac{1}{\sqrt{3}}\\
\hline

\end{array}
$$
\end{center}
\end{table}
\begin{table}[h!]
\begin{center}
\caption{ The one particle CFP's involving the spin color symmetry, needed in the case of a 3-quark state,  with color symmetry (11) and (0,0) with spin 1/2.}
\label{tab:spc6cfp11}
$$
\begin{array}{|c|c|c|}
\hline
[f_1](\lambda_1,\mu_1)S_1&[f](\lambda,\mu)S&\mbox{CFP}\\
\hline
\mbox{$[2](2,0)1$}&[2,1](1,1)\frac{1}{2}&\frac{1}{\sqrt{2}}\\
\mbox{$[2](0,1)0$}&[2,1](1,1)\frac{1}{2}&-\frac{1}{\sqrt{2}}\\
\hline
\end{array}\quad
\begin{array}{|c|r|c|}
\hline
[f_1](\lambda_1,\mu_1)S_1&[f](\lambda,\mu)S&\mbox{CFP}\\
\hline
\mbox{$[1^2](2,0)0$}&[2,1](1,1)\frac{1}{2}&\frac{1}{\sqrt{2}}\\
\mbox{$[1^2](0,1)1$}&[2,1](1,1)\frac{1}{2}&\frac{1}{\sqrt{2}}\\
\hline
\end{array}
$$
$$
\begin{array}{|c|c|c|}
\hline
[f_1](\lambda_1,\mu_1)S_1&[f](\lambda,\mu)S&\mbox{CFP}\\
\hline
\mbox{$[2](0,1)0$}&[2,1](0,0)\frac{1}{2}&1\\
\hline
\mbox{$[2](2,0)1$}&[2,1](1,1)\frac{3}{2}&1\\
\hline
\end{array}\quad
\begin{array}{|c|r|c|}
\hline
[f_1](\lambda_1,\mu_1)S_1&[f](\lambda,\mu)S&\mbox{CFP}\\
\hline
\mbox{$[1^2](0,1)1$}&[2,1](0,0)\frac{1}{2}&1\\
\hline
\mbox{$[1^2](0,1)1$}&[2,1](1,1)\frac{3}{2}&-1\\
\hline
\end{array}
$$
\end{center}
\end{table}

From these one can obtain the 2-particle CFP's in the usual way \cite{JDV68}. Those of interest for the pentaquarks are listed in table \ref{tab:spc4cfp15}. The 2-particle CFP's for three quark states coincide with the corresponding one particle CFP's.
\begin{table}[h!]
\begin{center}
\caption{ The two particle CFP's involving the spin color symmetry for 4 identical quarks , needed in the study of pentaquarks, corresponding to color state (0,1) and spin one.
\label{tab:spc4cfp15}}
$$
\begin{array}{|c|c|c|c|}
\hline
[f_1](\lambda_1,\mu_1)S_1&[f_2](\lambda_2,\mu_2)S_2&[f](\lambda,\mu)S&\mbox{CFP}\\
\hline
\mbox{$[2](0,1)0$}&[2](2,0)1&[2^2](10)1&-\frac{1}{\sqrt{2}}\\
\mbox{$[2](2,0)1$}&[2](0,1)0&[2^2](10)1&-\frac{1}{\sqrt{2}}\\
\hline
\end{array}
\begin{array}{|c|c|c|c|}
\hline
[f_1](\lambda_1,\mu_1)S_1&[f_2](\lambda_2,\mu_2)S_2&[f](\lambda,\mu)S&\mbox{CFP}\\
\hline
\mbox{$[2](0,1)0$}&[2](2,0)1&[2,1^2](10)1&0\\
\mbox{$[2](2,0)1$}&[2](0,1)0&[2,1^2](10)1&0\\
\hline
\hline
\end{array}
$$
$$
\begin{array}{|c|c|c|c|}
\hline
[f_1](\lambda_1,\mu_1)S_1&[f_2](\lambda_2,\mu_2)S_2&[f](\lambda,\mu)S&\mbox{CFP}\\
\hline
\mbox{$[1^2](2,0)0$}&[1^2](0,1)1&[2^2](10)1&\frac{1}{\sqrt{6}}\\
\mbox{$[1^2](0,1)1$}&[1^2](2,0)0&[2^2](10)1&\frac{1}{\sqrt{6}}\\
\mbox{$[1^2](0,1)1$}&[1^2](0,1)1&[2^2](10)1&\sqrt{\frac{2}{3}}\\
\hline
\end{array}
\begin{array}{|c|c|c|c|}
\hline
[f_1](\lambda_1,\mu_1)S_1&[f_2](\lambda_2,\mu_2)S_2&[f](\lambda,\mu)S&\mbox{CFP}\\
\hline
\mbox{$[1^2](2,0)0$}&[1^2](0,1)1&[2,1^2](10)1&\frac{1}{\sqrt{3}}\\
\mbox{$[1^2](0,1)1$}&[1^2](2,0)0&[2,1^2](10)1&\frac{1}{\sqrt{3}}\\
\mbox{$[1^2](0,1)1$}&[1^2](0,1)1&[2,1^2](10)1&-\frac{1}{\sqrt{3}}\\
\hline
\end{array}
$$
\end{center}
\end{table}

The CFPs are very useful in constructing the many body Hamiltonian matrix. For this one needs in addition the matrix elements of the interaction between two particle states, e.g. those discussed in section \ref{sec.potentials}.
\subsection{Evaluation of the many body matrix elements of the 2-body interaction}
\label{sec:meint}
These can be obtained by using the CFPs calculated above for the pentaquark states. The results are given as follows: \\ 
Case i):\\
$$
\langle q^4\bar{q}\mbox{[$f^{sc}$]}_1(0,1)S_1,I_1,S,I|V|q^4 \bar{q}[f^{sc}]'_1(0,1)S'_1,I'_1,S,I\rangle=$$ $$ \delta_{I_1,I'_1}\delta_{S_1,S'_1}\langle q^4\mbox{[$f^{sc}$]}_1(0,1)S_1,I_1|V|q^4[f^{sc}]'_1(0,1)S'_1,I'_1\rangle$$ $$+4\sum_{x,y} 
\frac{n_{[f_x]}}{\sqrt{n_{[f^{sc}]}n_{[f^{'sc}]}}}
\langle [f]_x(\lambda_x,\mu_x) S_x|[f^{sc}]_1(0,1)S_1\rangle \langle [f]_x(\lambda_x,\mu_x) S_x|[f^{sc}]'_1(0,1)S'_1\rangle$$  
$$U(I_x,\frac{1}{2}I\frac{1}{2},I_1,I_y)U(I_x,\frac{1}{2}I\frac{1}{2},I'_1,I_y)U(S_x,\frac{1}{2}S\frac{1}{2},S_1,S_y)U(S_x,\frac{1}{2}S\frac{1}{2},S'_1,S_y)$$ $$C^{I_1}_{I_x,\frac{1}{2}} C^{I'_1}_{I_x,\frac{1}{2}}\langle q\bar{q}I_y (\mu_x\lambda_x)S_y|V| q\bar{q}I_y (\mu_x,\lambda_x)S_y\rangle
$$
Where $n_{[f_x]},n_{[f^{sc}]},n_{[f^{'sc}]}$ are the dimensions of the corresponding representations of the symmetric group $S_4$, $ (\mu_x\lambda_x)=(1,1),\,(0,0)$, $U$=the usual unitary Racah functions, $\langle [f]_x(\lambda_x,\mu_x) S_x|[f^{sc}]_1(0,1)S_1\rangle $ etc are the one particle CFPs for 4 particles and the last factor is the elementary two particle interaction. Furthermore
$$\langle q^4 \mbox{[$f^{sc}$]}_1(0,1)S_1,I_1|V|q^2[f^{sc}]'_1(0,1)S_1,I_1\rangle=\frac{1}{2}4 \times 3$$
$$\sum_{x,y} 
\frac{n_{[f_x]}}{\sqrt{n_{[f^{sc}]}n_{[f^{'sc}]}}}
\langle [f_x]_2(\lambda_x,\mu_x) S_x[f_y]_2(\lambda_y,\mu_y) S_y|[f^{sc}]_1(0,1)S_1\rangle$$ $$C^{I_1}_{I_x,I_y} C^{I'_1}_{I_x,I_y}
\langle [f_x]_2(\lambda_x,\mu_x) S_x[f_y]_2(\lambda_y,\mu_y) S_y|[f^{'sc}]_1(0,1)S_1\rangle$$
$$\langle q^2I_y (\lambda_y,\mu_y)S_y|V| q^2I_y I_y (\lambda_y,\mu_y)S_y\rangle$$
where $[f_x]=[2]$ or $[1^2]$ and $[f_y]=[2]$ or $[1^2]$ as selected by the two paricle CFPs.
Case ii):\\
$$
\langle q^4\bar{q}_{\alpha}\mbox{[$f^{sc}$]}_1(0,1)S_1,S,I|V|q^4\bar{q}_{\beta} [f^{sc}]'_1(0,1)S'_1,S,I\rangle=$$ $$ \delta_{\bar{q}_{\alpha},\bar{q}_{\beta}}\delta_{S_1,S'_1}$$ $$\langle q^4\mbox{[$f^{sc}$]}_1(0,1)S_1,I|V|q^4|[f^{sc}]'_1(0,1)S'_1,I\rangle$$  $$+4\sum_{x,y} 
\frac{n_{[f_x]}}{\sqrt{n_{[f^{sc}]}n_{[f^{'sc}]}}}
\langle [f]_x(\lambda_x,\mu_x) S_x|[f^{sc}]_1(0,1)S_1\rangle \langle [f]_x(\lambda_x,\mu_x) S_x|[f^{sc}]'_1(0,1)S'_1\rangle$$ $$
C^{I_1}_{I_x,\frac{1}{2}} C^{I'_1}_{I_x,\frac{1}{2}} U(S_x,\frac{1}{2}S\frac{1}{2},S_1,S_y)U(S_x,\frac{1}{2}S\frac{1}{2},S'_1,S_y)$$ $$\langle q\bar{q}_{\alpha} (\mu_x\lambda_x)S_y|V| q\bar{q}_{\beta} (\mu_x,\lambda_x)S_y\rangle
$$
$\langle q^4\mbox{[$f^{sc}$]}_1(0,1)S_1,I|V|q^4|[f^{sc}]'_1(0,1)S'_1,I\rangle$ is given as in the previous case with the obvious substitutions.\\
Next we get case iii):\\
$$\langle q^4\bar{q}\mbox{[$f^{sc}$]}_1(0,1)S_1,I_1,S,I|V|q^4\bar{q}_{\beta} [f^{sc}]'_1(0,1)S'_1,S,I\rangle=0$$
Case iv):\\
$$\langle q^4\bar{q}\mbox{[$f^{sc}$]}_1(0,1)S_1,I_1,S,I|V|q^3[f^{sc}]'_1(\lambda,\lambda)S_1'I,\,q_{\alpha} \bar{q}_{\beta}(\lambda,\lambda)S'_2,S,I\rangle= $$ $$+4\sum_{x,y} 
\frac{n_{[f_x]}}{\sqrt{n_{[f^{sc}]}n_{[f^{'sc}]}}}\langle [f]_x(\lambda,\lambda) S'_1|[f^{sc}]_1(0,1)S_1\rangle  U(S'_1,\frac{1}{2}S\frac{1}{2},S_1,S'_2)$$ $$C^{I_1}_{I_x,\frac{1}{2}} C^{I'_1}_{I_x,\frac{1}{2}}\langle q\bar{q} (\lambda,\lambda)S'_2|V| q_{\alpha}\bar{q}_{\beta} (\lambda,\lambda)S'_2\rangle $$
Case v):\\
$$ q^3[f^{sc}]_1(\lambda,\lambda)S_1I,\,q_{\alpha} \bar{q}_{\beta}(\lambda,\lambda)S_2,S,I|V|q^3[f^{sc}]'_1(\lambda',\lambda')S_1'I,\,q'_{\alpha} \bar{q}'_{\beta}(\lambda,\lambda)S'_2,S,I\rangle=V_{1}\delta_{\beta,\beta'}+V_2\delta_{\alpha,\alpha'}$$ $$\delta_{S_1,S_1'}\delta_{S_2,S_2'}\delta_{\lambda,\lambda'}\left (\left (\delta_{[f^{sc}]_1,[f^{'sc}]_1}\langle (\lambda,\lambda)S_2|V| (\lambda,\lambda)S_2\rangle\right )+ \langle q^3[f^{sc}]_1(\lambda,\lambda)S_1|V|q^3[f^{'sc}]_1(\lambda,\lambda)S_1\rangle \right )$$
with
$$V_1=\sum_{x}\langle q^3[f^{sc}]_1(\lambda,\lambda)S_1I,\,q_{\alpha}(1,0)\frac{1}{2}S_x;(10)|V|\langle q^3[f^{'sc}]_1(\lambda,\lambda)S'_1I,\,q_{\alpha'}(1,0)\frac{1}{2}S'_x;(10)\rangle$$ $$ U(S_1,\frac{1}{2}S\frac{1}{2};S_x,S_2) U(S'_1,\frac{1}{2}S\frac{1}{2};S'_x,S'_2)=
$$ $$
3\sum_{x,y} 
\frac{1}{\sqrt{n_{[f^{sc}]}n_{[f^{'sc}]}}} U(S_1,\frac{1}{2}S\frac{1}{2};S_x,S_2) U(S'_1,\frac{1}{2}S\frac{1}{2};S'_x,S'_2) U(S_z,\frac{1}{2}S_x\frac{1}{2};S_1,S_y) $$ $$U(S_z,\frac{1}{2}S'_x\frac{1}{2};S'_1,S_y) U((\lambda_z,\mu_z)(1,0)(1,0)(1,0)(\lambda,\lambda)(\lambda_y,\mu_y))^2U(I_z,\frac{1}{2},I,0,I,\frac{1}{2})^2$$ $$\langle [f_1]_2(\lambda_z,\mu_z )S_z|[f^{sc}]_1(\lambda,\lambda)S_1\rangle \langle [f_1]_2(\lambda_z,\mu_z)S_z|
[f^{'sc}]_1(\lambda,\lambda)S'_1\rangle $$ $$C^{I_1}_{I_z,\frac{1}{2}} C^{I'_1}_{I_z,\frac{1}{2}}\langle q q_{\alpha}(\lambda_y,\mu_y)S_y|V| q q_{\alpha}(\lambda_y,\mu_y)S_y\rangle $$

The isospin $U$-function is unity, but it was introduced to indicate the angular momentum constraint on $I_z$, which in tern constrains the two particle spin isospin symmetry in the CFP. Clearly $(\lambda_y,\mu_y)=(2,0),(0,1)$
$$
V_2=\sum_{x}\langle q^3[f^{sc}]_1(\lambda,\lambda)S_1I,\,\bar{q}_{\beta}(0,1)\frac{1}{2}S_x;(10)|V|\langle q^3[f^{'sc}]_1(\lambda,\lambda)S'_1I,\,\bar{q}_{\beta}(0,1)\frac{1}{2}S'_x;(10)\rangle$$ $$ U(S_1,\frac{1}{2}S\frac{1}{2};S_x,S_2) U(S'_1,\frac{1}{2}S\frac{1}{2};S'_x,S'_2)=$$  $$3\sum_{x,z,y} \frac{1}{\sqrt{n_{[f^{sc}]}n_{[f^{'sc}]}}}U(S_1,\frac{1}{2}S\frac{1}{2};S_x,S_2) U(S'_1,\frac{1}{2}S\frac{1}{2};S'_x,S'_2) U(S_z,\frac{1}{2}S_x\frac{1}{2};S_1,S_y) $$ $$ U(S_z,\frac{1}{2}S'_x\frac{1}{2};S'_1,S_y) U((\lambda_z,\mu_z)(1,0)(0,1)(0,1)(\lambda,\lambda)(\lambda_y,\mu_y))^2U(I_z,\frac{1}{2},I,0,I,\frac{1}{2})^2$$ $$\langle [f_1]_2(\lambda_z,\mu_z )S_z|[f^{sc}]_1(\lambda,\lambda)S_1\rangle \langle [f_1]_2(\lambda_z,\mu_z)S_z|
[f^{'sc}]_1(\lambda,\lambda)S'_1\rangle $$ $$C^{I_1}_{I_z,\frac{1}{2}} C^{I'_1}_{I_z,\frac{1}{2}}\langle q \bar{q}_{\beta}(\lambda_y,\mu_y)S_y|V|q\bar{q}_{\beta}(\lambda_y,\mu_y)S_y\rangle $$
Now $(\lambda_y,\mu_y)=(1,1),(0,0)$

Finally in the simple case of three identical spin 1/2 quarks we have the following simple cases:\\
i) I=1/2\\
$$\langle [2,1](1,1)\frac{1}{2}|V|[2,1](1,1)\frac{1}{2}\rangle=3 \times \frac{1}{2}\times\frac{1}{2} $$ $$\left ( \langle (2,0)1|V| (2,0)1\rangle+\langle (2,0)0|V| (2,0)0\rangle+\frac{1}{3}\langle (0,1)1|V (0,1)1\rangle+\langle (0,1)0|V| (0,1)0\rangle\right ) $$
( 3 is the number of pairs one can make out of 3 particles, the first 1/2 is the ratio of the dimensions of the symmetry group involved, the last 1/2 is the square of the CFP, while 1/3 is the relevant isisp;in CFP). Similarly
$$\langle [2,1](1,1)\frac{3}{2}|V|[2,1](1,1)\frac{3}{2}\rangle= 3 \times \frac{1}{2}\times 1\left ( \langle (2,0)1|V| (2,0)1\rangle+\frac{1}{3}\langle (0,1)1|V (0,1)1\rangle\right ) $$
$$\langle [2,1](0,0)\frac{1}{2}|V|[2,1](0,0)\frac{1}{2}\rangle=  3\times \frac{1}{2} \times 1 \left (\frac{1}{3}\langle (0,1)1|V |(0,1)1\rangle +(0,1)0|V| (0,1)0\rangle \right )$$
i) I=3/2\\
$$\langle [1^3](1,1)\frac{1}{2}|V|[1^3](1,1)\frac{1}{2}\rangle= 3 \times 1 \times \frac{1}{2} \left (\langle (2,0)0|V| (2,0)0\rangle+\frac{1}{3}\langle (0,1)1|V| (0,1)1\rangle \right )$$
$$\langle [1^3](0,0)\frac{3}{2}|V|[1^3](0,0)\frac{3}{2}\rangle=  3\times 1 \times 1 \times \frac{1}{3}\langle (0,1)1|V| (0,1)1\rangle $$
\subsection{A pentaquark with the quantum numbers of the proton}
We will consider a pentaquatk with spin $1/2$ and isospin 1/2. We will demand that there will be at least three quarks of the type $q$ with isospin I=1/2. Then we have the following possibilities:\\
two states of the form 
$$q^4[2^2] (1,0)S_1={1},I_1=0\times \bar{q} \mbox{ and } q^4[2,1^2] (1,0)S_1={1},I_1=1\times \bar{q}$$
and five states of the form:
$$q^3[2,1] (\lambda,\lambda)S_1={1},I_1=\frac{1}{2}\times \left(q_{\alpha}\bar{q}_{\alpha}(\lambda,\lambda),S_2\right),\alpha=s,c,b,t$$
where $(\lambda,\lambda) $ are the self conjugate color functions (1,1), octet, and  (0,0), singlet. The order of the states is 
$$(1,1)S_1=\frac{1}{2},S_2=0,\,(1,1)S_1=\frac{1}{2},S_2=1,\,(1,1)S_1=\frac{3}{2},S_2=1,$$ $$(0,0)S_1=\frac{1}{2},S_2=0,\,(0,0)S_1=\frac{1}{2},S_2=1$$
states of the form 
$q^4[f^{sc}]\times \bar{q}_{\alpha}$ cannot yield an $I=1/2$ state and states of the form:
$$q^3[2,1] (\lambda,\lambda)S_1={1},I_1=\frac{1}{2}\times \left(q_{\alpha}\bar{q}_{\beta}(\lambda,\lambda),S_1\right),\alpha, \beta=s,c,b,t$$
essentially do not require any new symmetry input and they will not be considered in this discussion. Also states of the form 
$$\left [ q^2[f^{sc}] (\lambda_x,\mu_x)S_x,I_x\times q_{\beta}\right ]^{(\lambda,\lambda),S_1,I_1} \times \left(q_{\alpha}\bar{q}_{\alpha}(\lambda,\lambda),S_2\right), $$
like the pentaquark actually discovered,  involve only more complicated combinatorics in the evaluation of the relevant matrix elements  and do not require any additional CFP's. So these are not going to be included in our discussion, except to say that it will be interesting to see whether the charmonium component corresponds to one of the eigenstates of the type $ {\cal B}$ matrix discussed below. We should also mention that other types of symmetries maybe involved in evaluation  the energies resulting from realistic wave functions, obtained, e.g.,  with a linear potential, can be computed as perturbations of the Coulomb potential using the non compact SO[2,1] algebra \cite{Vergdos16}, using standard techniques \cite{Bied62},\cite{ACP88}$.$ 

Anyway the above  Hamiltonian matrix takes the form:
$${\cal H}=\left (\begin{array}{cc} {\cal A}&{\cal C}\\{\cal C}^T&{\cal B}\\ \end{array} \right ) \mbox { with }{\cal A} \,(2\times 2) \mbox{ matrix}, {\cal C} \,(2\times 5) \mbox{ matrix}, {\cal B}\, (5\times 5) \mbox{ matrix}$$
In evaluating the matrix ${\cal A}$ we need the following combined spin-color and isospin CFP's
$$\begin{array}{cccccc}
 [2^2](1,0)1:\quad &\frac{1}{\sqrt{2}},\,&-\frac{1}{\sqrt{2}},\,&\frac{1}{\sqrt{6}}\times \frac{1}{\sqrt{3}},\,&\frac{1}{\sqrt{6}}\times \frac{1}{\sqrt{3}},\,&\sqrt{\frac{2}{3}}\times \frac{1}{\sqrt{3}}\\
 \mbox{[2},1^2](1,0)1:\quad &0,\,&0,\,&\frac{1}{\sqrt{3}}\times \frac{1}{\sqrt{3}},\,&\frac{1}{\sqrt{3}}\times \frac{1}{\sqrt{3}},\,&-\frac{1}{\sqrt{3}}\times\frac{1}{\sqrt{3}}\\ \end{array}$$
with ratios of the dimensions of the symmetric group $\frac{1}{2}$ and $\frac{1}{3}$ respectively and 6 particle combinations allowed.
In the order of the pairs with spin color combinations:
$$(0,1)1,(2,0)1,\,(2,0)1,(0,1)0,\, (2,0)1,(0,1)0,\, (2,0)0,(0,1)1,\, (0,1)1,(2,0)0,\,(0,1)1,(0,1)1$$
We also need the spin-color and isospin 1-paricle CFP's:
$$\begin{array}{cccc}
 [2^2](1,0)1:\quad \frac{1}{\sqrt{3}}\times \frac{1}{\sqrt{2}},\,&-\frac{1}{\sqrt{3}}\times\frac{1}{\sqrt{2}},\,-\frac{1}{\sqrt{3}}\times \frac{1}{\sqrt{2}}\\
 \mbox{[2},1^2](1,0)1:\quad \frac{\sqrt{2}}{\sqrt{3}}\times \frac{1}{\sqrt{2}},\,&\frac{1}{\sqrt{6}}\times \frac{1}{\sqrt{2}},\,\frac{1}{\sqrt{6}}\times \frac{1}{\sqrt{2}}\\
 \end{array}$$
in the order of $(1,1) \frac{1}{2},\,(1,1) \frac{3}{2},\,(0,0) \frac{1}{2}$ states of spin-color symmetry [2,1].
with dimensions of the representation of the symmetric group 1 and $1/\sqrt{3}$ respectively and number combinations 3. With this information and using the formulas  of case i)  section \ref{sec:meint} we can construct the matrix  ${\cal A}$. We can also constrct the matrix  ${\cal C}$, using the formulas of case  iii).

We come next to the evaluation of matrix  ${\cal B}$, using the formulas pf case v) of section \ref{sec:meint}. We now need the following spin-color and isospin 1-particle CFP's:
$$\begin{array}{crrrr}
 [2,1](1,1)\frac{1}{2}:\quad & \frac{1}{\sqrt{2}},\,&-\frac{1}{\sqrt{2}},\,&\frac{1}{\sqrt{2}}\frac{1}{\sqrt{3}},\,&\frac{1}{\sqrt{2}}\times \frac{1}{\sqrt{3}}\\
 \mbox{[2},1](1,1)\frac{3}{2}:\quad & 1,\,&0,\,&0,\,&-1\times \frac{1}{\sqrt{3}}\\
 \mbox{[2},1](0,0)\frac{3}{2}:\quad & 0,\,&1,\,&0,\,&1\times \frac{1}{\sqrt{3}}\\
	\end{array}$$
\begin{table}[h!]
\begin{center}
\caption{ The one particle CFP's involving the spin color symmetry, needed in the case of a 4-quark state,  with color symmetry (10) and spin 1.}
\label{tab:MatrixABC}
$$
{\cal A}=\left(
\begin{array}{cc}
 \frac{8}{3} & \frac{2}{3 \sqrt{3}} \\
 \frac{2}{3 \sqrt{3}} & -\frac{26}{9} \\
\end{array}
\right),\,{\cal C}=\left(
\begin{array}{ccccc}
 -\sqrt{\frac{2}{3}} & \frac{\sqrt{\frac{2}{3}}}{3} &
   -\frac{\sqrt{\frac{2}{3}}}{3} & \frac{8 \sqrt{\frac{2}{3}}}{3} & 0 \\
 -\frac{2}{\sqrt{3}} & \frac{2}{3 \sqrt{3}} & \frac{1}{3 \sqrt{3}} & -\frac{8}{3
   \sqrt{3}} & 0 \\
\end{array}
\right)
$$
$$
{\cal B}=\left(
\begin{array}{ccccc}
 -\frac{214}{81} & -\frac{14}{81 \sqrt{3}} & \frac{14}{81 \sqrt{3}} & 0 & 0 \\
 -\frac{14}{81 \sqrt{3}} & -\frac{290}{243} & \frac{14}{243} & 0 & 0 \\
 \frac{14}{81 \sqrt{3}} & \frac{14}{243} & -\frac{1276}{243} & -\frac{56}{81
   \sqrt{3}} & -\frac{56}{243} \\
 0 & 0 & -\frac{56}{81 \sqrt{3}} & -\frac{40}{81} & -\frac{8}{81 \sqrt{3}} \\
 0 & 0 & -\frac{56}{243} & -\frac{8}{81 \sqrt{3}} & -\frac{752}{243} \\
\end{array}
\right)
$$
\end{center}
\end{table}
These matrices are expressed in units of the relevant radial integrals, e.g. $I_{\mbox{{\tiny HO}}}$ or  $I_{\mbox{{\tiny bag}}}$ (see Eqs \ref{Eq:IHO} and \ref{Eq:MITbag}).

To proceed further one must also add tp the Hamiltonian the 1-body terms arsing from the quark masses. So in the case of heavy quarks the mixing is small. So we will not proceed further in this direction. So we will draw some conclusions based on the form of ${\cal A}$ and ${\cal B}$. In the firsr place we see that in the case of ${\cal A}$ there is very little mixing. The lowest state is almost pure  $[2,1^2](1,0)1$ with the $[2^2](1,0)1$ being higher at  5.5 times the radial integral.

Regarding the spectrum of of ${\cal B}$ we find the eigenvalues
$$ \{-5.31,-3.07,-2.64,-1.19,-0.46\}$$ in units of the radial integrals. The lowest state is predominantly $|q^3([2,1](1,1)S_1=\frac{3}{2}\times (11)S_2=1 \rangle$
\section{symmetries of six-quark clusters}
To further illustrate the role of the symmetries  involved in multi-quark structures, we will discuss here in some detail the case of the  six-quark clusters, which can be present in the nucleus. We will restrict ourselves in the case that the six quark cluster has the same quantum numbers with those of the two nucleon system, i.e. being colorless, $(\lambda,\mu)=(0,0)$, and with  isospin 0 or 1.

 In describing the six quark clusters we will use standard group theory. This involves:
\begin{enumerate}
\item The isospin structure.\\
Here both the initial and the final cluster have isospin $I=1$, with $I_3=-1$ (initial state with the quantum numbers of two neutrons) and $I_{3}=1$ (final state with the quantum numbers
of two protons). The corresponding $SU(2)$ symmetry is described by a Young Tableaux, which has at most 2 rows
$\tilde{f}=[42],\,[3,3]$ corresponding to isospin $I=1,\,0$ respectively.
\item{The orbital spin color structure.}\\
In order to get a totally antisymmetric wave function the overall symmetry must be the conjugate of the above, namely $f=[2^2,1^2]$. This symmetry can be described by the product
of an orbital symmetry$f_L$ associated with $SU(6)$ and a color spin symmetry $f_{cs}$
also associated with $SU(6)$.
\item Possible $f_{cs}$ structures.\\
Clearly $f_{cs}$ can be further analyzed \cite{SOSTRO} in terms of
the symmetries $SU_{s}(2)$ for the spin and $SU_c(3)$ for the color group. Only the trivial
representations of the latter symmetry, namely the $s=0,1$ and the color singlet $(0,0)$ of
$SU_c(3)$ are relevant.
\item The possible $f_L$ structures.\\
We will consider the quarks moving in a confining potential. We will choose a basis set
obtained in a harmonic oscillator potential considering positive parity states with
$E\leq 2\hbar \omega$ excitations. We thus encounter:
\begin{itemize}
\item $0s^6$ configurations. The allowed symmetry is $f_L=[6]$ with $L=0$
\item $0s^51s$ configurations. The allowed symmetries are $f_L=[6]$ and $[5,1]$ with $L=0$.
\item $0s^2,0p^2$ configurations. The allowed symmetries are $f_{L}=[6,0],[5,1]$ and $[4,2]$
with $L=0,2$.
\end{itemize}
Using the standard group theory  techniques we find the following color singlets\cite{SOSTRO}:
$$f_{L}=[6]~,~f_{cs}=[2^2,1^2]\mbox{ with } (\lambda,\mu)=(0,0) \mbox{ and }s=0,2,\, I=1$$
$$f_{L}=[6]~,~f_{cs}=[2^3]\mbox{ with } (\lambda,\mu)=(0,0) \mbox{ and }s=1,3, I=0$$

 $$f_{L}=[51]~,~f_{cs}=[321]\mbox{ with } (0,0) \mbox{ and }s=1,2$$
It is not relevant since it has more than two columns.
 $$f_{L}=[51]~,~f_{cs}=[2^3]\mbox{ with } (\lambda,\mu)=(0,0) \mbox{ and } s=1,3,\, I=0$$
 $$f_{L}=[51]~,~f_{cs}=[2^21^2]\mbox{ with } (\lambda,\mu)=(0,0) \mbox{ and }s=0,2. \,I=1$$
 $$f_{L}=[51]~,~f_{cs}=[31^3]\mbox{ with } (\lambda,\mu)=(0,0) \mbox{ and }s=1$$
It is not relevant since it has more than two columns.
  $$f_{L}=[51]~,~f_{cs}=[21^4]\mbox{ with } (\lambda,\mu)=(0,0) \mbox{ and }s=1$$
	It is not relevant since it is characterized by isospin I=2.
	
 $$f_{L}=[42]~,~f_{cs}=[321]\mbox{ with } (\lambda,\mu)=(0,0) \mbox{ and }s=1,2$$
It is not relevant since it has more than two columns.
  $$f_{L}=[42]~,~f_{cs}=[41^2]\mbox{ with } (\lambda,\mu)=(0,0) \mbox{ and }s=0$$
	It is not relevant since it has more than two columns.
 $$f_{L}=[42]~,~f_{cs}=[2^3]\mbox{ with } (\lambda,\mu)=(0,0) \mbox{ and }s=1,3,\,I=0$$
  $$f_{L}=[42]~,~f_{cs}=[31^3]\mbox{ with } (\lambda,\mu)=(0,0) \mbox{ and }s=1$$
	It is not relevant since it has more than two columns.
   $$f_{L}=[42]~,~f_{cs}=[2^21^2]\mbox{ with } (\lambda,\mu)=(0,0) \mbox{ and }s=0,2,\,I=1$$
 $$f_{L}=[42]~,~f_{cs}=[21^4]\mbox{ with } (\lambda,\mu)=(0,0) \mbox{ and }s=1,\,I=2$$
It is not relevant since it does not have an isospin  quantum number corresponding to a  two nucleon system .
 $$f_{L}=[42]~,~f_{cs}=[33]\mbox{ with } (\lambda,\mu)=(0,0) \mbox{ and }s=0$$
It is not relevant since it has more than two columns.
\end{enumerate}

\section{Obtaining the lowest  six quark cluster.}
In order to achieve this goal,  one must diagonalize the QCD Hamiltonian in the above basis
\cite{YBFA}, using the standard
two particle coefficients of fractional parentage  for the
orbital part, for the spin color part  and for isospin (for notation see sec. \ref{section.notation} below).

 Those involved in the symmetry $f_{cs}$ to $SU_c(3)\times SU_I(2)$ can be obtained from the one particle CFP's 
calculated by  So and Strottman \cite{SOSTRO}.

\begin{table}[h!]
\begin{center}
\caption{ The one particle CFP's involving the spin color symmetry, needed in the case of a 6-quark cluster,  colorless with  spins 0,1,2  and 3.}
\label{tab:spc6cfp5}
$$
\begin{array}{|c|c|c|}
\hline
[f_1](\lambda_1,\mu_1)S_1&[f](\lambda,\mu)S&\mbox{CFP}\\
\hline
\mbox{$[2^2,1](0,1)\frac{1}{2}$}&[2^3](00)1&\frac{\sqrt{5}}{3}\\
\mbox{$[2^2,1](0,1)\frac{3}{2}$}&[2^3](00)1&\frac{2}{3}\\
\hline
\end{array}\quad
\begin{array}{|c|c|c|}
\hline
[f_1](\lambda_1,\mu_1)S_1&[f](\lambda,\mu)S&\mbox{CFP}\\
\hline
\mbox{$[2^2,1](0,1)(0,1)\frac{5}{2}$}&[2^3](00)3&1\\
\hline
\end{array}
$$
$$
\begin{array}{|c|c|c|}
\hline
[f_1](\lambda_1,\mu_1)S_1&[f](\lambda,\mu)S&\mbox{CFP}\\
\hline
\mbox{$[2^2,1](0,1)\frac{1}{2}$}&[2^21^2](00)0&1\\
\hline
\end{array}\quad
\begin{array}{|c|c|r|}
\hline
[f_1](\lambda_1,\mu_1)S_1&[f](\lambda,\mu)S&\mbox{CFP}\\
\hline
\mbox{$[2^2,1](0,1)\frac{3}{2}$}&[2^21^2](00)2&\frac{4}{5}\\
\mbox{$[2^2,1](0,1)\frac{5}{2}$}&[2^2,1^2](00)2&-\frac{3}{5}\\
\hline
\end{array}
$$
\end{center}
\end{table} 
\begin{table}[h!]
\begin{center}
\caption{ The two particle CFP's involving the spin color symmetry, needed in the case of a 6-quark cluster,  colorless with  spin zero and 2.}
\label{tab:spc6cfp3}
$$
\begin{array}{|c|c|c|c|}
\hline
[f_1](\lambda_1,\mu_1)S_1&[f_2](\lambda_2,\mu_2)S_2&[f](\lambda,\mu)S&\mbox{CFP}\\
\hline
\mbox{$[2,2](0,2)0$}&[1^2](2,0)0&[2^21^2](00)0&-\frac{\sqrt{3}}{2}\\
\mbox{$[2,2](0,2)0$}&[1^2](0,1)1&[2^21^2](00)0&\frac{1}{2}\\
\hline
\end{array}
\begin{array}{|c|c|c|c|}
\hline
[f_1](\lambda_1,\mu_1)S_1&[f_2](\lambda_2,\mu_2)S_2&[f](\lambda,\mu)S&\mbox{CFP}\\
\hline
\mbox{$[1^4](0,2)0$}&[1^2](2,0)0&[2^21^2](00)0&-\sqrt{\frac{3}{5}}\\
\mbox{$[1^4](1,0)1$}&[1^2](0,1)1&[2^21^2](00)0&-\sqrt{\frac{2}{5}}\\
\hline
\end{array}
$$
$$
\begin{array}{|c|c|c|c|}
\hline
[f_1](\lambda_1,\mu_1)S_1&[f_2](\lambda_2,\mu_2)S_2&[f](\lambda,\mu)S&\mbox{CFP}\\
\hline
\mbox{$[2,1^2](0,2)1$}&[2](2,0)1&[2^21^2](00)0&\frac{1}{\sqrt{2}}\\
\mbox{$[2,1^2](0,2)1$}&[2](2,0)1&[2^21^2](00)0&-\frac{1}{\sqrt{2}}\\
\hline
\end{array}
\begin{array}{|c|c|c|c|}
\hline
[f_1](\lambda_1,\mu_1)S_1&[f_2](\lambda_2,\mu_2)S_2&[f](\lambda,\mu)S&\mbox{CFP}\\
\hline
\mbox{$[2,1^2](0,2)1$}&[1^2](0,1)1&[2^21^2](00)2&1\\
\hline
\end{array}
$$
\end{center}
\end{table}

\begin{table}[h!]
\begin{center}
\caption{ The two particle CFP's involving the spin color symmetry, needed in the case of a 6-quark cluster,  colorless with  spin two.}
\label{tab:spc6cfp4}
$$
\begin{array}{|c|c|c|c|}
\hline
[f_1](\lambda_1,\mu_1)S_1&[f_2](\lambda_2,\mu_2)S_2&[f](\lambda,\mu)S&\mbox{CFP}\\
\hline
\mbox{$[2,2](0,2)2$}&[1^2](2,0)0&[2^21^2](00)2&\sqrt{\frac{3}{5}}\\
\mbox{$[2,2](1,0)1$}&[1^2](0,1)1&[2^21^2](00)2&-\sqrt{\frac{2}{5}}\\
\hline
\end{array}
\begin{array}{|c|c|c|c|}
\hline
[f_1](\lambda_1,\mu_1)S_1&[f_2](\lambda_2,\mu_2)S_2&[f](\lambda,\mu)S&\mbox{CFP}\\
\hline
\mbox{$[2,1^2](0,2)1$}&[2](2,0)1&[2^21^2](00)2&-\sqrt{\frac{4}{5}}\\
\mbox{$[2,1^2](1,0)2$}&[2](0,1)0&[2^21^2](00)0&-\sqrt{\frac{1}{5}}\\
\hline
\end{array}
$$
$$
\begin{array}{|c|c|c|c|}
\hline
[f_1](\lambda_1,\mu_1)S_1&[f_2](\lambda_2,\mu_2)S_2&[f](\lambda,\mu)S&\mbox{CFP}\\
\hline
\mbox{$[2,1^2](1,0)1$}&[1^2](0,1)1&[2^21^2](00)2&\sqrt{\frac{2}{5}}\\
\mbox{$[2,1^2](1,0)2$}&[1^2](0,1)2&[2^21^2](00)2&-\sqrt{\frac{3}{5}}\\
\hline
\end{array}
\begin{array}{|c|c|c|c|}
\hline
[f_1](\lambda_1,\mu_1)S_1&[f_2](\lambda_2,\mu_2)S_2&[f](\lambda,\mu)S&\mbox{CFP}\\
\hline
\mbox{$[1^4](1,0)1$}&[1^2](0,1)1&[2^21^2](00)2&1\\
\hline
\end{array}
$$
\end{center}
\end{table}

\begin{table}[h!]
\begin{center}
\caption{ The two particle CFP's involving the spin color symmetry, needed in the case of a 6-quark cluster,  colorless with  spin two.}
\label{tab:spc6cfp4}
$$
\begin{array}{|c|c|c|c|}
\hline
[f_1](\lambda_1,\mu_1)S_1&[f_2](\lambda_2,\mu_2)S_2&[f](\lambda,\mu)S&\mbox{CFP}\\
\hline
\mbox{$[2,2](0,2)2$}&[1^2](2,0)0&[2^21^2](00)2&\sqrt{\frac{3}{5}}\\
\mbox{$[2,2](1,0)1$}&[1^2](0,1)1&[2^21^2](00)2&-\sqrt{\frac{2}{5}}\\
\hline
\end{array}
\begin{array}{|c|c|c|c|}
\hline
[f_1](\lambda_1,\mu_1)S_1&[f_2](\lambda_2,\mu_2)S_2&[f](\lambda,\mu)S&\mbox{CFP}\\
\hline
\mbox{$[2,1^2](0,2)1$}&[2](2,0)1&[2^21^2](00)2&-\sqrt{\frac{4}{5}}\\
\mbox{$[2,1^2](1,0)2$}&[2](0,1)0&[2^21^2](00)0&-\sqrt{\frac{1}{5}}\\
\hline
\end{array}
$$
$$
\begin{array}{|c|c|c|c|}
\hline
[f_1](\lambda_1,\mu_1)S_1&[f_2](\lambda_2,\mu_2)S_2&[f](\lambda,\mu)S&\mbox{CFP}\\
\hline
\mbox{$[2,1^2](1,0)1$}&[1^2](0,1)1&[2^21^2](00)2&\sqrt{\frac{2}{5}}\\
\mbox{$[2,1^2](1,0)2$}&[1^2](0,1)2&[2^21^2](00)2&-\sqrt{\frac{3}{5}}\\
\hline
\end{array}
\begin{array}{|c|c|c|c|}
\hline
[f_1](\lambda_1,\mu_1)S_1&[f_2](\lambda_2,\mu_2)S_2&[f](\lambda,\mu)S&\mbox{CFP}\\
\hline
\mbox{$[1^4](1,0)1$}&[1^2](0,1)1&[2^21^2](00)2&1\\
\hline
\end{array}
$$
\end{center}
\end{table}

\begin{table}[h!]
\begin{center}
\caption{ The two particle CFP's involving the spin color symmetry, needed in the case of a 6-quark cluster,  colorless with  spins 1 and 3.}
\label{tab:spc6cfp5}
$$
\begin{array}{|c|c|c|c|}
\hline
[f_1](\lambda_1,\mu_1)S_1&[f_2](\lambda_2,\mu_2)S_2&[f](\lambda,\mu)S&\mbox{CFP}\\
\hline
\mbox{$[2,2](0,2)0$}&[2](2,0)1&[2^3](00)1&\sqrt{\frac{5}{12}}\\
\mbox{$[2,2](0,2)2$}&[2](2,0)1&[2^3](00)1&\frac{1}{\sqrt{6}}\\
\mbox{$[2,2](1,0)0$}&[2](0,1)0&[2^3](00)1&\sqrt{\frac{5}{12}}\\
\hline
\end{array}
\begin{array}{|c|c|c|c|}
\hline
[f_1](\lambda_1,\mu_1)S_1&[f_2](\lambda_2,\mu_2)S_2&[f](\lambda,\mu)S&\mbox{CFP}\\
\hline
\mbox{$[2,1^2](0,2)0$}&[1^2](2,0)0&[2^3](00)1&-\sqrt{\frac{5}{18}}\\
\mbox{$[2,1^2](1,0)0$}&[1^2](0,1)1&[2^3](00)1&-\sqrt{\frac{5}{54}}\\
\mbox{$[2,1^2](1,0)1$}&[1^2](0,1)1&[2^3](00)1&-\sqrt{\frac{5}{9}}\\
\mbox{$[2,1^2](1,0)2$}&[1^2](0,1)1&[2^3](00)1&-\sqrt{\frac{2}{27}}\\
\hline
\hline
\end{array}
$$
$$
\begin{array}{|c|c|c|c|}
\hline
[f_1](\lambda_1,\mu_1)S_1&[f_2](\lambda_2,\mu_2)S_2&[f](\lambda,\mu)S&\mbox{CFP}\\
\hline
\mbox{$[2^2](0,2)2$}&[2](2,0)1&[2^3](00)3&1\\
\hline
\end{array}
\begin{array}{|c|c|c|c|}
\hline
[f_1](\lambda_1,\mu_1)S_1&[f_2](\lambda_2,\mu_2)S_2&[f](\lambda,\mu)S&\mbox{CFP}\\
\hline
\mbox{$[2,1^2](0,2)2$}&[1^2](0,1)1&[2^3](00)3&1\\
\hline
\end{array}
$$
\end{center}
\end{table}

 From those we obtained the needed 2-particle CFP's which are given in tables \ref{tab:spc6cfp3} and \ref{tab:spc6cfp4}

In some cases the corresponding 1-particle cfp's for the orbital part may
be needed. For the cases of interest in the present study they are shown in tables
\ref{table.onecfp1},\ref{table.onecfp2} and \ref{table.onecfp3}.
 \begin{table}[t]
\caption{
 The  2-particle coefficients of fractional parentage (cfp) involved in the orbital symmetries [6] and [51] for the configurations $0s^5\ell $ for $\ell =1s$ and  $0d$. See the main text for notation.}
 \label{table.cfp1}
\begin{center}
\begin{tabular}{rrrrrrrr}
$[f_1]$&$[f_2]$&$[6]$&$[51]_1$&$[51]_2$&$[51]_3$&$[51]_4$&$[51]_5$\\
\hline
 $s^3\ell [3,1]_1$&[2]&0&1 & 0 & 0 & 0 & 0 \\
 $s^3\ell [3,1]_2$&[2]& 0&1 &0 & 0 & 0 & 0 \\
 $s^3\ell [3,1]_3$&[2]&0& 0 & 1  & 0 & 0 &0\\
$s^3\ell [4]$&[2]&$\sqrt{\frac{2}{3}}$& 0 & 0 & 0 &$-\frac{1}{\sqrt{5}}$&$-\sqrt{\frac{2}{15}}$      \\
 $4$&$s\ell$ [2]&$\frac{1}{\sqrt{3}}$&0&0&0&$\sqrt{\frac{2}{5}}$ &$\frac{2}{\sqrt{15}}$\\
 $4$& $s\ell$&0& 0 &0&0& $\sqrt{\frac{2}{5}}$&$-\frac{3}{\sqrt{15}}$   \\
 \end{tabular}
\end{center}
\end{table}
 \begin{table}[t]
\caption{
 The  2-particle coefficients of fractional parentage (cfp) involved in the orbital symmetries [6] and [51] for the configurations $0s^40p^2$. See the main text for notation.}
 \label{table.cfp2}
\begin{center}
\begin{tabular}{rrrrrrrr}
$[f_1]$&$[f_2]$&$[6]$&$[51]_1$&$[51]_2$&$[51]_3$&$[51]_4$&$[51]_5$\\
\hline
  $s^2p^2[2,2]_1$&[2]&0&0 & 0 & 0 & 0 & 0 \\
 $s^2p^2[2,2]_2$&[2]& 0&0 & 0 & 0 & 0 & 0 \\
 $s^2p^2[3,1]_1$&[2]&0&$\frac{1}{\sqrt{2}}$ & 0 & 0 & 0 & 0 \\
$s^2p^2[3,1]_2$&[2]& 0&0 & $\frac{1}{\sqrt{2}}$ & 0 & 0 & 0 \\
$s^2p^2[3,1]_3$&[2]&0& 0 & 0 & $\frac{1}{\sqrt{2}}$ & 0 & 0 \\
$s^2p^2[4]$&[2]&$\sqrt{\frac{2}{3}}$& 0 & 0 & 0 &$ -\sqrt{\frac{3}{10}}$ & $-\frac{1}{\sqrt{5}}$
   \\
 $s^3p[3,1]_1$&sp[2]&0& $\frac{1}{\sqrt{2}}$ & 0 & 0 & 0 & 0 \\
 $s^3p[3,1]_2$&sp[2]&0& 0 &$\sqrt{\frac{2}{3}}$ & 0 & 0 & 0 \\
 $s^3p[3,1]_3$&sp[2]&0& 0 & 0 & $\frac{1}{\sqrt{2}}$ & 0 & 0 \\
 $s^3p[4]$&sp[2]&$\sqrt{\frac{8}{15}}$&0 & 0 & 0 & $\frac{1}{\sqrt{10}} $& $\frac{1}{\sqrt{15}}$
   \\
$s^3p[3,1]_1$&sp[11]&0& 0 & 0 & 0 & 0 & 0 \\
$s^3p[3,1]_2$&sp[11]&0& 0 & 0 & 0 & 0 & 0 \\
$s^3p[3,1]_3$&sp[11]&0&0 & 0 & 0 & 0 & 0 \\
$s^3p[4]$&sp[11]&0& 0 & 0 & 0 &$ -\sqrt{\frac{2}{5}}$ & $\sqrt{\frac{3}{5}}$
   \\
 $[4]$&$p^2$[2]& $\sqrt{\frac{1}{15}}$&0& 0 & 0 & $\frac{1}{\sqrt{5}}$ &$ \sqrt{\frac{2}{15}}$
 \end{tabular}
\end{center}
\end{table}
\begin{table}[t]
\caption{
 The 2-particle coefficients of fractional parentage (cfp) involved in the orbital symmetry [42] for the configurations $0s^40p^2$. See the main text for notation.}
 \label{table.cfp3}
\begin{center}
\begin{tabular}{rrrrrrrrrrr}
$[f_1]$&$[f_2]$&$[42]_1$&$[42]_2$&$[42]_3$&$[42]_4$&$[42]_5$&$[42]_6$&$[42]_7$&$[42]_8$&$[42]_9$\\
\hline
 $s^2p^2[2,2]_1$&[2]& $\frac{1}{2}$ & 0 & 0 & $-\frac{\sqrt{3}}{4}$ & $-\frac{1}{4}$ & 0 & $\frac{1}{\sqrt{2}}$ & 0 & 0 \\
 $s^2p^2[2,2]_2$&[2]& $ \frac{\sqrt{3}}{2}$ & 0 & 0 & $\frac{1}{4}$ & $\frac{1}{4 \sqrt{3}}$ & 0 & $-\frac{1}{\sqrt{6}}$ & 0 & 0 \\
$s^2p^2[3,1]_1$&[2]& 0 & $-\frac{1}{2 \sqrt{3}}$ & $-\frac{1}{2 \sqrt{6}}$ &$ -\frac{1}{4 \sqrt{3}}$ & $-\frac{1}{12}$ & $\frac{1}{2
   \sqrt{6}}$ & $-\frac{1}{6 \sqrt{2}}$ & $-\frac{\sqrt{3}}{4}$ & $-\frac{\sqrt{\frac{5}{3}}}{4}$ \\
$s^2p^2[3,1]_2$&[2]& 0 & $-\frac{1}{2} $& $-\frac{1}{2 \sqrt{2}}$ & $\frac{1}{12}$ & $\frac{1}{12 \sqrt{3}}$ & $-\frac{1}{6 \sqrt{2}} $&
   $\frac{1}{6 \sqrt{6}}$ & $\frac{1}{4}$ & $\frac{\sqrt{5}}{12}$ \\
 $s^2p^2[3,1]_3$&[2]& 0 & 0 & 0 &$ -\frac{\sqrt{2}}{3}$ & $-\frac{\sqrt{\frac{2}{3}}}{3}$ &$ -\frac{1}{12}$ &$ -\frac{2}{3 \sqrt{3}}$ &
  $ \frac{1}{4 \sqrt{2}}$ & $\frac{\sqrt{\frac{5}{2}}}{12}$ \\
$s^2p^2[4]$&[2]& 0 & 0 & 0 & 0 & 0 & $\frac{1}{4}$ & 0 & $\frac{1}{4 \sqrt{2}}$ &$ -\frac{1}{4 \sqrt{10}}$ \\
 $s^3p[3,1]_1$&sp[2]&  0 &$ \frac{1}{2 \sqrt{3}}$ & $\frac{1}{2 \sqrt{6}}$ & $\frac{1}{4 \sqrt{3}}$ & $\frac{1}{12}$ & $-\frac{1}{2
   \sqrt{6}}$ & $\frac{1}{6 \sqrt{2}}$ & $\frac{\sqrt{3}}{4}$ & $\frac{\sqrt{\frac{5}{3}}}{4}$ \\
 $s^3p[3,1]_2$&sp[2]& 0 &$ \frac{1}{2}$ & $\frac{1}{2 \sqrt{2}}$ &$ -\frac{1}{12}$ &$ -\frac{1}{12 \sqrt{3}}$ & $\frac{1}{6 \sqrt{2}}$ &
  $ -\frac{1}{6 \sqrt{6}}$ &$ -\frac{1}{4} $& $-\frac{\sqrt{5}}{12}$ \\
$s^3p[3,1]_3$&sp[2]& 0 & 0 & 0 & $\frac{\sqrt{2}}{3}$ &$ \frac{\sqrt{\frac{2}{3}}}{3}$ & $\frac{1}{12} $& $\frac{2}{3 \sqrt{3}}$ &
  $ -\frac{1}{4 \sqrt{2}}$ & $-\frac{\sqrt{\frac{5}{2}}}{12}$ \\
$s^3p[4]$&sp[2]&0 & $-\frac{1}{2 \sqrt{3}}$ & $\frac{1}{\sqrt{6}}$ & $-\frac{1}{4 \sqrt{3}}$ & $\frac{1}{4}$ &
   $\frac{1}{\sqrt{6}}$ & 0 & $-\frac{1}{2 \sqrt{3}}$ &$ \frac{\sqrt{\frac{5}{3}}}{2}$ \\
$s^3p[3,1]_1$&sp[11]& 0 &$ -\frac{1}{2}$ &$ \frac{1}{\sqrt{2}}$ & $\frac{1}{12}$ & $-\frac{1}{4 \sqrt{3}}$ & $-\frac{1}{3 \sqrt{2}}$ & 0 & $\frac{1}{6}$ &
  $ -\frac{\sqrt{5}}{6}$ \\
$s^3p[3,1]_2$&sp[11]& 0 & 0 & 0 &$ -\frac{\sqrt{2}}{3}$ & $\sqrt{\frac{2}{3}} $&$ -\frac{1}{6} $& 0 & $\frac{1}{6 \sqrt{2}}$ &
   $-\frac{\sqrt{\frac{5}{2}}}{6}$ \\
 $s^3p[3,1]_3$&sp[11]&0 & 0 & 0 & 0 & 0 & $-\frac{\sqrt{3}}{4}$ & 0 &$ -\frac{\sqrt{\frac{3}{2}}}{4}$ & $\frac{\sqrt{\frac{3}{10}}}{4}$ \\
  $s^3p[4]$&sp[11]& 0 & 0 & 0 & 0 & 0 & 0 & 0 & 0 & 0 \\
 $[4]$&$p^2$[2]&0 & 0 & 0 & 0 & 0 & $\frac{\sqrt{\frac{3}{2}}}{2} $& 0 & $\frac{\sqrt{3}}{4}$ &
 $  -\frac{\sqrt{\frac{3}{5}}}{4}$
 \end{tabular}
\end{center}
\end{table}

 \begin{table}[t]
\caption{
 The 1-particle coefficients of fractional parentage (cfp) involved in the orbital symmetries [6] and [51] for the configurations $0s^5\ell $ with $\ell =$ 0s or 0d. The rows are labeled by the five particle symmetry $[f_1]$ and the angular momentum of the last particle. See the main text for notation.}
 \label{table.onecfp1}
\begin{center}
\begin{tabular}{rrrrrrrr}
$[f_1]$&$\ell $&$[6]$&$[51]_1$&$[51]_2$&$[51]_3$&$[51]_4$&$[51]_5$\\
\hline
 $[41]_1$&0s& 0 & 1 & 0 & 0 & 0 & 0 \\
 $[41]_1$&0s& 0 & 0 & 1 & 0 & 0 & 0 \\
 $[41]_1$&0s& 0 & 0 & 0 & 1 & 0 & 0 \\
 $[41]_1$&0s& 0 & 0 & 0 & 0 & 1 & 0 \\
 $[5]$&0s& $\frac{1}{\sqrt{6}}$ & 0 & 0 & 0 & 0 & $\sqrt{\frac{5}{6}}$ \\
 $[5]$&1s or 0d& $\sqrt{\frac{5}{6}}$ & 0 & 0 & 0 & 0 & $-\frac{1}{\sqrt{6}}$
  \end{tabular}
\end{center}
\end{table}
 \begin{table}[t]
\caption{
 The  1-particle coefficients of fractional parentage (cfp) involved in the orbital symmetries [6] and [51] for the configurations $0s^40p^2$. The rows are labeled by the five particle symmetry $[f_1]$ and the angular momentum of the last particle. See the main text for notation.}
 \label{table.onecfp2}
\begin{center}
\begin{tabular}{rrrrrrrr}
$[f_1]$&$\ell $&$[6]$&$[51]_1$&$[51]_2$&$[51]_3$&$[51]_4$&$[51]_5$\\
\hline
 $[41]_1$&0s&0 & $\frac{\sqrt{3}}{2} $& 0 & 0 & 0 & 0 \\
 $[41]_2$&0s& 0 & 0 & $\frac{\sqrt{3}}{2}$ & 0 & 0 & 0 \\
 $[41]_3$&0s& 0 & 0 & 0 &$ \frac{\sqrt{3}}{2} $& 0 & 0 \\
 $[41]_4$&0s& 0 & 0 & 0 & 0 &$ \frac{\sqrt{3}}{2}$ & 0 \\
  $[41]_1$&0p&0 & $\frac{1}{2}$ & 0 & 0 & 0 & 0 \\
  $[41]_2$&0p& 0 & 0 & $\frac{1}{2}$ & 0 & 0 & 0 \\
   $[41]_3$&0p&0 & 0 & 0 & $\frac{1}{2}$ & 0 & 0 \\
  $[41]_4$&0p& 0 & 0 & 0 & 0 &$ \frac{1}{2}$ & 0 \\
 $[5]$&0s& $\frac{1}{\sqrt{3}}$ & 0 & 0 & 0 & 0 &$ \sqrt{\frac{2}{3}}$ \\
  $[5]$&0p&$\sqrt{\frac{2}{3}}$ & 0 & 0 & 0 & 0 &$-\frac{1}{\sqrt{3}}$
  \end{tabular}
\end{center}
\end{table}
 \begin{table}[ht]
\caption{
 The  1-particle coefficients of fractional parentage (cfp) involved in the orbital symmetry [42] for the configurations $0s^40p^2$. The rows are labeled by the five particle symmetry $[f_1]$ and the angular momentum of the last particle. See the main text for notation.}
 \label{table.onecfp3}
\begin{center}
\begin{tabular}{rrrrrrrrrrr}
$[f_1]$&$\ell $&$[42]_1$&$[42]_2$&$[42]_3$&$[42]_4$&$[42]_5$&$[42]_6$&$[42]_7$&$[42]_8$&$[42]_9$\\
\hline
$[32]_1$&0s&1 & 0 & 0 & 0 & 0 & 0 & 0 & 0 & 0 \\
$[32]_2$&0s&0 & 1 & 0 & 0 & 0 & 0 & 0 & 0 & 0 \\
$[32]_3$&0s& 0 & 0 & 0 & 1 & 0 & 0 & 0 & 0 & 0 \\
$[32]_4$&0s& 0 & 0 & 0 & 0 &$ -\frac{1}{3}$ & 0 & $\frac{2 \sqrt{2}}{3}$ & 0 & 0 \\
$[32]_5$&0s& 0 & 0 & 0 & 0 & 0 &$ -\frac{1}{\sqrt{3}}$ & 0 & $\sqrt{\frac{2}{3}}$ &
   0 \\
$[41]_1$&0s&0 & 0 & $-\frac{1}{4} $& 0 & $-\frac{1}{3 \sqrt{6}}$ & $-\frac{1}{12}$ &
   $-\frac{1}{12 \sqrt{3}}$ &$-\frac{1}{12 \sqrt{2}}$ &
  $ -\frac{\sqrt{\frac{5}{2}}}{4}$ \\
$[41]_2$&0s& 0 & 0 & $-\frac{\sqrt{3}}{4}$ & 0 &$\frac{1}{9 \sqrt{2}}$ &
  $ \frac{1}{12 \sqrt{3}}$ & $\frac{1}{36}$ & $\frac{1}{12 \sqrt{6}}$ &
   $\frac{\sqrt{\frac{5}{6}}}{4}$ \\
$[41]_3$&0s& 0 & 0 & 0 & 0 & $-\frac{4}{9}$ & $\frac{1}{12 \sqrt{6}}$ &
   $-\frac{\sqrt{2}}{9} $& $\frac{1}{24 \sqrt{3}} $&
   $\frac{\sqrt{\frac{5}{3}}}{8}$ \\
$[41]_4$&0s& 0 & 0 & 0 & 0 & 0 &$ -\frac{\sqrt{\frac{5}{2}}}{4} $& 0 &
  $ -\frac{\sqrt{5}}{8}$ & $\frac{1}{8}$ \\
$ [41]_1$&0p&0 & 0 & $\frac{\sqrt{3}}{4}$ & 0 & $\frac{1}{3 \sqrt{2}}$ & $\frac{1}{4
   \sqrt{3}}$ &$ \frac{1}{12}$ & $\frac{1}{4 \sqrt{6}}$ &
  $ \frac{\sqrt{\frac{15}{2}}}{4}$ \\
$[41]_2$&0p& 0 & 0 & $\frac{3}{4}$ & 0 & $-\frac{1}{3 \sqrt{6}}$ & $-\frac{1}{12}$ &
  $ -\frac{1}{12 \sqrt{3}}$ & $-\frac{1}{12 \sqrt{2}}$ &
  $ -\frac{\sqrt{\frac{5}{2}}}{4}$ \\
$[41]_3$&0p& 0 & 0 & 0 & 0 & $\frac{4}{3 \sqrt{3}}$ & $-\frac{1}{12 \sqrt{2}}$ &
  $ \frac{\sqrt{\frac{2}{3}}}{3}$ & $-\frac{1}{24}$ &
   $-\frac{\sqrt{5}}{8}$ \\
$[41]_4$&0p& 0 & 0 & 0 & 0 & 0 & $\frac{\sqrt{\frac{15}{2}}}{4}$ & 0 &
   $\frac{\sqrt{15}}{8}$ & $-\frac{\sqrt{3}}{8}$
 \end{tabular}
\end{center}
\end{table}
 \section{Computing the mixing of the lowest six quark cluster and with the two nucleon wf.}
In this case we will concentrate on the orbital symmetry and transform the original wf
into the Jacobi coordinates defined as follows:
$$\xi_1=\frac{1}{\sqrt{2}}(\bf{r}_1-\bf{r}_2)$$
$$\eta_1=\frac{1}{\sqrt{6}}(\bf{r}_1+\bf{r}_2-2\bf{r}_3)$$
$$\xi_2=\frac{1}{\sqrt{2}}(\bf{r}_4-\bf{r}_5)$$
$$\eta_2=\frac{1}{\sqrt{6}}(\bf{r}_4+\bf{r}_5-2\bf{r}_6)$$
$$\bf{r}=\frac{1}{\sqrt{6}}(\bf{r}_1+\bf{r}_2+\bf{r}_3-\bf{r}_4-\bf{r}_5-\bf{r}_6)$$
$$\bf{r}=\frac{1}{\sqrt{6}}(\bf{r}_1+\bf{r}_2+\bf{r}_3+\bf{r}_4+\bf{r}_5+\bf{r}_6)$$
\section{Expressions regarding the orbital symmetry}
\label{section.notation}
\begin{itemize}
\item The representation $f=[6]$ is one-dimensional.
\item The representation $f=[5,1]$ is five-dimensional.\\
A basis can be found using the Yamanouchi symbols \cite{HAMMERMESH}. We chose them in the order $Y_1$=[111121],
$Y_2$=[111211] $Y_3$=[112111], $Y_4$=[121111], $Y_5$=[211111].
The corresponding states are not orthogonal. We orthogonalize them using the Gram-Schmidt procedure in the above i.e we start with first state $[5,1]_1\Leftrightarrow Y_1$, the second state $[5,1]_2$ is an orthogonal combination of the first two associated with $Y_1$ and $Y_2$
etc. The orthogonal states will be indicated as  $[51]_i$, i=1,2,...,5. Thus:
\beq
[51]_1=Y_1~,~[51]_2=\frac{1}{\sqrt{3}}\left (-Y_1+2Y_2 \right ),
[51]_3=\frac{\left (-Y_1-Y_2 +3Y_3 \right )}{\sqrt{6}},
\eeq
\beq
[51]_4=\frac{\left (-Y_1-Y_2-Y_3+4Y_4 \right )}{{\sqrt{10}}},~
[51]_5=\frac{\left (-Y_1-Y_2 -Y_3 -Y_4+5Y_5\right )}{\sqrt{15}}.
\eeq
\item The representation $f=[4,2]$ is nine-dimensional.\\
A basis is found using the Yamanouchi symbols: $Y_1$=[112211], $Y_2$=[121211],
$Y_3$=[211211], $Y_4$=[122111],
$Y_5$=[212111], $Y_6$=[221111], $Y_7$=[112121], $Y_8$=[121121], and $Y_9$=[211121].
 An orthonormal basis is obtained
as in the previous case. The obtained states will be denoted as $[4,2]_i$, i=1,2,...,9. Thus:
\beq
[42]_1=Y_1~,~[42]_2=\frac{\left (-Y_1+2Y_2 \right )}{\sqrt{3}},
~[42]_3=\frac{\left (-Y_1-Y_2 +3Y_3 \right )}{\sqrt{6}}
\eeq
\beq
[42]_4=\frac{\left (-Y_2+2Y_4 \right )}{\sqrt{3}},~
[42]_5=\frac{\left (Y_2-2 Y_3-2 Y_4+4Y_5 \right )}{3},
\eeq
\beq
~[42]_6=\frac{\left (-Y_3-Y_5 -3Y_6 \right )}{\sqrt{6}},[42]_7=\frac{\left (-3Y_1-Y_2-Y_3+2Y_4+2Y_5-6Y_7 \right )}{3\sqrt{2}},
\eeq
\beq
[42]_8=\frac{\left (2 Y_1-2Y_2-Y_3-2Y_4-Y_5+3Y_6-4Y_7 +6Y_8\right )}{2\sqrt{3}},
\eeq
\beq
[42]_9=\frac{\left (2 Y_1+2 Y_2-3Y_3+2Y_4-3Y_5-3Y_6-4Y_7 -6Y_8+12Y_9\right )}{2 \sqrt{15}}.
\eeq
\end{itemize}
 In computing the cfp's we need a description of the two particle and four particle states.
 If only $0s$ states are involved we use the standard labels [4],[31],[22] for the four
 particles and [2] and [11] for the two particles. If configurations other than $0s$
 are involved we attach to them suitable configuration labels. If the representation
 is more than one dimensional we specify the relevant Yamanouchi symbols.
 \begin{itemize}
 \item The representation [2,2].\\
 The relevant Yamanouchi symbols are $Y_1=[2121]$ and $Y_2$=[2211]. The first state $[2,2]_1$
 is associated with $Y_1$ and the other is orthogonal to it. Thus:
  \beq
[22]_1=Y_1~,~[22]_2=\frac{1}{\sqrt{3}}\left (-Y_1+2Y_2 \right ).
\eeq
 \item The representation [3,1].\\
 Now we have $Y_1$=[1121], $Y_1$=[1211], $Y_1$=[2111]. We orthogonalize them in this order
 $([3,1]\Leftrightarrow Y_1$ etc). Thus:
 \beq
[31]_1=Y_1~,~[31]_2=\frac{1}{\sqrt{3}}\left (-Y_1+2Y_2 \right ),
~[31]_3=\frac{1}{\sqrt{6}}\left (-Y_1-Y_2 +3Y_3 \right )
\eeq
 \end{itemize}
 In the case of 1-particle cfp's we need the 5-particle symmetries [5], [41] and [32]. The first is
 uniquely specified. In the case of [4,1] we follow a procedure analogous to that for [5,1] discussed above. In the case of [32] the five orthogonal states were obtained as follows:
$$ [32]_1=Y_1~,~[32]_2=(-Y_1+2Y_2)/\sqrt{3}~,~[32]_3=(-Y_2+2Y_3)/\sqrt{3}~,~$$
$$[32]_4=(-2Y1-Y_2+2Y_3+4Y_4)/3~,~ [32]_5=\frac{\sqrt{2}}{3}(Y_1-Y_2-Y_3-2Y_4+3Y_5)$$
where the Yamanouchi symbols are:
$Y_1=[12121]$, $Y_2=[21211]$, $Y_3=[22111]$, $Y_4=[12211]$, $Y_1=[21121]$
\subsection{Separating out the two nucleon like component of the orbital symmetry.}
We distinguish the following cases:
\begin{enumerate}
\item The symmetry $f=[6].$\\
We have the following possibilities:
\begin{itemize}
\item The configuration $0s^6$.
\beq
|0s^6~f_L=[6]~L=0>=\psi(\xi_1,\eta_1)\psi(\xi_2,\eta_2)\phi_{0s}(r)\phi_{0s}(R)
 \eeq
 where $\psi$ is the nucleon like w.f. (in the indicated relative
 coordinates) and $\phi$ the one particle harmonic oscillator w.f. in the standard
 notation.They are characterized by an oscillation parameter
 $a\approx1$fm, as opposed to the size parameter $b$ of the shell
 model harmonic oscillator which depends on the nuclear mass
 number A.
\item The configuration $0s^5~ 1s$.
\begin{eqnarray}
 |0s^5~1s~f_L&=&[6]~L=0>=\frac{1}{\sqrt{6}}\sum_{i=1}^4 \chi(x_i)\phi_{0s}(r)\phi_{0s}(R)
 \nonumber\\
 &+&\frac{1}{\sqrt{6}} \psi(\xi_1,\eta_1)\psi(\xi_2,\eta_2)
 (\phi_{1s}(r)\phi_{0s}(R)+\phi_{0s}(r)\phi_{1s}(R))\nonumber\\
 \end{eqnarray}
 In the above expression $ \chi(x_i)$ is the product of a $1s$ harmonic oscillator
 wave function in the coordinate $\xi_1,\eta_1,\xi_2,\eta_2$
 respectively  for $i=1,2,3,4$ with $0s$ in all other coordinates.
 \item  The configuration $0s^5~0d$.
 \begin{eqnarray}
 |0s^5~0d~f_L&=&[6]~L>=\frac{1}{\sqrt{6}}\sum_{i=1}^4 \chi(x_i)\phi_{0s}(r)\phi_{0s}(R)
 \nonumber\\
 &+& \frac{1}{\sqrt{6}} \psi(\xi_1,\eta_1)\psi(\xi_2,\eta_2)
 (\phi_{0d}(r)\phi_{0s}(R)+\phi_{0s}(r)\phi_{0d}(R))\nonumber\\
 \end{eqnarray}
 Now  $ \chi(x_i)$ is $1d$ harmonic oscillator
 wave function in the coordinate $\xi_1,\eta_1,\xi_2,\eta_2$
 respectively with $0s$ in all other coordinates.
  \item  The configuration $0s^4~0p^2$.
  The resulting expression is now more complicated. We find:
\begin{eqnarray}
 0s^40p^2[6]L>&=&A_1(\xi_1,\eta_1,\xi_2,\eta_2)\nonumber\\&+&
  \bf{B}_1(\xi_1,\eta_1,\xi_2,\eta_2)\otimes \bf{r}+
  \bf{C}_1(\xi_1,\eta_1,\xi_2,\eta_2)\otimes \bf{R}
  \nonumber\\
  &+&\frac{1}{2\sqrt{15}} \psi(\xi_1,\eta_1)\psi(\xi_2,\eta_2)
 \left(5 (\phi^2_{0p}(R)L)-(\phi^2_{0p}(r)L)\right )\nonumber\\
 \end{eqnarray}
 The functions $A_1,\bf{B}_1,\bf{C}_1$ are not expected to
 contribute significantly to the coupling to the nucleon system.
\end{itemize}
\item The symmetry $f_L=[5,1].$\\
 Omitting those components in
which we do not have a nucleon like structure in the internal
coordinates we find:
\begin{itemize}
\item The configuration $0s^5~ 1s$.
\beq | 0s^5~1s~f_L=[5,1]~L=0>_i \sim 0~~,~~ i=1,2
 \eeq
 \beq
  | 0s^5~1s~f_L=[5,1]~L=0>_i \sim \psi(\xi_1,\eta_1)\psi(\xi_2,\eta_2)
 C_i [\phi_{1p}(r)\otimes\phi_{1p}(R)]L=0
 \eeq
 with
 $C_i=\frac{1}{\sqrt{6}},\frac{1}{\sqrt{10}},\frac{1}{\sqrt{15}}$
 for $i=3,4,5$
\item The configuration $0s^5~ 1d$.
\beq | 0s^5~0d~f_L=[5,1]~L=2>_i \sim 0~~,~~ i=1,2
 \eeq
\beq
  | 0s^5~0d~f_L=[5,1]~L=2>_i \sim \psi(\xi_1,\eta_1)\psi(\xi_2,\eta_2)
 C_i[\phi_{1p}(r)\otimes\phi_{1p}(R)]L=2
 \eeq
 with
 $C_i=-\frac{1}{\sqrt{6}},-\frac{1}{\sqrt{10}},\frac{1}{\sqrt{15}}$
 for $i=3,4,5$
 \item  The configuration $0s^4~0p^2$.
 \beq
 | 0s^4~0p^2~f_L=[5,1]~L=0,2>_i \sim=0~~,~~i=1,2
 \eeq
 \beq
  | 0s^4~0p^2~f_L=[5,1]~L>_i \sim \psi(\xi_1,\eta_1)\psi(\xi_2,\eta_2)
 C_i[\phi_{1p}(r)\otimes\phi_{1p}(R)]L
 \eeq
 with $L=0,2$ and
 $C_i=-\frac{2}{\sqrt{3}},-\frac{2}{\sqrt{5}},-\frac{2 \sqrt{2}}{\sqrt{15}}$
 for $i=3,4,5$
\end{itemize}
\item The symmetry  $f=[4,2].$\\
 The only allowed configuration is $0s^40p^2$.
 we have:
 \beq
  | 0s^4~0p^2~f_L=[4,2]~L=0,2>_i \sim0~~,~~i=1,2,3,8
 \eeq
 \beq
  | 0s^4~0p^2~f_L=[4,2]~L=0,2>_i \sim \psi(\xi_1,\eta_1)\psi(\xi_2,\eta_2)
 C_i[\phi_{1p}^2(r)]L=0,2
 \eeq
 with
 $C_i=\frac{2}{\sqrt{3}},\frac{2}{9},-\frac{ \sqrt{2}}{3\sqrt{3}},\frac{ 2\sqrt{2}}{9},-\frac{2}{3\sqrt{15}}$
 for $i=4,5,6,7,9$
 \end{enumerate}
From the above expressions we can see that the the 6 quark states configurations involving $1s$, $0d$ or $0p^2$
 contain spurious components, i.e. the CM is not in a $0s$ state. We could, of course, project them out, but we do not want to do that. The reason being that the 6-quark cluster is going to be admixed in the two nucleon component
of the shell model basis (see below). In the two nucleon w.f. the center of mass component of the two particles
is not necessarily in $0s$ mode.
\subsection{Transformation into the two nucleon relative and CM coordinates.}
As it was mentioned above we must transform the above cluster orbital w.f. into those
encountered in the shell model. Clearly the wave functions associated with different
angular momenta are orthogonal. Thus
\beq
\phi_{n_0 \ell}(a)=\sum_n |n,\ell > C_{n_0,n,\ell }(c), \mbox{ with } C_{n_0,n,\ell }(c)= <n,\ell (b)|\phi_{n_0,\ell }(a)>~,c~=\frac{b}{a}
\eeq
Limiting ourselves to two nucleon harmonic oscillator wave functions with $\hbar \omega\leq 5$
we get $C_{n_0,n,\ell }(c)$ as follows:\\
For $n_0,\ell =0s$ in the order $(0s,1s,2s,3s,4s,5s,6s,7s,8s,9s,10s)$:\\
For  $c=3/2$
$$({0.886864},{0.417762},{0.127026},{0.030467},{0.006214},{0.001121},{0.000183},{0.000027}$$ $$,
{3.\times 10^{-6},0,0})$$
For  $c=2$
$$
({0.715541},{0.525813},{0.249415},{0.093322},{0.029695},{0.008356},{0.00213},{0.000500}$$ $$,{0.0001
   09},{0.000022},0)$$
   For  $c=5/2$
   $$(0.572727, 0.507942, 0.290787, 0.131313, 0.050428, 0.017128, 0.00527,
0.001493, 0.000394$$ $$, 0.000097, 0.000022 )$$
   For   $c=7/2$
   $$(0.383992, 0.399305, 0.268029, 0.141916, 0.063902, 0.025448, 0.009181,
0.003049, 0.000943$$ $$, 0.000274, 0.000075) $$
For $n_0,\ell=1s$  in the above order we get:\\
For  $c=3/2$
$$(-0.417762, 0.558881, 0.399709, 0.150981, 0.042037,
    0.009611, 0.001902, 0.000336, 0.000054, 0,
    0) $$
    For  $c=2$
$$ (-0.525813, 0.071554, 0.251164, 0.175254, 0.081627, 0.03025, 0.009567,
0.002681, 0.000681$$ $$, 0.000159, 0.000034)$$
For $c=5/2$
$$ (-0.507942, -0.178083, 0.053996, 0.094805, 0.063452, 0.030737, 0.012284,
0.00428, 0.001341$$ $$, 0.000384, 0.000102)$$
For $c=7/2$
$$ (-0.399305, -0.308055, -0.134839, -0.033304, 0.002154, 0.007688, 0.005238,
0.002558, 0.001044$$ $$, 0.000377, 0.000124)$$
For $n_0,\ell=1p$ in the order $(0p,1p,2p,3p,4p,5p,6p,7p,8p,9p,10p)$ we get:\\
For  $c=3/2$
$$(0.818643, 0.497841, 0.17911, 0.048711, 0.010984,
    0.002154, 0.000378, 0.00006, 0$$ $$,
    0, 0) $$
    For  $c=2$
$$(0.572433, 0.543058, 0.30479, 0.129311,
    0.045489, 0.013916, 0.003811,
    0.000952, 0.00022$$ $$, 0.000047, 0) $$
  For  $c=5/2$
$$ (0.394984, 0.452242, 0.306334, 0.156856, 0.066595, 0.024589, 0.008127,
0.002451$$ $$, 0.000683, 0.000178, 0.000043)$$
For  $c=7/2$
$$(0.202864, 0.27234, 0.216298, 0.129859, 0.064644, 0.027986, 0.010846,
0.003835, 0.001254$$ $$, 0.000383, 0.00011) $$
For $n_0,\ell=1d$ in the order  $(0d,1d,2d,3d,4d,5d,6d,7d,8d,9d,10d)$ we get:\\
For  $c=3/2$
$$(0.75567, 0.543742, 0.221817, 0.066693, 0.016349, 0.003444,
    0.000643, 0.000109, 0.000016$$ $$, 0, 0)$$
    For  $c=2$
$$(0.457946, 0.514043, 0.327135, 0.15344, 0.058679, 0.019284, 0.005622,
0.001485, 0.00036$$ $$, 0.000081, 0.000017) $$
For  $c=5/2$
$$ (0.272403, 0.369034, 0.283442, 0.160452, 0.074056, 0.029372, 0.010335,
0.003295, 0.000966$$ $$, 0.000263, 0.000067)$$
For  $c=7/2$
$$(0.107173, 0.170238, 0.15331, 0.101757, 0.055068, 0.025609, 0.010565,
0.003949, 0.001358$$ $$, 0.000434, 0.00013) $$
We notice that for $0s$ and $1s$ the convergence is reasonably good even for high $c$. For
$0p$ and $0d$ the overlap for high $c$ is not as good.
\section{The combined orbit-spin-color symmetry}
In the case of $I=1$ states the isospin state has symmetry $[4,2]$, the orbital and spin-color symmetries can be combined to yield the orbital-spin-color symmetry $[f^{osc}]=[2^21^2]$ of the
group $S_6$, using the relevant Clebsch-Gordan coefficients. Similarly for I=0 one gets $[f^{osc}]=[2^3]$. There exist now routines that provided these
coefficients \cite{SM78}. Since the above computed cfp's do not depend on the selected Yamanouchi symbols, the symmetry $[f^{osc}]$ is merely a
label  to select the allowed states. So we do not need either the C-G coefficients or the two particle cfp's
associated with $[f^{osc}]$. 

In fact it is easy to show the following:\\
i) The configuration  $(0s)^6$ is associated with the 1-dimensional completely symmetric state. So in this case $f^{0sc}=f^{sc}$.
 and the two-particle CFPs provided above are adequate. \\
ii) Configurations of the type $(0s)^5 1s$. In this case, using the 1-particle CFPs given in appropriate the tables, the corresponding symmetries can be expressed in terms of antisymmetric states of a given $J$ and $I$ of the form:
$$\left \{(0s)^5 [f^{sc)}]=[2^21](0,1)S_1,I_1=1/2\otimes 1s(10)s_2=\frac{1}{2},I_2=\frac{1}{2}\right \}^{J,I=0,1}$$
$s_1=|J-1/2|,J+1/2$. In this case one needs the CFP of the spin color symmetry $[2^21](0,1)s_1,s_1=1/2,3/2,5/2$.
$$\left \{(0s)^5 [f^{sc)}]=[21^3](0,1)S_1,I_3=3/2\times 1s(10)s_2=\frac{1}{2},I_2=\frac{1}{2}\right \}^{J,I=1,2}$$
$s_1=|J-1/2|,J+1/2$. Now one needs the CFP of the spin color symmetry $[2,1^3](0,1)s_1,s_1=1/2,3/2$.\\
The needed one partcle spin-color CFPs are given in table \ref{tab:spc6cfp36}.
\begin{table}[h!]
\begin{center}
\caption{ The one particle CFP's involving the spin color symmetry, needed in the case of a 5-quark state,  with color symmetry (0,1) and spin 1/2, 3/2 and 5/2.}
\label{tab:spc6cfp36}
$$
\begin{array}{|c|c|c|}
\hline
[f_1](\lambda_1,\mu_1)S_1&[f](\lambda,\mu)S&\mbox{CFP}\\
\hline
\mbox{$[2^2](0,2)0$}&[2^2,1](0,1)\frac{1}{2}&\frac{\sqrt{3}}{2}\\
\hline
\mbox{$[2^2](1,0)1$}&[2^2,1](0,1)\frac{1}{2}&-\frac{\sqrt{1}}{2}\\
\hline
\end{array}
\quad
\begin{array}{|c|c|c|}
\hline
[f_1](\lambda_1,\mu_1)S_1&[f](\lambda,\mu)S&\mbox{CFP}\\
\hline
\mbox{$[2^2](0,2)2$}&[2^2,1](0,1)\frac{3}{2}&\frac{\sqrt{3}}{8}\\
\mbox{$[2^2](1,0)1$}&[2^2](0,1)\frac{3}{2}&\frac{\sqrt{5}}{8}\\
\hline
\mbox{$[2^2](0,2)2$}&[2^2](0,1)\frac{5}{2}&1\\
\hline
\end{array}
$$
$$
\begin{array}{|c|c|c|}
\hline
[f_1](\lambda_1,\mu_1)S_1&[f](\lambda,\mu)S&\mbox{CFP}\\
\hline
\mbox{$[21^2](0,2)1$}&[2,1^3](0,1)\frac{1}{2}&\frac{1}{\sqrt{3}}\\
\mbox{$[21^2](1,0)0$}&[2,1^3](0,1)2&-\frac{1}{\sqrt{3}}\\
\hline
\mbox{$[21^2](1,0)1$}&[2,1^3](0,1)2&\frac{1}{\sqrt{3}}\\
\hline
\end{array}
\quad
\begin{array}{|c|c|c|}
\hline
[f_1](\lambda_1,\mu_1)S_1&[f](\lambda,\mu)S&\mbox{CFP}\\
\hline
\mbox{$[21^2](0,2)1$}&[2,1^3](0,1)\frac{3}{2}&\sqrt{\frac{8}{15}}\\
\mbox{$[21^2](1,0)1$}&[2,1^3](0,1)\frac{3}{2}&-\sqrt{\frac{2}{15}}\\
\mbox{$[21^2](1,0)2$}&[2,1^3](0,1)\frac{3}{2}&\sqrt{\frac{1}{3}}\\
\hline
\end{array}
$$
$$
\begin{array}{|c|c|c|}
\hline
[f_1](\lambda_1,\mu_1)S_1&[f](\lambda,\mu)S&\mbox{CFP}\\
\hline
\mbox{$[1^4](0,2)0$}&[1^5](0,1)\frac{1}{2}&\sqrt{\frac{2}{5}}\\
\mbox{$[1^4](0,1)1$}&1^5](0,1)\frac{1}{2}&-\sqrt{\frac{3}{5}}\\
\hline
\end{array}
$$
\end{center}
\end{table}
Using the results of table \ref{tab:spc6cfp36} one can compute the corresponding  2-particle CFPs given in table \ref{tab:spc4cfp35}.

\begin{table}[h!]
\begin{center}
\caption{ The two particle CFP's involving the spin color symmetry for 5 identical quarks , needed in the study of the six quark clusters, corresponding to color state (0,1) and spin 1/2, and 3/2.
\label{tab:spc4cfp35}}
$$
\begin{array}{|c|c|c|r|}
\hline
[f_1](\lambda_1,\mu_1)S_1&[f_2](\lambda_2,\mu_2)S_2&[f](\lambda,\mu)S&\mbox{CFP}\\
\hline
\mbox{$[2,1](1,1)\frac{1}{2}$}&[2](2,0)1&[2^21](0,1)\frac{1}{2}&\frac{1}{2}\\
\mbox{$[2,1](1,1)\frac{3}{2}$}&[2](2,0)1&[2^21](0,1)\frac{1}{2}&\frac{1}{2}\\
\mbox{$[2,1](1,1)\frac{1}{2}$}&[2](2,0)&[2^21](0,1)\frac{1}{2}&-\frac{1}{2}\\
\mbox{$[2,1](0,0)\frac{1}{2}$}&[2](2,0)0&[2^21](0,1)\frac{1}{2}&\frac{1}{2}\\
\hline
\end{array}
\quad
\begin{array}{|c|c|c|c|}
\hline
[f_1](\lambda_1,\mu_1)S_1&[f_2](\lambda_2,\mu_2)S_2&[f](\lambda,\mu)S&\mbox{CFP}\\
\hline
\mbox{$[2,1](1,1)\frac{1}{2}$}&[2](2,0)1&[2^21](0,1)\frac{3}{2}&\sqrt{\frac{5}{8}}\\
\mbox{$[2,1](1,1)\frac{3}{2}$}&[2](0,1)0&[2^21](0,1)\frac{3}{2}&\frac{1}{4}\\
\mbox{$[2,1](1,1)\frac{3}{2}$}&[1^2](0,1)0&[2^21](0,1)\frac{3}{2}&\sqrt{\frac{5}{4}}\\
\mbox{$[2,1](1,1)\frac{3}{2}$}&[1^2](2,0)0&[2^21](0,1)\frac{3}{2}&-\sqrt{\frac{5}{4}}\\
\mbox{$[2,1](1,1)\frac{3}{2}$}&[2](0,1)1&[2^21](0,1)\frac{3}{2}&\frac{1}{12}\\
\mbox{$[2,1](0,0)\frac{1}{2}$}&[2](0,1)1&[2^21](0,1)\frac{3}{2}&\sqrt{\frac{5}{18}}\\
\hline
\end{array}
$$
$$
\begin{array}{|c|c|c|c|}
\hline
[f_1](\lambda_1,\mu_1)S_1&[f_2](\lambda_2,\mu_2)S_2&[f](\lambda,\mu)S&\mbox{CFP}\\
\hline
\mbox{$[1^3](1,1)\frac{1}{2}$}&[1^2](0.1)1&[2^21](0,1)\frac{3}{2}&\frac{\sqrt{5}}{3}\\
\mbox{$[1^3](1,1)\frac{3}{2}$}&[1^2](0,1)1&[2^21](0,1)\frac{3}{2}&\frac{2}{3}\\
\hline
\end{array}
\quad
\begin{array}{|c|c|c|c|}
\hline
[f_1](\lambda_1,\mu_1)S_1&[f_2](\lambda_2,\mu_2)S_2&[f](\lambda,\mu)S&\mbox{CFP}\\
\hline
\mbox{$[1^3](1,1)\frac{1}{2}$}&[1^2](2,0)0&[2^21](0,1)\frac{1}{2}&\frac{1}{\sqrt{2}}\\
\mbox{$[1^3](1,1)\frac{3}{2}$}&[1^2](0,1)1&[2^21](0,1)\frac{1}{2}&\frac{1}{3\sqrt{2}}\\
\mbox{$[1^3](0,0)\frac{3}{2}$}&[1^2](0,1)1&[2^21](0,1)\frac{1}{2}&-\frac{2}{3}\\
\hline
\end{array}
$$
\end{center}
\end{table}

iii) Configurations of the type $(0s)^4 (0p)^2$. Then, using the 2-particle CFPs given in appropriate the tables,  the corresponding symmetries can be expressed in terms of the form:
$$\left \{ \left [(0s)^4[f^{sc}]_1(\mu,\lambda)S_1=J_1,I_1\right ]\times \left [(0p)^2L[f]_{2L}[f^{sc}]_2(\lambda,\mu),S_2,J_2,I_2\right]^{J_2}\right \}^{J,I}$$
where
$$[f^{sc}]_1=2^2,[2,1^2],[1^4],\, \mbox{ for} I_1=0,1,2 \mbox{ respectively} $$
 and
$$I_2=0:\, L=\mbox{even}\leftrightarrow [f]_{2L}=[2],[f^{sc}]_2=[2],L=\mbox{odd}\leftrightarrow [f]_{2L}=[1^2],[f^{sc}]_2=[1^2]$$
$$I_2=1:\, L=\mbox{even}\leftrightarrow [f]_{2L}=[2],[f^{sc}]_2=[1^2],L=\mbox{odd}\leftrightarrow [f]_{2L}=[1^2],[f^{sc}]_2=[2]$$
In this case one needs the CFP of the spin color symmetry $$[f^{sc}](\mu,\lambda)S_1,\,[f^{sc}]_1=2^2,[2,1^2],[1^4]$$
Those with  color spin $(1,0),1$ were given above in connection with the pentaquarks. The ones with  $((0,2),1$ are given in table \ref{tab:spc4cfp25}:
\begin{table}[h!]
\begin{center}
\caption{ The two particle CFP's involving the spin color symmetry for 4 identical quarks , needed in the study of the six quark clusters, corresponding to color state (0,2) and spin 0,1, and 2.
\label{tab:spc4cfp25}}
$$
\begin{array}{|c|c|c|c|}
\hline
[f_1](\lambda_1,\mu_1)S_1&[f_2](\lambda_2,\mu_2)S_2&[f](\lambda,\mu)S&\mbox{CFP}\\
\hline
\mbox{$[2](2,0)1$}&[2](2,0)1&[2^2](0,2)0&\frac{1}{\sqrt{2}}\\
\mbox{$[2](0,1)0$}&[2](0,1)0&[2^2](0,2)0&-\frac{1}{\sqrt{2}}\\
\hline
\mbox{$[2](2,0)1$}&[2](2,0)1&[2^2](0,2)2&1\\
\hline
\end{array}
\begin{array}{|c|c|c|c|}
\hline
[f_1](\lambda_1,\mu_1)S_1&[f_2](\lambda_2,\mu_2)S_2&[f](\lambda,\mu)S&\mbox{CFP}\\
\hline
\mbox{$[1^2](2,0)0$}&[1^2](0,1)1&[2,1^2](0,2)1&\frac{\sqrt{3}}{2}\\
\mbox{$[1^2](0,1)1$}&[1^2](2,0)0&[2,1^2](10)1&\frac{1}{2}\\
\hline
\hline
\end{array}
$$
$$
\begin{array}{|c|c|c|c|}
\hline
[f_1](\lambda_1,\mu_1)S_1&[f_2](\lambda_2,\mu_2)S_2&[f](\lambda,\mu)S&\mbox{CFP}\\
\hline
\mbox{$[1^2](0,1)1$}&[1^2](0,1)1&[2,1^2](02)1&1\\
\hline
\end{array}
\begin{array}{|c|c|c|c|}
\hline
[f_1](\lambda_1,\mu_1)S_1&[f_2](\lambda_2,\mu_2)S_2&[f](\lambda,\mu)S&\mbox{CFP}\\
\hline
\mbox{$[1^2](2,0)0$}&[1^2](2,0)0&[1^4](0,2)0&\frac{1}{\sqrt{2}}\\
\mbox{$[1^2](0,1)1$}&[1^2](0,1)1&[1^4](0,2)0&-\frac{1}{\sqrt{2}}\\
\hline
\end{array}
$$
\end{center}
\end{table}
These were constructed from the one particle CFPs needed in the pentaquark case as well as those given in table \ref{tab:spc6cfp26}.
\begin{table}[h!]
\begin{center}
\caption{ The one particle CFP's involving the spin color symmetry, needed in the case of a 4-quark state,  with color symmetry (0,2) and spin 0,1, and 2.}
\label{tab:spc6cfp26}
$$
\begin{array}{|c|c|c|}
\hline
[f_1](\lambda_1,\mu_1)S_1&[f](\lambda,\mu)S&\mbox{CFP}\\
\hline
\mbox{$[2,1](1,1)\frac{1}{2}$}&[2^2](0,2)0&1\\
\hline
\mbox{$[2,1](1,1)\frac{3}{2}$}&[2^2](0,2)2&1\\
\hline
\end{array}
\quad
\begin{array}{|c|c|c|}
\hline
[f_1](\lambda_1,\mu_1)S_1&[f](\lambda,\mu)S&\mbox{CFP}\\
\hline
\mbox{$[2,1]((1,1)\frac{1}{2}$}&[2,1^2](0,2)1&\frac{1}{\sqrt{2}}\\
\mbox{$[2,1]((1,1)\frac{3}{2}$}&[2,1^2](0,2)1&-\frac{1}{\sqrt{2}}\\
\hline
\mbox{$[1^3]((1,1)\frac{1}{2}$}&[1^4](0,2)0&1\\
\hline

\end{array}
$$
\end{center}
\end{table}

Even though, however, in the present calculation we do not explicitly make use of the coefficients involving explicitly $[f^{osc}$, we have constructed them and they will appear in the appendix.
\section{Evaluation of the matrix elements of the two body interaction}
We begin by considering the matrix element involving the orbital $(06)^6$ configurations:

\begin{eqnarray}
&&< (0s)^6\, [6]_r\; [2211]_{cs}(0 0)_c \;S=J,
\;I=1\Big|\;\sum_{i,j}V_{ij}\;\Big|
(0s)^6\, [6]_r\; [2211]_{cs}(0 0)_c \;S=J, \;I=1>\;
\nonumber\\
&&= \frac{6\cdot
5}{2}\sum_{f'',(\lambda'',\mu''),I'',[f_2],S_2,[f'_2],S'_2,I_2}
\frac{n_{f''}}{n_{[2211]}}
\;\;\Bigg\langle\begin{array}{cc}[f'']_{cs} & [f_2]_{cs}\\
(\lambda'' \mu'')_c \,S'' &(\lambda_2 \mu_2)_c\,S_2\\
\end{array}\Bigg|\begin{array}{c}[2211]_{cs}\\(0 0)_c \;S \end{array}
\Bigg\rangle \nonumber\\
&&\times\;\;\Bigg\langle\begin{array}{cc}[f'']_{cs}&
[f_2']_{cs}\\
(\lambda'' \mu'')_c \,S'' &(\lambda_2' \mu_2')_c\,S_2'\\
\end{array}\Bigg|\begin{array}{c}[2211]_{cs}\\(0 0)_c \;S \end{array}
\Bigg\rangle \;\;\;\;
\nonumber\\
&&\;\times\;\;\; \Big< (0s)^2\, [2]_r\;
[f_2']_{cs}(\lambda_2' \mu_2')_c \;S_2' \;I_2 \Big| \;V_{56} \;\Big|
(0s)^2\, [2]_r\; [f_2]_{cs}(\lambda_2 \mu_2)_c \;S_2 \;I_2
\Big>\;.\label{eq:first1}
\end{eqnarray}
In the above expression the first factor represents the number of ways one can select two pricles oit of 6. $n_{[f'']}$ and $n_{[2211}]$ are the dimensions of the representations $[f'']$ and $[2^21^2]$ of the symmetric group $S_6$ respectively. The spin-color two paricle CFP's are shown explicitely. The orbital and isospin two particle CFPs are trivial in this example. The last term is the elementary 2-body interation matrix element, which is diagonal in isospin.\\
For the other matrix elements the reader is referred to the literature, see  Strottman-Vergados \cite{StrotVer18} and references there in.
\section{Discussion} 
In the present paper we discussed and developed  the formalism of extending Quantum Chromodynamics (QCD) at low energy processes. Then using symmetries we applied this in the case of multiquark-antiquarks. The essential symmetries involved for $n$ identical quarks were the orbital symmetry $SU^o(n)$, the isospin symmetry $SU^I(2)$ and the combined spin color-symmetry $SU^{sc}(6)$. It was shown how these symmetries can be combined to yield a wave function with the proper symmetry, e.g. totally antisymmetric in the case of fermions. The formalism has been applied in the case of pentaquarks, which have recently been found to exist. Also the essential tools have been given to obtain the structure of six quark clusters up to $2\hbar \omega$ excitations. Such clusters may be admixed with the standard nuclear wave function and thus they may become useful in estimating the matrix elements of short range operators, such as those appearing in the case of neutrinoless double beta decay mediated by the exchange of heavy particles.  

\begin{thebibliography}{10}

\bibitem{Henley74}
H.~Frauenfelder and E.~M. Henley, {\em Subatomic Physics} (Prentice Hall,
  1974).

\bibitem{JDV85}
J.~Vergados, {\em Nuc. Phys. B} {\bf 250}  (1985)   418.

\bibitem{Langacker16}
P.~Langacker, {\em The Standard Model and Beyond, Second Edition} (CRC Press,
  2016).

\bibitem{Vergdos17}
J.~Vergados, {\em The Standard Model and Beyond} (World Scientific, 2017).

\bibitem{Vergdos16}
J.~Vergados, {\em Group and Representation Theory} (World Scientific, 2016).

\bibitem{HOV86}
E.~M. Henley, T.~Oka and J.~D. Vergados, {\em Phys. Lett.} {\bf 166 B}  (1986)
   274.

\bibitem{HOV88}
E.~M. Henley, T.~Oka and J.~D. Vergados, {\em Nuc. Phys.} {\bf 476 A}  (1988)
  589.

\bibitem{Perkins71}
D.~H. Perkins, {\em Introduction to High Energy physics} (Addison Wesley,
  1971).

\bibitem{CJJKW74}
A.~Chodos, R.~Jaffe, K.Johnson, J.~Kiskis and V.~F. Weisskoph, {\em Phys. Rev.
  D}   (1974)   3471.

\bibitem{Eichten80}
E.~Eichten {\em et~al.}, {\em Phys. Rev. D} {\bf 21}  (1980)   203.

\bibitem{MITbag1}
A.~Chodos, R.~L. Jaffee, K.~Johnson and C.~Thorn, {\em Phys. Rev D} {\bf 10}
  (1974).

\bibitem{MITbag2}
R.~L. Jaffee, {\em Phys. Rev. D} {\bf 14}  (1977).

\bibitem{Stro78}
D.~Strottman, {\em Phys. Rev. D.} {\bf 18}  (1978)   2716.

\bibitem{Stro79}
D.~Strottman, {\em Phys. Rev. D.} {\bf 20}  (1979)   748.

\bibitem{multiq80a}
R.~P. Bickerstaff and B.~G. Wybourne, {\em Austr. Jour. Phys.} {\bf 33}  (1980)
    951.

\bibitem{multiq80b}
S.~Y. Tsai, {\em Prog. Theor. Phys.} {\bf 64}  (1980)   1710.

\bibitem{pentaq03}
C.~E. Carlson {\em et~al.}, {\em Phys. Lett B} {\bf 573}  (2003)   101.

\bibitem{pentaq05}
F.~Wang, J.~Ping, D.~Qing and T.~Goldman, {\em Phys. Atom. Nucl} {\bf 68}
  (2005)   808, Yad. Fiz. 68:841-846:2005; arXiv:nucl-th/0406036.

\bibitem{pentaq15}
 LHCb Collaboration Collaboration (R.~Aaij {\em et~al.}), {\em Phys. Rev. Lett}
  {\bf 115}  (2015)   072001, arXiv:1507.03414.

\bibitem{pentaq16a}
 LHCb Collaboration Collaboration (R.~Aaij {\em et~al.}), {\em Phys. Rev. Lett}
  {\bf 17}  (2016)   082003.

\bibitem{fourq10}
 Belle Experiment Collaboration (R.~Ali, C.~Hambrock and M.~J. Aslam), {\em
  Phys. Rev. Lett.} {\bf 104}  (2010)   162001.

\bibitem{fourq14}
 LHCb Collaboration Collaboration (R.~Aaij {\em et~al.}), {\em Phys. Rev.
  Lett.} {\bf 112}  (2014)   222002.

\bibitem{EPP16}
A.~Esposito, A.~Pilloni and A.~D. Polosa, {\em Phys. Rep.} {\bf 668}  (2016)
  ~1.

\bibitem{CCLZ16}
H.~X. Chen, W.~Chen, X.~Liu and S.~L. Zhu, {\em Phys. Rep.} {\bf 639}  (2016)
  ~1.

\bibitem{ALS16}
A.~Ali, J.~S. Lange and S.~Stone, {\em Prog. Part. Nuc. Phys.} {\bf 97}  (2017)
    123.

\bibitem{ARS17}
M.~Karliner, J.~L. Rosner and T.~Skwarnicki, {\em Ann. Rev. Nuc. Sci} {\bf 10}
  (2017) arXiv:1711.10626 [hep-ph].

\bibitem{SOSTRO}
S.~I. SO and D.~Strottman, {\em J. Math. Phys.} {\bf 20}  (1976)   153.

\bibitem{JDV68}
J.~D. Vergados, {\em Nuc. Phys. A} {\bf 111}  (1968)   681.

\bibitem{Bied62}
L.~C. Biedenharn, {\em Phys. Rev.} {\bf 120}  (1962).

\bibitem{ACP88}
G.~Adams, J.~Cizek and J.~Paldus, {\em Adv. Quant. Chem} {\bf 19}  (1988).

\bibitem{YBFA}
Y.~Yamaguchi, A.~Buchman, A.~Faessler and A.~Arima, {\em Nuc. Phys.} {\bf 526}
  (1991)   495.

\bibitem{HAMMERMESH}
M.~Hammermesh, {\em Group Theory} 1964.

\bibitem{SM78}
S.~Schindler and R.~Mirman, {\em Comp.Phys. Commun.} {\bf 15}  (1978)   147.

\bibitem{StrotVer18}
D.~Strottman and J.~D. Vergados, {\em arXiv:1808.06042 [nucl-th]}   (2018).

\bibitem{Vergados15}
J.~Vergados  (2015) arXiv:1505.04200 (math-ph).

\end{thebibliography}

\section{Appendix: The two particle CFP's in the symmetric group $S_6$}

Even though, however, in the present calculation we did not explicitly make use of the coefficients involving explicitly $[f^{osc}$, we have constructed them and they will be discussed below.\\
In the special case of $[f_o]=[6]$ the spin color symmetry is uniquely specified to be  $[2^21^2]$. In the case of $f_o=[51]$ and
$f_{cs}=[42]$, one possibility encountered in present calculation, we present them in table \ref{table:cgc2}.
The same coefficients result for the case  $f_o=[42]$ and $f_{cs}=[5,1].$

In reading this table we mention that the states are order according with the Yamanouch symbols as follows :
\begin{enumerate}
\item for the [51]: 211111, 121111, 112111, 111211, 111121
\item for the [42]: 221111, 212111, 211211, 211121, 122111, 121211, 121121, 112211, 112121
\item For the $[2^2,1^2]$: 432211, 432121, 423211, 423121, 421321, 243211, 243121, 241321, 214321
\end{enumerate}
\begin{table}
\caption{The Clebsch-Gordan coefficients entering in the product $[51]\times [42]\rightarrow [2^21^2]$. The columns are labeled by the Yamanouchi symbols   of the combined symmetry $[2^21^2]$. The rows are labeled by two integers $p$ and
$q$, i.e. $pq\Leftrightarrow (p,q)\Leftrightarrow (Y_p,Y_q)$, where $Y_p,Y_q$ are the
Yamanouchi symbols of the [51] and  $[42]$ respectively. (see text).
\label{table:cgc2}
}
{\tiny
$$\left(
\begin{array}{llllllllllllllllll}
 13 & \frac{1}{2 \sqrt{3}} & 11 & -\frac{1}{\sqrt{15}} & 11 &
   \frac{1}{\sqrt{30}} & 12 & \frac{1}{\sqrt{30}} & 11 &
   -\sqrt{\frac{2}{15}} & 11 & -\frac{1}{\sqrt{15}} & 12 &
   -\frac{1}{\sqrt{15}} & 13 & -\frac{1}{\sqrt{15}} & 14 &
   -\frac{1}{\sqrt{15}} \\
 16 & -\frac{1}{2 \sqrt{6}} & 13 & \frac{1}{6 \sqrt{5}} & 13
   & \frac{1}{3 \sqrt{10}} & 14 & \frac{1}{3 \sqrt{10}} & 13
   & -\frac{\sqrt{\frac{2}{5}}}{3} & 13 & \frac{2}{3
   \sqrt{5}} & 14 & \frac{2}{3 \sqrt{5}} & 16 &
   \frac{1}{\sqrt{30}} & 15 & -\sqrt{\frac{2}{15}} \\
 24 & -\frac{1}{2 \sqrt{3}} & 16 &
   \frac{\sqrt{\frac{5}{2}}}{6} & 16 & \frac{\sqrt{5}}{6} &
   15 & -\frac{1}{6 \sqrt{5}} & 16 & -\frac{1}{6 \sqrt{5}} &
   16 & \frac{1}{3 \sqrt{10}} & 15 &
   -\frac{\sqrt{\frac{2}{5}}}{3} & 24 & -\frac{1}{\sqrt{15}}
   & 17 & \frac{1}{\sqrt{30}} \\
 27 & \frac{1}{2 \sqrt{6}} & 22 & \frac{1}{\sqrt{15}} & 22 &
   \frac{1}{\sqrt{30}} & 17 & \frac{\sqrt{5}}{6} & 22 &
   \sqrt{\frac{2}{15}} & 22 & -\frac{1}{\sqrt{15}} & 17 &
   \frac{1}{3 \sqrt{10}} & 25 & \sqrt{\frac{2}{15}} & 18 &
   -\frac{1}{\sqrt{15}} \\
 35 & \frac{1}{2 \sqrt{3}} & 24 & -\frac{1}{6 \sqrt{5}} & 24
   & \frac{1}{3 \sqrt{10}} & 18 &
   \frac{\sqrt{\frac{5}{2}}}{6} & 24 &
   \frac{\sqrt{\frac{2}{5}}}{3} & 24 & \frac{2}{3 \sqrt{5}} &
   18 & \frac{1}{6 \sqrt{5}} & 27 & \frac{1}{\sqrt{30}} & 23
   & \frac{1}{\sqrt{15}} \\
 38 & \frac{1}{2 \sqrt{6}} & 27 &
   -\frac{\sqrt{\frac{5}{2}}}{6} & 25 & \frac{1}{6 \sqrt{5}}
   & 19 & -\frac{1}{2 \sqrt{6}} & 27 & \frac{1}{6 \sqrt{5}} &
   25 & \frac{\sqrt{\frac{2}{5}}}{3} & 19 & \frac{1}{2
   \sqrt{3}} & 28 & \frac{1}{\sqrt{15}} & 26 &
   -\frac{1}{\sqrt{30}} \\
 49 & \frac{\sqrt{\frac{3}{2}}}{2} & 35 & -\frac{1}{3
   \sqrt{5}} & 27 & \frac{\sqrt{5}}{6} & 21 &
   -\frac{1}{\sqrt{30}} & 35 & \frac{2 \sqrt{\frac{2}{5}}}{3}
   & 27 & \frac{1}{3 \sqrt{10}} & 21 & \frac{1}{\sqrt{15}} &
   34 & -\sqrt{\frac{2}{15}} & 33 & -\sqrt{\frac{2}{15}} \\
 59 & \frac{1}{2} & 38 & \frac{\sqrt{\frac{5}{2}}}{3} & 28 &
   -\frac{\sqrt{\frac{5}{2}}}{6} & 23 & -\frac{1}{3
   \sqrt{10}} & 38 & -\frac{1}{3 \sqrt{5}} & 28 & -\frac{1}{6
   \sqrt{5}} & 23 & -\frac{2}{3 \sqrt{5}} & 37 &
   \frac{1}{\sqrt{15}} & 36 & \frac{1}{\sqrt{15}} \\
 0 & 0 & 39 & \frac{1}{2 \sqrt{6}} & 29 & \frac{1}{2
   \sqrt{6}} & 26 & -\frac{\sqrt{5}}{6} & 39 & \frac{1}{2
   \sqrt{3}} & 29 & -\frac{1}{2 \sqrt{3}} & 26 & -\frac{1}{3
   \sqrt{10}} & 42 & -\frac{\sqrt{6}}{5} & 41 &
   -\frac{\sqrt{6}}{5} \\
 0 & 0 & 45 & -\frac{1}{2 \sqrt{3}} & 32 &
   \frac{1}{\sqrt{15}} & 31 & \frac{1}{\sqrt{15}} & 45 &
   \frac{2 \sqrt{\frac{2}{3}}}{5} & 32 & -\sqrt{\frac{2}{15}}
   & 31 & -\sqrt{\frac{2}{15}} & 52 & -\frac{2}{5} & 51 &
   -\frac{2}{5} \\
 0 & 0 & 48 & -\frac{1}{2 \sqrt{6}} & 34 & -\frac{1}{6
   \sqrt{5}} & 33 & -\frac{1}{6 \sqrt{5}} & 48 & -\frac{1}{2
   \sqrt{3}} & 34 & -\frac{\sqrt{\frac{2}{5}}}{3} & 33 &
   -\frac{\sqrt{\frac{2}{5}}}{3} & 0 & 0 & 0 & 0 \\
 0 & 0 & 58 & \frac{1}{2} & 37 &
   -\frac{\sqrt{\frac{5}{2}}}{6} & 36 &
   -\frac{\sqrt{\frac{5}{2}}}{6} & 55 & -\frac{2}{5} & 37 &
   -\frac{1}{6 \sqrt{5}} & 36 & -\frac{1}{6 \sqrt{5}} & 0 & 0
   & 0 & 0 \\
 0 & 0 & 0 & 0 & 44 & \frac{1}{2 \sqrt{3}} & 43 & \frac{1}{2
   \sqrt{3}} & 0 & 0 & 44 & \frac{2 \sqrt{\frac{2}{3}}}{5} &
   43 & \frac{2 \sqrt{\frac{2}{3}}}{5} & 0 & 0 & 0 & 0 \\
 0 & 0 & 0 & 0 & 47 & -\frac{1}{2 \sqrt{6}} & 46 &
   -\frac{1}{2 \sqrt{6}} & 0 & 0 & 47 & \frac{1}{2 \sqrt{3}}
   & 46 & \frac{1}{2 \sqrt{3}} & 0 & 0 & 0 & 0 \\
 0 & 0 & 0 & 0 & 57 & \frac{1}{2} & 56 & \frac{1}{2} & 0 & 0
   & 54 & -\frac{2}{5} & 53 & -\frac{2}{5} & 0 & 0 & 0 & 0
\end{array}
\right)
$$
}
\end{table}
In general the two particle cfp's for $S_6$ are very complicated. One has to deal with $6!$ states \cite{Vergados15}
For completeness, however, we will briefly include in the appendix the two particle cfp's involved, which may enter in other
applications.
In general the two particle cfp's for $S_6$ are very complicated. One has to deal with $6!$ states \cite{Vergados15}. We will classify them in terms of two particle basis states, which are either symmetric $S\Leftrightarrow [f]=[2,0]$ and antisymmetric
$A\Leftrightarrow [f]=[1,1]$. Thus the 24 four particle orthonormal states are of the form:
$$(S\times S)[f_4]~,~[f_4]=[4],[3,1]_i,[2,2]_j~,~i=1,2,3~;~j=1,2$$
$$(A\times A)[f_4]~,~[f_4]=[1^4],[2,1^2]_i,[2,2]_j~,~i=1,2,3~;~j=1,2$$
$$(S\times A)[f_4]~,~[f_4]=[3,1]_i,[2,1^2]_j~,~i=1,2,3~;~j=1,2,3$$
$$(A\times S)[f_4]~,~[f_4]=[3,1]_i,[2,1^2]_j~,~i=1,2,3~;~j=1,2,3$$
The coefficients of fractional parentage take the form:
$$\left <(S_1\times S_2)[f_4]_k;[2]_{\ell }\}|[f_6]_n \right >~,~\left <(S_1\times S_2)[f_4]_k;[11]_{\ell }\}|[f_6]_n \right >~,~S_1=S,A~;~S_2=S,A~$$
where the indices $k,\ell ,n$ are merely labels to completely specify the states of the given symmetry. The index
$k$ takes values $1,2,...,15$ associated with the  allowed ordered pairs of six particles. The associated $[f_4]$ symmetry is understood to characterize the remaining particles. The orbital-spin-color symmetries $[f]=[2^2,1^2]~,~[2^3]$ are needed, when six quark clusters with the quantum numbers of two nucleons
are considered.

All cfp's have been obtained. We can summarize our results as follows:
$$(A\times A)[1^4]\times [1^2]\Rightarrow[1^6]+[2,1^4]+[2^2,1^2],$$
$$(S\times S)[4]\times [2]\Rightarrow[6]+[5,1]+[4,2],$$
$$(A\times A)[1^4]\times [2]\Rightarrow[3,1^3]+[2,1^4],$$
$$(S\times S)[4]\times [1^2]\Rightarrow[4,1^2]+[5,1],$$
$$(A\times A)[2,1^2]\times [1^2]\Rightarrow[2,1^4]+[2^2,1^2]+[3^2]+[3,1^3]+[3,2,1],$$
$$(S\times S)[3,1]\times [2]\Rightarrow[5,1]+[4,2]+[2^3]+[4,1^2]+[3,2,1],$$
$$(A\times A)[2,1^2]\times [2]\Rightarrow[5,1]+[4,2]+[2^3]+[4,1^2]+[3,2,1],$$
$$(S\times S)[3,1]\times [1^2]\Rightarrow[2,1^4]+[2^2,1^2]+[3^2]+[3,1^3]+[3,2,1],$$
 $$(A\times S)[3,1]\times [2]\Rightarrow[5,1]+[4,2]+[2^3]+[4,1^2]+[3,2,1],$$
 $$(S\times A)[3,1]\times [2]\Rightarrow[5,1]+[4,2]+[2^3]+[4,1^2]+[3,2,1],$$
 $$(S\times A)[3,1]\times [1^2]\Rightarrow[2,1^4]+[2^2,1^2]+[3^2]+[3,1^3]+[3,2,1],$$
 $$(A\times S)[3,1]\times [1^2]\Rightarrow[2,1^4]+[2^2,1^2]+[3^2]+[3,1^3]+[3,2,1],$$
 $$(A\times S)[2,1^2]\times [2]\Rightarrow[5,1]+[4,2]+[2^3]+[4,1^2]+[3,2,1],$$
 $$(S\times A)[2,1^2]\times [2]\Rightarrow[5,1]+[4,2]+[2^3]+[4,1^2]+[3,2,1],$$
 $$(A\times S)[2,1^2]\times [1^2]\Rightarrow[2,1^4]+[2^2,1^2]+[3^2]+[3,1^3]+[3,2,1],$$
 $$(S\times A)[2,1^2]\times [1^2]\Rightarrow[2,1^4]+[2^2,1^2]+[3^2]+[3,1^3]+[3,2,1],$$
 $$(A\times A)[2,2]\times [1^2]\Rightarrow[2^2,1^2]+[3^2]+[3,2,1],$$
 $$(S\times S)[2,2]\times [1^2]\Rightarrow[2^2,1^2]+[3^2]+[3,2,1],$$
 $$(A\times A)[2,2]\times [2]\Rightarrow[4,2]+[2^3]+[3,2,1],$$
 $$(S\times S)[2,2]\times [2]\Rightarrow[4,2]+[2^3]+[3,2,1],$$
 Taking into account that:
 $$\mbox{dim}([6])=\mbox{dim}([1^6])=1,~ \mbox{dim}[5,1])=\mbox{dim}([2,1^4])=5,\mbox{dim}[4,2])=\mbox{dim}([2^2,1^2])=9$$
$$  \mbox{dim}[4,1^2])=\mbox{dim}([3,1^3])=10 ~, ~ \mbox{dim}[3,3])=\mbox{dim}([2^6])=5 ~,$$ $$ \mbox{ dim}([321])=16\mbox{ (self-conjugate)}$$
 we find that the first four couplings yield $15\times 15$ matrices of cfp's,
 the next 12 yield $45\times 45$ matrices and in the last 4 the cfp matrices are
 $30\times 30$. These matrices were chosen to be orthogonal. Thus the $6!$=720-dimensional space has been reduced
to the above subspaces. No further reduction in
 dimensions seems possible. There exist, of course, symmetry relations among some of these matrices, but
in our formalism we did not bother to relate them this way.
  For lack of space  we present only the $30\times 30$ matrices of cfp's in the  case of the two simple one
dimensional  $[f_4]=[4]$ and $[f_4]=[1^4]$
  in tables \ref{table:ssym2}-\ref{table:asym2}
\newpage
\begin{table}
\caption{The coefficients of fractional parentage with $[f_4]=[4]$ coupled to the symmetric pairs
$$(p,q)=(1, 2), (1, 3), (1, 4), (1, 5), (1, 6), (2, 3), (2, 4), (2, 5), (2, 6), (3,
4), (3, 5), (3, 6),$$ $$ (4, 5), (4, 6), (5, 6),$$
 which label the rows. The resulting six particle symmetries, labeling the columns,
are given in the order
$[f_6]=[6],[5,1]_i,[4,2]_j~;~i=1,2,..,5~,~j=1,2,...,9$. The orbital-spin-color
 symmetry $[f]=[2^2,1^2]$, encountered in the present calculation, does not arise this way.
\label{table:ssym2}
}
\end{table}
{\tiny
$$
\left(
\begin{array}{lllllllllllllll}
 \frac{1}{\sqrt{15}} & 0 & 0 & -\frac{1}{\sqrt{5}} & -\frac{3}{4
   \sqrt{5}} & \frac{1}{4 \sqrt{3}} & \frac{1}{\sqrt{6}} &
   \frac{1}{\sqrt{30}} & \frac{1}{\sqrt{70}} &
   \sqrt{\frac{3}{35}} & \sqrt{\frac{3}{65}} &
   \frac{\sqrt{\frac{3}{26}}}{2} & -\frac{1}{2 \sqrt{2}} &
   -\frac{1}{4} & -\frac{\sqrt{\frac{3}{5}}}{4} \\
 \frac{1}{\sqrt{15}} & 0 & 0 & -\frac{1}{\sqrt{5}} & \frac{1}{2
   \sqrt{5}} & -\frac{1}{2 \sqrt{3}} & \frac{1}{\sqrt{6}} &
   \frac{1}{\sqrt{30}} & \frac{1}{\sqrt{70}} &
   -\frac{4}{\sqrt{105}} & -\frac{4}{\sqrt{195}} &
   -\sqrt{\frac{2}{39}} & \frac{1}{3 \sqrt{2}} & \frac{1}{6} &
   \frac{1}{2 \sqrt{15}} \\
 \frac{1}{\sqrt{15}} & 0 & \frac{\sqrt{3}}{4} & -\frac{1}{4
   \sqrt{5}} & -\frac{1}{2 \sqrt{5}} & -\frac{1}{2 \sqrt{3}} & 0
   & -\sqrt{\frac{3}{10}} & -\frac{3}{\sqrt{70}} &
   \frac{\sqrt{\frac{3}{35}}}{2} & \frac{\sqrt{\frac{3}{65}}}{2}
   & -\frac{5}{2 \sqrt{78}} & \frac{1}{6 \sqrt{2}} & \frac{1}{12}
   & -\frac{\sqrt{\frac{3}{5}}}{4} \\
 \frac{1}{\sqrt{15}} & \frac{1}{\sqrt{6}} & -\frac{1}{4 \sqrt{3}}
   & -\frac{1}{4 \sqrt{5}} & -\frac{1}{2 \sqrt{5}} & -\frac{1}{2
   \sqrt{3}} & -\frac{1}{\sqrt{6}} & \sqrt{\frac{2}{15}} &
   -\frac{3}{\sqrt{70}} & \frac{\sqrt{\frac{3}{35}}}{2} &
   -\frac{7}{2 \sqrt{195}} & \frac{\sqrt{\frac{3}{26}}}{2} &
   \frac{1}{6 \sqrt{2}} & -\frac{1}{6} & \frac{1}{2 \sqrt{15}} \\
 \frac{1}{\sqrt{15}} & \frac{1}{\sqrt{6}} & \frac{1}{2 \sqrt{3}}
   & -\frac{1}{2 \sqrt{5}} & \frac{1}{4 \sqrt{5}} & \frac{1}{4
   \sqrt{3}} & -\frac{1}{\sqrt{6}} & -\frac{1}{\sqrt{30}} & 2
   \sqrt{\frac{2}{35}} & -\frac{2}{\sqrt{105}} &
   \sqrt{\frac{3}{65}} & \frac{\sqrt{\frac{3}{26}}}{2} &
   -\frac{1}{6 \sqrt{2}} & \frac{1}{6} & \frac{1}{2 \sqrt{15}} \\
 \frac{1}{\sqrt{15}} & -\frac{1}{\sqrt{6}} & -\frac{1}{2
   \sqrt{3}} & -\frac{1}{2 \sqrt{5}} & \frac{1}{4 \sqrt{5}} &
   \frac{1}{4 \sqrt{3}} & -\frac{1}{\sqrt{6}} &
   -\frac{1}{\sqrt{30}} & -\frac{1}{\sqrt{70}} &
   -\sqrt{\frac{3}{35}} & -\sqrt{\frac{3}{65}} &
   -\frac{\sqrt{\frac{3}{26}}}{2} & -\frac{1}{2 \sqrt{2}} &
   -\frac{1}{4} & -\frac{\sqrt{\frac{3}{5}}}{4} \\
 \frac{1}{\sqrt{15}} & -\frac{1}{\sqrt{6}} & \frac{1}{4 \sqrt{3}}
   & \frac{1}{4 \sqrt{5}} & -\frac{3}{4 \sqrt{5}} & \frac{1}{4
   \sqrt{3}} & 0 & 0 & 0 & 0 & 0 & 0 & 0 & 0 & \sqrt{\frac{3}{5}}
   \\
 \frac{1}{\sqrt{15}} & 0 & -\frac{\sqrt{3}}{4} & \frac{1}{4
   \sqrt{5}} & -\frac{3}{4 \sqrt{5}} & \frac{1}{4 \sqrt{3}} & 0 &
   0 & 0 & 0 & 0 & 0 & 0 & \frac{3}{4} &
   -\frac{\sqrt{\frac{3}{5}}}{4} \\
 \frac{1}{\sqrt{15}} & 0 & 0 & 0 & 0 & \frac{1}{\sqrt{3}} & 0 & 0
   & 0 & 0 & 0 & 0 & \frac{1}{\sqrt{2}} & -\frac{1}{4} &
   -\frac{\sqrt{\frac{3}{5}}}{4} \\
 \frac{1}{\sqrt{15}} & -\frac{1}{\sqrt{6}} & \frac{1}{4 \sqrt{3}}
   & \frac{1}{4 \sqrt{5}} & \frac{1}{2 \sqrt{5}} & -\frac{1}{2
   \sqrt{3}} & 0 & 0 & 0 & 0 & 0 & \frac{\sqrt{\frac{13}{6}}}{2}
   & \frac{1}{6 \sqrt{2}} & \frac{1}{12} &
   -\frac{\sqrt{\frac{3}{5}}}{4} \\
 \frac{1}{\sqrt{15}} & 0 & -\frac{\sqrt{3}}{4} & \frac{1}{4
   \sqrt{5}} & \frac{1}{2 \sqrt{5}} & -\frac{1}{2 \sqrt{3}} & 0 &
   0 & 0 & 0 & 2 \sqrt{\frac{5}{39}} &
   -\frac{\sqrt{\frac{3}{26}}}{2} & \frac{1}{6 \sqrt{2}} &
   -\frac{1}{6} & \frac{1}{2 \sqrt{15}} \\
 \frac{1}{\sqrt{15}} & 0 & 0 & 0 & \frac{\sqrt{5}}{4} &
   \frac{1}{4 \sqrt{3}} & 0 & 0 & 0 & \sqrt{\frac{7}{15}} &
   -\sqrt{\frac{3}{65}} & -\frac{\sqrt{\frac{3}{26}}}{2} &
   -\frac{1}{6 \sqrt{2}} & \frac{1}{6} & \frac{1}{2 \sqrt{15}} \\
 \frac{1}{\sqrt{15}} & 0 & 0 & \frac{1}{\sqrt{5}} & -\frac{1}{2
   \sqrt{5}} & -\frac{1}{2 \sqrt{3}} & 0 & 0 &
   \sqrt{\frac{5}{14}} & \frac{1}{\sqrt{105}} &
   -\frac{4}{\sqrt{195}} & -\sqrt{\frac{2}{39}} & 0 &
   -\frac{1}{4} & -\frac{\sqrt{\frac{3}{5}}}{4} \\
 \frac{1}{\sqrt{15}} & 0 & \frac{\sqrt{3}}{4} & \frac{3}{4
   \sqrt{5}} & \frac{1}{4 \sqrt{5}} & \frac{1}{4 \sqrt{3}} & 0 &
   \sqrt{\frac{3}{10}} & -\sqrt{\frac{2}{35}} &
   -\frac{\sqrt{\frac{5}{21}}}{2} & \frac{\sqrt{\frac{5}{39}}}{2}
   & -\sqrt{\frac{2}{39}} & -\frac{1}{3 \sqrt{2}} & \frac{1}{12}
   & -\frac{\sqrt{\frac{3}{5}}}{4} \\
 \frac{1}{\sqrt{15}} & \frac{1}{\sqrt{6}} & -\frac{1}{4 \sqrt{3}}
   & \frac{3}{4 \sqrt{5}} & \frac{1}{4 \sqrt{5}} & \frac{1}{4
   \sqrt{3}} & \frac{1}{\sqrt{6}} & -\sqrt{\frac{2}{15}} &
   -\sqrt{\frac{2}{35}} & -\frac{\sqrt{\frac{5}{21}}}{2} &
   -\frac{\sqrt{\frac{5}{39}}}{2} & \sqrt{\frac{2}{39}} &
   -\frac{1}{3 \sqrt{2}} & -\frac{1}{6} & \frac{1}{2 \sqrt{15}}
\end{array}
\right)$$
}
\newpage
\begin{table}
\caption{The coefficients of fractional parentage with $[f_4]=[4]$ with the antisymmetric pairs
$$(p,q)=(1, 2), (1, 3), (1, 4), (1, 5), (1, 6), (2, 3), (2, 4), (2, 5), (2, 6), (3,
4), (3, 5), (3, 6),$$ $$ (4, 5), (4, 6), (5, 6),$$
 which label the rows. The resulting six particle symmetries, labeling the columns,
are given in the order
$[f_6]=[5,1]_i,[4,1^2]_j~;~i=1,2,..,5~,~j=1,2,...,10$. The orbital-spin-color
 symmetry $[f]=[2^2,1^2]$, encountered in the present calculation, does not arise this way.
\label{table:santisym2}
}
\end{table}
{\tiny
$$
\left(
\begin{array}{lllllllllllllll}
 0 & 0 & 0 & -\frac{1}{2} & \frac{1}{2 \sqrt{3}} & 0 & 0 & 0 & 0
   & 0 & 0 & \frac{\sqrt{\frac{5}{3}}}{2} & \frac{\sqrt{5}}{6} &
   \frac{\sqrt{\frac{5}{2}}}{6} & \frac{1}{2 \sqrt{6}} \\
 0 & 0 & -\frac{\sqrt{2}}{3} & -\frac{1}{6} & \frac{1}{2
   \sqrt{3}} & 0 & 0 & 0 & \sqrt{\frac{2}{5}} &
   \sqrt{\frac{2}{15}} & \frac{1}{\sqrt{15}} & -\frac{1}{2
   \sqrt{15}} & -\frac{1}{6 \sqrt{5}} & -\frac{1}{6 \sqrt{10}} &
   -\frac{1}{2 \sqrt{6}} \\
 0 & -\frac{\sqrt{\frac{5}{6}}}{2} & -\frac{1}{6 \sqrt{2}} &
   -\frac{1}{6} & \frac{1}{2 \sqrt{3}} & 0 &
   \frac{\sqrt{\frac{3}{2}}}{2} & \frac{1}{2 \sqrt{2}} &
   -\frac{1}{2 \sqrt{10}} & -\frac{1}{2 \sqrt{30}} &
   -\frac{1}{\sqrt{15}} & -\frac{1}{2 \sqrt{15}} & -\frac{1}{6
   \sqrt{5}} & -\frac{\sqrt{\frac{2}{5}}}{3} & 0 \\
 -\frac{1}{\sqrt{5}} & -\frac{1}{2 \sqrt{30}} & -\frac{1}{6
   \sqrt{2}} & -\frac{1}{6} & \frac{1}{2 \sqrt{3}} &
   \frac{1}{\sqrt{3}} & -\frac{1}{2 \sqrt{6}} & -\frac{1}{2
   \sqrt{2}} & -\frac{1}{2 \sqrt{10}} &
   -\frac{\sqrt{\frac{3}{10}}}{2} & 0 & -\frac{1}{2 \sqrt{15}} &
   -\frac{1}{2 \sqrt{5}} & 0 & 0 \\
 0 & 0 & 0 & 0 & \frac{1}{\sqrt{3}} & -\frac{1}{\sqrt{3}} &
   -\frac{1}{\sqrt{6}} & 0 & -\frac{1}{\sqrt{10}} & 0 & 0 &
   -\frac{1}{\sqrt{15}} & 0 & 0 & 0 \\
 0 & 0 & -\frac{\sqrt{2}}{3} & \frac{1}{3} & 0 & 0 & 0 & 0 & 0 &
   0 & 0 & 0 & 0 & 0 & \sqrt{\frac{2}{3}} \\
 0 & -\frac{\sqrt{\frac{5}{6}}}{2} & -\frac{1}{6 \sqrt{2}} &
   \frac{1}{3} & 0 & 0 & 0 & 0 & 0 & 0 & 0 & 0 & 0 &
   \frac{\sqrt{\frac{5}{2}}}{2} & -\frac{1}{2 \sqrt{6}} \\
 -\frac{1}{\sqrt{5}} & -\frac{1}{2 \sqrt{30}} & -\frac{1}{6
   \sqrt{2}} & \frac{1}{3} & 0 & 0 & 0 & 0 & 0 & 0 & 0 & 0 &
   \frac{\sqrt{5}}{3} & -\frac{\sqrt{\frac{5}{2}}}{6} &
   -\frac{1}{2 \sqrt{6}} \\
 0 & 0 & 0 & \frac{1}{2} & \frac{1}{2 \sqrt{3}} & 0 & 0 & 0 & 0 &
   0 & 0 & \frac{\sqrt{\frac{5}{3}}}{2} & -\frac{\sqrt{5}}{6} &
   -\frac{\sqrt{\frac{5}{2}}}{6} & -\frac{1}{2 \sqrt{6}} \\
 0 & -\frac{\sqrt{\frac{5}{6}}}{2} & \frac{1}{2 \sqrt{2}} & 0 & 0
   & 0 & 0 & 0 & 0 & 0 & \sqrt{\frac{3}{5}} & 0 & 0 & -\frac{1}{2
   \sqrt{10}} & \frac{1}{2 \sqrt{6}} \\
 -\frac{1}{\sqrt{5}} & -\frac{1}{2 \sqrt{30}} & \frac{1}{2
   \sqrt{2}} & 0 & 0 & 0 & 0 & 0 & 0 & 2 \sqrt{\frac{2}{15}} &
   -\frac{1}{\sqrt{15}} & 0 & -\frac{1}{3 \sqrt{5}} & \frac{1}{6
   \sqrt{10}} & \frac{1}{2 \sqrt{6}} \\
 0 & 0 & \frac{\sqrt{2}}{3} & \frac{1}{6} & \frac{1}{2 \sqrt{3}}
   & 0 & 0 & 0 & \sqrt{\frac{2}{5}} & -\sqrt{\frac{2}{15}} &
   -\frac{1}{\sqrt{15}} & -\frac{1}{2 \sqrt{15}} & \frac{1}{6
   \sqrt{5}} & \frac{1}{6 \sqrt{10}} & \frac{1}{2 \sqrt{6}} \\
 -\frac{1}{\sqrt{5}} & \sqrt{\frac{2}{15}} & 0 & 0 & 0 & 0 & 0 &
   \frac{1}{\sqrt{2}} & 0 & -\frac{1}{\sqrt{30}} &
   \frac{1}{\sqrt{15}} & 0 & -\frac{1}{3 \sqrt{5}} &
   \frac{\sqrt{\frac{2}{5}}}{3} & 0 \\
 0 & \frac{\sqrt{\frac{5}{6}}}{2} & \frac{1}{6 \sqrt{2}} &
   \frac{1}{6} & \frac{1}{2 \sqrt{3}} & 0 &
   \frac{\sqrt{\frac{3}{2}}}{2} & -\frac{1}{2 \sqrt{2}} &
   -\frac{1}{2 \sqrt{10}} & \frac{1}{2 \sqrt{30}} &
   \frac{1}{\sqrt{15}} & -\frac{1}{2 \sqrt{15}} & \frac{1}{6
   \sqrt{5}} & \frac{\sqrt{\frac{2}{5}}}{3} & 0 \\
 \frac{1}{\sqrt{5}} & \frac{1}{2 \sqrt{30}} & \frac{1}{6
   \sqrt{2}} & \frac{1}{6} & \frac{1}{2 \sqrt{3}} &
   \frac{1}{\sqrt{3}} & -\frac{1}{2 \sqrt{6}} & \frac{1}{2
   \sqrt{2}} & -\frac{1}{2 \sqrt{10}} &
   \frac{\sqrt{\frac{3}{10}}}{2} & 0 & -\frac{1}{2 \sqrt{15}} &
   \frac{1}{2 \sqrt{5}} & 0 & 0
\end{array}
\right)
$$
}
\newpage
\begin{table}
\caption{The cfp's with $[f_4]=[1^4]$ involving the antisymmetric pairs
$$(p,q)=(1, 2), (1, 3), (1, 4), (1, 5), (1, 6), (2, 3), (2, 4), (2, 5), (2, 6), (3,
4), (3, 5), (3, 6),$$ $$ (4, 5), (4, 6), (5, 6),$$
 which label the rows. The resulting six particle symmetries, labeling the columns,
are given in the order
$[f_6]=[1^6],[2,1^4]_i,[2^2,1^2]_j~;~i=1,2,..,5~,~j=1,2,...,9.$
\label{table:aantisym2}
}
{\tiny
$$
\left(
\begin{array}{lllllllllllllll}
 \frac{1}{\sqrt{15}} & 0 & 0 & -\frac{1}{\sqrt{5}} & -\frac{3}{4
   \sqrt{5}} & -\frac{1}{4 \sqrt{3}} & \frac{1}{\sqrt{6}} &
   -\frac{1}{\sqrt{30}} & \frac{1}{\sqrt{70}} &
   \sqrt{\frac{3}{35}} & -\sqrt{\frac{3}{65}} &
   \frac{\sqrt{\frac{3}{26}}}{2} & \frac{1}{2 \sqrt{2}} &
   -\frac{1}{4} & \frac{\sqrt{\frac{3}{5}}}{4} \\
 -\frac{1}{\sqrt{15}} & 0 & 0 & \frac{1}{\sqrt{5}} & -\frac{1}{2
   \sqrt{5}} & -\frac{1}{2 \sqrt{3}} & -\frac{1}{\sqrt{6}} &
   \frac{1}{\sqrt{30}} & -\frac{1}{\sqrt{70}} &
   \frac{4}{\sqrt{105}} & -\frac{4}{\sqrt{195}} &
   \sqrt{\frac{2}{39}} & \frac{1}{3 \sqrt{2}} & -\frac{1}{6} &
   \frac{1}{2 \sqrt{15}} \\
 \frac{1}{\sqrt{15}} & 0 & -\frac{\sqrt{3}}{4} & -\frac{1}{4
   \sqrt{5}} & -\frac{1}{2 \sqrt{5}} & \frac{1}{2 \sqrt{3}} & 0 &
   \sqrt{\frac{3}{10}} & -\frac{3}{\sqrt{70}} &
   \frac{\sqrt{\frac{3}{35}}}{2} & -\frac{\sqrt{\frac{3}{65}}}{2}
   & -\frac{5}{2 \sqrt{78}} & -\frac{1}{6 \sqrt{2}} &
   \frac{1}{12} & \frac{\sqrt{\frac{3}{5}}}{4} \\
 -\frac{1}{\sqrt{15}} & -\frac{1}{\sqrt{6}} & -\frac{1}{4
   \sqrt{3}} & \frac{1}{4 \sqrt{5}} & \frac{1}{2 \sqrt{5}} &
   -\frac{1}{2 \sqrt{3}} & \frac{1}{\sqrt{6}} &
   \sqrt{\frac{2}{15}} & \frac{3}{\sqrt{70}} &
   -\frac{\sqrt{\frac{3}{35}}}{2} & -\frac{7}{2 \sqrt{195}} &
   -\frac{\sqrt{\frac{3}{26}}}{2} & \frac{1}{6 \sqrt{2}} &
   \frac{1}{6} & \frac{1}{2 \sqrt{15}} \\
 \frac{1}{\sqrt{15}} & \frac{1}{\sqrt{6}} & -\frac{1}{2 \sqrt{3}}
   & -\frac{1}{2 \sqrt{5}} & \frac{1}{4 \sqrt{5}} & -\frac{1}{4
   \sqrt{3}} & -\frac{1}{\sqrt{6}} & \frac{1}{\sqrt{30}} & 2
   \sqrt{\frac{2}{35}} & -\frac{2}{\sqrt{105}} &
   -\sqrt{\frac{3}{65}} & \frac{\sqrt{\frac{3}{26}}}{2} &
   \frac{1}{6 \sqrt{2}} & \frac{1}{6} & -\frac{1}{2 \sqrt{15}} \\
 \frac{1}{\sqrt{15}} & -\frac{1}{\sqrt{6}} & \frac{1}{2 \sqrt{3}}
   & -\frac{1}{2 \sqrt{5}} & \frac{1}{4 \sqrt{5}} & -\frac{1}{4
   \sqrt{3}} & -\frac{1}{\sqrt{6}} & \frac{1}{\sqrt{30}} &
   -\frac{1}{\sqrt{70}} & -\sqrt{\frac{3}{35}} &
   \sqrt{\frac{3}{65}} & -\frac{\sqrt{\frac{3}{26}}}{2} &
   \frac{1}{2 \sqrt{2}} & -\frac{1}{4} &
   \frac{\sqrt{\frac{3}{5}}}{4} \\
 -\frac{1}{\sqrt{15}} & \frac{1}{\sqrt{6}} & \frac{1}{4 \sqrt{3}}
   & -\frac{1}{4 \sqrt{5}} & \frac{3}{4 \sqrt{5}} & \frac{1}{4
   \sqrt{3}} & 0 & 0 & 0 & 0 & 0 & 0 & 0 & 0 & \sqrt{\frac{3}{5}}
   \\
 \frac{1}{\sqrt{15}} & 0 & \frac{\sqrt{3}}{4} & \frac{1}{4
   \sqrt{5}} & -\frac{3}{4 \sqrt{5}} & -\frac{1}{4 \sqrt{3}} & 0
   & 0 & 0 & 0 & 0 & 0 & 0 & \frac{3}{4} &
   \frac{\sqrt{\frac{3}{5}}}{4} \\
 -\frac{1}{\sqrt{15}} & 0 & 0 & 0 & 0 & \frac{1}{\sqrt{3}} & 0 &
   0 & 0 & 0 & 0 & 0 & \frac{1}{\sqrt{2}} & \frac{1}{4} &
   -\frac{\sqrt{\frac{3}{5}}}{4} \\
 \frac{1}{\sqrt{15}} & -\frac{1}{\sqrt{6}} & -\frac{1}{4
   \sqrt{3}} & \frac{1}{4 \sqrt{5}} & \frac{1}{2 \sqrt{5}} &
   \frac{1}{2 \sqrt{3}} & 0 & 0 & 0 & 0 & 0 &
   \frac{\sqrt{\frac{13}{6}}}{2} & -\frac{1}{6 \sqrt{2}} &
   \frac{1}{12} & \frac{\sqrt{\frac{3}{5}}}{4} \\
 -\frac{1}{\sqrt{15}} & 0 & -\frac{\sqrt{3}}{4} & -\frac{1}{4
   \sqrt{5}} & -\frac{1}{2 \sqrt{5}} & -\frac{1}{2 \sqrt{3}} & 0
   & 0 & 0 & 0 & 2 \sqrt{\frac{5}{39}} &
   \frac{\sqrt{\frac{3}{26}}}{2} & \frac{1}{6 \sqrt{2}} &
   \frac{1}{6} & \frac{1}{2 \sqrt{15}} \\
 \frac{1}{\sqrt{15}} & 0 & 0 & 0 & \frac{\sqrt{5}}{4} &
   -\frac{1}{4 \sqrt{3}} & 0 & 0 & 0 & \sqrt{\frac{7}{15}} &
   \sqrt{\frac{3}{65}} & -\frac{\sqrt{\frac{3}{26}}}{2} &
   \frac{1}{6 \sqrt{2}} & \frac{1}{6} & -\frac{1}{2 \sqrt{15}} \\
 \frac{1}{\sqrt{15}} & 0 & 0 & \frac{1}{\sqrt{5}} & -\frac{1}{2
   \sqrt{5}} & \frac{1}{2 \sqrt{3}} & 0 & 0 & \sqrt{\frac{5}{14}}
   & \frac{1}{\sqrt{105}} & \frac{4}{\sqrt{195}} &
   -\sqrt{\frac{2}{39}} & 0 & -\frac{1}{4} &
   \frac{\sqrt{\frac{3}{5}}}{4} \\
 -\frac{1}{\sqrt{15}} & 0 & \frac{\sqrt{3}}{4} & -\frac{3}{4
   \sqrt{5}} & -\frac{1}{4 \sqrt{5}} & \frac{1}{4 \sqrt{3}} & 0 &
   \sqrt{\frac{3}{10}} & \sqrt{\frac{2}{35}} &
   \frac{\sqrt{\frac{5}{21}}}{2} & \frac{\sqrt{\frac{5}{39}}}{2}
   & \sqrt{\frac{2}{39}} & -\frac{1}{3 \sqrt{2}} & -\frac{1}{12}
   & -\frac{\sqrt{\frac{3}{5}}}{4} \\
 \frac{1}{\sqrt{15}} & \frac{1}{\sqrt{6}} & \frac{1}{4 \sqrt{3}}
   & \frac{3}{4 \sqrt{5}} & \frac{1}{4 \sqrt{5}} & -\frac{1}{4
   \sqrt{3}} & \frac{1}{\sqrt{6}} & \sqrt{\frac{2}{15}} &
   -\sqrt{\frac{2}{35}} & -\frac{\sqrt{\frac{5}{21}}}{2} &
   \frac{\sqrt{\frac{5}{39}}}{2} & \sqrt{\frac{2}{39}} &
   \frac{1}{3 \sqrt{2}} & -\frac{1}{6} & -\frac{1}{2 \sqrt{15}}
\end{array}
\right)
$$
}
\end{table}
\begin{table}
\caption{The cfp's with $[f_4]=[1^4]$ involving the symmetric pairs
$$(p,q)=(1, 2), (1, 3), (1, 4), (1, 5), (1, 6), (2, 3), (2, 4), (2, 5), (2, 6), (3,
4), (3, 5), (3, 6),$$ $$ (4, 5), (4, 6), (5, 6),$$
 which label the rows. The resulting six particle symmetries, labeling the columns,
are given in the order
$[f_6]=[2,1^4]_i,[3,1^3]_j~;~i=1,2,..,5~,~j=1,2,...,10.$ The orbital-spin-color
 symmetry $[f]=[2^2,1^2]$, encountered in the present calculation, does not arise this way.}
\label{table:asym2}
{\tiny
$$
\left(
\begin{array}{lllllllllllllll}
 0 & 0 & 0 & \frac{1}{2} & \frac{1}{2 \sqrt{3}} & 0 & 0 & 0 & 0 &
   0 & 0 & -\frac{\sqrt{\frac{5}{3}}}{2} & \frac{\sqrt{5}}{6} &
   -\frac{\sqrt{\frac{5}{2}}}{6} & \frac{1}{2 \sqrt{6}} \\
 0 & 0 & \frac{\sqrt{2}}{3} & -\frac{1}{6} & -\frac{1}{2
   \sqrt{3}} & 0 & 0 & 0 & -\sqrt{\frac{2}{5}} &
   \sqrt{\frac{2}{15}} & -\frac{1}{\sqrt{15}} & -\frac{1}{2
   \sqrt{15}} & \frac{1}{6 \sqrt{5}} & -\frac{1}{6 \sqrt{10}} &
   \frac{1}{2 \sqrt{6}} \\
 0 & \frac{\sqrt{\frac{5}{6}}}{2} & -\frac{1}{6 \sqrt{2}} &
   \frac{1}{6} & \frac{1}{2 \sqrt{3}} & 0 &
   -\frac{\sqrt{\frac{3}{2}}}{2} & \frac{1}{2 \sqrt{2}} &
   -\frac{1}{2 \sqrt{10}} & \frac{1}{2 \sqrt{30}} &
   -\frac{1}{\sqrt{15}} & \frac{1}{2 \sqrt{15}} & -\frac{1}{6
   \sqrt{5}} & \frac{\sqrt{\frac{2}{5}}}{3} & 0 \\
 \frac{1}{\sqrt{5}} & -\frac{1}{2 \sqrt{30}} & \frac{1}{6
   \sqrt{2}} & -\frac{1}{6} & -\frac{1}{2 \sqrt{3}} &
   -\frac{1}{\sqrt{3}} & -\frac{1}{2 \sqrt{6}} & \frac{1}{2
   \sqrt{2}} & \frac{1}{2 \sqrt{10}} &
   -\frac{\sqrt{\frac{3}{10}}}{2} & 0 & -\frac{1}{2 \sqrt{15}} &
   \frac{1}{2 \sqrt{5}} & 0 & 0 \\
 0 & 0 & 0 & 0 & \frac{1}{\sqrt{3}} & -\frac{1}{\sqrt{3}} &
   \frac{1}{\sqrt{6}} & 0 & -\frac{1}{\sqrt{10}} & 0 & 0 &
   \frac{1}{\sqrt{15}} & 0 & 0 & 0 \\
 0 & 0 & -\frac{\sqrt{2}}{3} & -\frac{1}{3} & 0 & 0 & 0 & 0 & 0 &
   0 & 0 & 0 & 0 & 0 & \sqrt{\frac{2}{3}} \\
 0 & -\frac{\sqrt{\frac{5}{6}}}{2} & \frac{1}{6 \sqrt{2}} &
   \frac{1}{3} & 0 & 0 & 0 & 0 & 0 & 0 & 0 & 0 & 0 &
   \frac{\sqrt{\frac{5}{2}}}{2} & \frac{1}{2 \sqrt{6}} \\
 -\frac{1}{\sqrt{5}} & \frac{1}{2 \sqrt{30}} & -\frac{1}{6
   \sqrt{2}} & -\frac{1}{3} & 0 & 0 & 0 & 0 & 0 & 0 & 0 & 0 &
   \frac{\sqrt{5}}{3} & \frac{\sqrt{\frac{5}{2}}}{6} &
   -\frac{1}{2 \sqrt{6}} \\
 0 & 0 & 0 & \frac{1}{2} & -\frac{1}{2 \sqrt{3}} & 0 & 0 & 0 & 0
   & 0 & 0 & \frac{\sqrt{\frac{5}{3}}}{2} & \frac{\sqrt{5}}{6} &
   -\frac{\sqrt{\frac{5}{2}}}{6} & \frac{1}{2 \sqrt{6}} \\
 0 & \frac{\sqrt{\frac{5}{6}}}{2} & \frac{1}{2 \sqrt{2}} & 0 & 0
   & 0 & 0 & 0 & 0 & 0 & \sqrt{\frac{3}{5}} & 0 & 0 & \frac{1}{2
   \sqrt{10}} & \frac{1}{2 \sqrt{6}} \\
 \frac{1}{\sqrt{5}} & -\frac{1}{2 \sqrt{30}} & -\frac{1}{2
   \sqrt{2}} & 0 & 0 & 0 & 0 & 0 & 0 & 2 \sqrt{\frac{2}{15}} &
   \frac{1}{\sqrt{15}} & 0 & \frac{1}{3 \sqrt{5}} & \frac{1}{6
   \sqrt{10}} & -\frac{1}{2 \sqrt{6}} \\
 0 & 0 & \frac{\sqrt{2}}{3} & -\frac{1}{6} & \frac{1}{2 \sqrt{3}}
   & 0 & 0 & 0 & \sqrt{\frac{2}{5}} & \sqrt{\frac{2}{15}} &
   -\frac{1}{\sqrt{15}} & \frac{1}{2 \sqrt{15}} & \frac{1}{6
   \sqrt{5}} & -\frac{1}{6 \sqrt{10}} & \frac{1}{2 \sqrt{6}} \\
 -\frac{1}{\sqrt{5}} & -\sqrt{\frac{2}{15}} & 0 & 0 & 0 & 0 & 0 &
   \frac{1}{\sqrt{2}} & 0 & \frac{1}{\sqrt{30}} &
   \frac{1}{\sqrt{15}} & 0 & -\frac{1}{3 \sqrt{5}} &
   -\frac{\sqrt{\frac{2}{5}}}{3} & 0 \\
 0 & \frac{\sqrt{\frac{5}{6}}}{2} & -\frac{1}{6 \sqrt{2}} &
   \frac{1}{6} & -\frac{1}{2 \sqrt{3}} & 0 &
   \frac{\sqrt{\frac{3}{2}}}{2} & \frac{1}{2 \sqrt{2}} &
   \frac{1}{2 \sqrt{10}} & \frac{1}{2 \sqrt{30}} &
   -\frac{1}{\sqrt{15}} & -\frac{1}{2 \sqrt{15}} & -\frac{1}{6
   \sqrt{5}} & \frac{\sqrt{\frac{2}{5}}}{3} & 0 \\
 \frac{1}{\sqrt{5}} & -\frac{1}{2 \sqrt{30}} & \frac{1}{6
   \sqrt{2}} & -\frac{1}{6} & \frac{1}{2 \sqrt{3}} &
   \frac{1}{\sqrt{3}} & \frac{1}{2 \sqrt{6}} & \frac{1}{2
   \sqrt{2}} & -\frac{1}{2 \sqrt{10}} &
   -\frac{\sqrt{\frac{3}{10}}}{2} & 0 & \frac{1}{2 \sqrt{15}} &
   \frac{1}{2 \sqrt{5}} & 0 & 0
\end{array}
\right)
$$
}
\end{table}
\end{document}